\documentstyle[12pt,psfig]{article}

\renewcommand{\thefootnote}{\fnsymbol{footnote} }
\newcommand{\s}{\\ \vspace*{-3mm} }
\newcommand{\nn}{\noindent}
\newcommand{\non}{\nonumber}
\newcommand{\ee}{e^+ e^-}
\newcommand{\ra}{\rightarrow}

\newcommand{\beq}{\begin{eqnarray}}
\newcommand{\eeq}{\end{eqnarray}}
\newcommand{\tb}{\mbox{\rm tg} \beta}
\newcommand{\lsim}{\raisebox{-0.13cm}{~\shortstack{$<$ \\[-0.07cm] $\sim$}}~}
\newcommand{\gsim}{\raisebox{-0.13cm}{~\shortstack{$>$ \\[-0.07cm] $\sim$}}~}

\begin{document}

\begin{flushright}
KA--TP--16--96\\
May 1996 \\
\end{flushright}

\begin{center}

{\large\sc {\bf  HIGGS PARTICLES}$^*$} 

\vspace{.5cm}

\nn {\it Conveners}: \\
\vspace*{0.2cm}

A.~Djouadi$^{1,2}$, H.E. Haber$^3$, P. Igo--Kemenes$^4$, P. Janot$^5$ 
and P.M.~Zerwas$^1$

\vspace*{3mm}

\nn {\it Working Group}: \\
\vspace*{0.2cm}

A. Arhrib$^6$, T. Binoth$^7$, E. Chopin$^8$, V. Driesen$^2$, W.
Hollik$^2$, C. J\"unger$^2$, J. Kalinowski$^9$, W. Kilian$^1$, B.R.
Kim$^{10}$, M. Kr\"amer$^1$, G. Kreyerhoff$^{10}$, G. Moultaka$^6$, S.K.
Oh$^{11}$, P. Ohmann$^{12}$, J. Rosiek$^2$ and J.J. van der Bij$^7$. 

\vspace{.7cm}

\begin{small}

$^1$ Deutsches Elektronen--Synchrotron DESY, D-22603 Hamburg, Germany. \\
\vspace{0.2cm}

$^2$ Institut f\"ur Theoretische Physik, Universit\"at Karlsruhe, \\
D-76128 Karlsruhe, Germany.\\
\vspace{0.2cm}

$^3$ Physikalisches Institut, Universit\"at Heidelberg, \\
D-6900 Heidelberg, Germany.\\
\vspace{0.2cm}

$^4$ Santa Cruz Institute for Particle Physics, University of California, 
\\ Santa Cruz, CA 95064, USA. \\
\vspace{0.2cm}

$^5$ PPE Division, CERN, CH--1211, Geneva 23, Switzerland.\\
\vspace{0.2cm}

$^6$ Physique Math\'ematique et Th\'eorique, E.S.A. du CNRS N$^o$ 5032, \\
Universit\'e Montpellier II, F--34095 Montpellier, France\\
\vspace{0.2cm}

$^7$ Albert--Ludwigs--Universit\"at Freiburg, Fakult\"at f\"ur Physik, \\
Hermann--Herder--Strasse 3, D--79104 Freiburg, Germany. 
\vspace{0.2cm}

$^8$ Laboratoire de Physique Th\'eorique ENSLAPP 
\\ B.P.110, 74941 Annecy-Le-Vieux Cedex, France \\
\vspace{0.2cm}

$^9$ Institute of Theoretical Physics,  Warsaw University, \\
PL--00681 Warsaw, Poland.
\vspace{0.2cm}

$^{10}$ III Physikalisches Institute A, RWTH Aachen, \\
D--52056 Aachen, Germany. 
\vspace{0.2cm}

$^{11}$ Institute for Advanced Physics, Kon--Kuk University, \\
Seoul 143--701, Korea. 
\vspace{0.2cm}

$^{12}$ Department of Theoretical Physics, Oxford University, \\
OX1 3NP Oxford, UK. 

\end{small}

\end{center}

\vspace*{.5cm}

\nn $^*$ To appear in the Proceedings of the Workshop {\it Physics with 
$\ee$ Linear Colliders}, Annecy--Gran Sasso--Hamburg, Feb. 4 -- Sept. 1,
1995, P.M. Zerwas (editor).

\newpage

\begin{center}

{\large\sc {\bf  HIGGS PARTICLES}}

\vspace*{0.3cm}

{\large\sc {\bf  Introduction and Summary}}

\vspace{.7cm}

{\sc A.~Djouadi$^{1,2}$} and {\sc P.M.~Zerwas$^1$}

\vspace{.5cm}

{\small

$^1$ Deutsches Elektronen--Synchrotron DESY, D-22603 Hamburg, FRG. \\
\vspace{0.2cm}

$^2$ Institut f\"ur Theoretische Physik, Universit\"at Karlsruhe,
D-76128 Karlsruhe, FRG. }

\end{center}

\vspace{.5cm}

The search for scalar Higgs particles and the exploration of the
mechanism which breaks the electroweak symmetry, will be one of the major
tasks at future high--energy $\ee$ colliders. In previous studies it has
been shown that $\ee$ linear colliders operating in the energy range
$\sqrt{s} \sim 300$ to $500$ GeV with a luminosity of $\int {\cal L}
\sim 20$ fb$^{-1}$ are   ideal machines to search for light Higgs
particles. \s 

In the Standard Model (SM) the Higgs mass range  $M_H \lsim 200$ GeV is
easy to cover at these energies. This intermediate Higgs mass range is one of
the theoretically most favored ranges, allowing the particles to remain
weakly interacting up to the GUT scale $\Lambda \sim
10^{16}$ GeV [a prerequisite for the perturbative renormalization of the
electroweak mixing angle from the GUT symmetry value 3/8 down to the
experimentally observed value at low energies]. The search of
intermediate-mass Higgs bosons can be carried out in three different
channels: the Higgs--strahlung process $\ee \ra ZH$ and the fusion
mechanisms $WW/ZZ \ra H$. The cross sections are large and the
properties of the Higgs boson, in particular spin--parity quantum
numbers and couplings to gauge bosons and fermions, can be thoroughly
investigated, allowing for crucial tests of the Higgs mechanism.  \s 

In the Minimal Supersymmetric Standard Model (MSSM) the Higgs sector is
extended to three neutral scalar and pseudoscalar particles $h/H$, $A$
and a pair of charged particles $H^\pm$. The lightest Higgs boson $h$
has a mass $M_h \lsim 140$ GeV and can be detected in the entire MSSM
parameter space either in the Higgs--strahlung process, $\ee \ra hZ$, or
by the complementary mechanism of associated production with the
pseudoscalar particle, $\ee \ra hA$. Moreover, there is a substantial
area in the MSSM parameter space where the heavy Higgs bosons can be
also found; for a total energy of 500 GeV this is possible if the $H,A$
and $H^\pm$ masses are less than 230 GeV. Similar to the SM, various
properties of these Higgs bosons can be investigated. \s 

Higher energies are required to sweep the entire mass range of the SM
Higgs particle, $M_H \lsim 1$ TeV. The high energies will also be needed
to produce and to study the heavy scalar particles in extensions of the SM,
such as the MSSM, if their masses are larger than $\sim 250$ GeV. Masses
of the heavy Higgs bosons in this range are suggested by grand unified
supersymmetric theories. In $\ee$ collisions, these experiments can be
performed in the second phase of the colliders with a c.m. energy up to
1.5 to 2 TeV. In this report, we analyze the potential of a 1.5 TeV
$\ee$ linear collider, with an integrated luminosity of $\int {\cal L}
\sim 200$ fb$^{-1}$ {\it per annum} to compensate for the drop of the
cross sections at high energies. We will study the heavy Higgs particles
in the Standard Model, in the minimal supersymmetric extension and in
other more speculative scenarios.

\bigskip

In the Standard Model, the main production mechanisms of Higgs
particles, $\ee \ra HZ$ and $\ee \ra \nu \bar{\nu}  H / \ee H$, will be
discussed and the cross sections, including the interference between the
Higgs--strahlung and the fusion processes, will be given [2]. The double
Higgs production process, in which the trilinear Higgs coupling can be
determined [therefore leading to the first non--trivial test of the
Higgs potential], will be discussed in Ref.[3]. Finally, possible
effects of New Physics beyond the SM on production cross sections and
angular distributions of Higgs bosons, will be summarized in Ref.[4].
Consequences of a model in which the Higgs boson interacts strongly with
scalar singlet fields in a hidden sector, are described in Ref.[5]. \s

Subsequently, we will investigate the properties of the heavy Higgs
particles in supersymmetric extensions of the SM. We will restrict
ourselves first to the minimal extension which is highly constrained,
parameterized by only two free parameters at the tree--level: a Higgs
mass parameter [generally the mass the pseudoscalar Higgs boson $M_A$]
and the ratio of the vacuum expectation values of the two doublet fields
responsible for the symmetry breaking, $\tb$, which in grand unified
supersymmetric models with Yukawa coupling unification is forced to be
either small, $\tb \sim 1.5$, or large, $\tb \sim 50$.  \s 

The various decay modes of the heavy CP--even Higgs boson $H$, the
pseudoscalar boson $A$ and the charged Higgs particles $H^\pm$ will be
analyzed in Ref.[6], in particular the decays into supersymmetric
particles, charginos, neutralinos, squarks and sleptons. The production
of the heavy Higgs particles, primarily through the channels $\ee \ra
HA$ and $H^+H^-$, will be also discussed in Ref.[6] and the complete
one--loop electroweak radiative corrections of the cross sections will
be summarized [7,8]. We will finally discuss the multiple production of
the SM and the light MSSM Higgs bosons in Refs.[9,10]. Some of these processes
will allow us to determine the fundamental Higgs trilinear couplings.\s 

A brief discussion of the Higgs sector in the next--to--minimal
supersymmetric extension of the Standard Model, Ref.[11], concludes 
this report.

\newpage

\makeatletter
\def\shortletter{%
  \setcounter{secnumdepth}{5}
  \def\paragraph{%
    \@startsection{paragraph}{4}{\parindent}%
      {3.25ex \@plus1ex \@minus.2ex}{-.5em}%
      {\reset@font\normalsize\bfseries}}%
  \renewcommand\theparagraph{\arabic{paragraph}.\hskip-.5em}
  \def\subparagraph{%
    \@startsection{subparagraph}{5}{\parindent}%
      {3.25ex \@plus1ex \@minus.2ex}{-.5em}%
      {\reset@font\normalsize\bfseries}}%
  \renewcommand\thesubparagraph{(\alph{subparagraph})\hskip-.5em}
}
%
\def\@citex[#1]#2{\if@filesw\immediate\write\@auxout{\string\citation{#2}}\fi
  \def\@citea{}\@cite{\@for\@citeb:=#2\do
    {\@citea\def\@citea{,\penalty\@m}\@ifundefined
       {b@\@citeb}{{\bf ?}\@warning
       {Citation `\@citeb' on page \thepage \space undefined}}%
\hbox{\csname b@\@citeb\endcsname}}}{#1}}
\def\citerange{\@ifnextchar [{\@tempswatrue\@citexr}{\@tempswafalse\@citexr[]}}
\def\@citexr[#1]#2{\if@filesw\immediate\write\@auxout{\string\citation{#2}}\fi
  \def\@citea{}\@cite{\@for\@citeb:=#2\do
    {\@citea\def\@citea{--\penalty\@m}\@ifundefined
       {b@\@citeb}{{\bf ?}\@warning
       {Citation `\@citeb' on page \thepage \space undefined}}%
\hbox{\csname b@\@citeb\endcsname}}}{#1}}
%
\long\def\@makecaption#1#2{%
  \vskip\abovecaptionskip
  \sbox\@tempboxa{#1: \emph{#2}}%
  \ifdim \wd\@tempboxa >\hsize
    #1: \emph{#2}\par
  \else
    \hbox to\hsize{\hfil\box\@tempboxa\hfil}%
  \fi
  \vskip\belowcaptionskip}
%
\def\fmslash{\@ifnextchar[{\fmsl@sh}{\fmsl@sh[0mu]}}
\def\fmsl@sh[#1]#2{%
  \mathchoice
    {\@fmsl@sh\displaystyle{#1}{#2}}%
    {\@fmsl@sh\textstyle{#1}{#2}}%
    {\@fmsl@sh\scriptstyle{#1}{#2}}%
    {\@fmsl@sh\scriptscriptstyle{#1}{#2}}}
\def\@fmsl@sh#1#2#3{\m@th\ooalign{$\hfil#1\mkern#2/\hfil$\crcr$#1#3$}}
\makeatother

\def\fmfL(#1,#2,#3)#4{\put(#1,#2){\makebox(0,0)[#3]{#4}}}

\def\hc{\mbox{h.c.}}
\def\op{{\cal O}}

\newcommand{\jnp}{Nucl.~Phys.~B}
\newcommand{\jpl}{Phys.~Lett.~}
\newcommand{\jzp}{Z.~Phys.~C}
\newcommand{\jpr}{Phys.~Rev.~D}
\newcommand{\jrmp}{Rev.~Mod.~Phys. }
\newcommand{\jprl}{Phys.~Rev.~Lett. }

\newcommand{\np}[3]{Nucl.~Phys.~B#1 (19#2) #3}
\newcommand{\npb}[3]{Nucl.~Phys.~#1 (19#2) #3}
\newcommand{\prl}[3]{Phys.~Rev.~Lett.~#1 (19#2) #3}
\newcommand{\pl}[3]{Phys.~Lett.~B#1 (19#2) #3}
\newcommand{\zp}[3]{Z.~Phys.~C#1 (19#2) #3}
\newcommand{\pr}[3]{Phys.~Rev.~D#1 (19#2) #3}
\newcommand{\ptp}[3]{Prog.~Theor.~Phys.~#1 (19#2) #3}
\newcommand{\apn}[3]{Ann.~Phys.~(N.Y.) #1 (19#2) #3}
\newcommand{\prp}[3]{Phys.~Rep.~#1C (19#2) #3}
\newcommand{\sj}[3]{Sov.~J.~Part.~Nucl.~#1 (19#2) #3}
\newcommand{\spj}[3]{Sov.~Phys.~JETP~#1 (19#2) #3}
\newcommand{\ij}[3]{Int.~J.~Mod.~Phys.~A#1 (19#2) #3}
\newcommand{\cpc}[3]{Comput.~Phys.~Commun.~#1 (19#2) #3}

\newcommand{\qt}[1]{``{\em #1}''}

\newcommand{\dx}{\mbox{\rm d}}
\newcommand{\SM}{\mbox{${\cal SM}$}}
\newcommand{\SUSY}{\mbox{${\cal SUSY}~$}}
\newcommand{\MSSM}{\mbox{${\cal MSSM}$}}
\newcommand{\SMG}{ {\rm SU(3) \times SU(2)\times U(1)}}
\newcommand{\tg}{\mbox{tg}}
\newcommand{\ctb}{\mbox{ctg}\beta}
\newcommand{\?}{ {\bf ???}}
\newcommand{\ctg}{\mbox{ctg}}
\def\inpb{\mbox{$\hbox{pb}^{-1}$}}
\def\Gcs{\hbox{GeV}/\mbox{$c^2$}}

\def\thefootnote{\fnsymbol{footnote}}

\begin{center}
 {\large\bf
 Higgs-strahlung and Vector Boson Fusion\\
 in $e^+e^-$ Collisions\\[\baselineskip]}
 {\sc W.~Kilian, M.~Kr\"amer, and P.M.~Zerwas}
 \\
 {\small Deutsches Elektronen-Synchrotron DESY, 
 D-22603 Hamburg/FRG}
\end{center}
\begin{abstract}
\nn Higgs-strahlung $e^+e^-\to ZH$ and $WW$ ($ZZ$) fusion
$e^+e^-\to\bar\nu_e\nu_e H$ ($e^+e^- H$) are the most important
mechanisms for the production of Higgs bosons at future $e^+e^-$
linear colliders.  We have calculated the cross sections and
energy/angular distributions of the Higgs boson for these production
mechanisms.  When the $Z$ boson decays into (electron-)neutrinos or
$e^+e^-$, the two production amplitudes interfere.  In the cross-over
region between the two mechanisms the interference term is positive
(negative) for $\bar\nu_e\nu_e$ ($e^+e^-$) decays, respectively, 
thus enhancing (reducing) the production rate.
\end{abstract}

\def\thefootnote{\arabic{footnote}}

\bigskip

\paragraph{}
The analysis of the mechanism which breaks the electroweak gauge
symmetry $SU(2)_{\rm L}\times U(1)_{\rm Y}$ down to $U(1)_{\rm EM}$,
is one of the key problems in particle physics.  If the gauge fields
involved remain weakly interacting up to high energies -- a
prerequisite for the (perturbative) renormalization of
$\sin^2\theta_W$ from the symmetry value $3/8$ of grand-unified
theories down to a value near $0.2$ at low energies -- fundamental
scalar Higgs bosons~\cite{Higgs} must exist which damp the rise of the
scattering amplitudes of massive gauge particles at high energies.  In
the Standard Model (SM) an isoscalar doublet field is introduced to
accomodate the electroweak data, leading to the prediction of a single
Higgs boson.  Supersymmetric extensions of the Standard Model expand
the scalar sector to a spectrum of Higgs particles~\cite{SUSY}.  The
Higgs particles have been searched for, unsuccessfully so far, at
LEP1, setting a lower limit on the SM Higgs mass of $m_H>65.2$
GeV~\cite{Limit}.  The search for these particles and, if found, the
exploration of their profile, will continue at LEP2~\cite{LEP2}, the
LHC~\cite{LHC}, and future $e^+e^-$ linear colliders~\cite{FLC}.

\begin{figure}[bht]
\vspace*{1.5cm}
\caption{Higgs-strahlung and vector boson fusion of (CP--even) Higgs
bosons in $e^+e^-$ collisions.}
\label{fig:graphs}
\end{figure}
In this note (see also~\cite{KKZ}) we will focus on the production of
scalar Higgs bosons in $e^+e^-$ collisions.  The main production
mechanisms for these particles are Higgs-strahlung~\cite{hst} and $WW$
($ZZ$) fusion~\citerange{wwfi,AMP} [supplemented in supersymmetric
theories by associated scalar/pseudo\-sca\-lar Higgs pair production].  In
particular, we will present a comprehensive analysis of the interplay
between the production mechanisms%
\footnote{
  We will concentrate
  first on the Standard Model (SM); the extension to the Minimal
  Supersymmetric Standard Model (MSSM) is trivial as will be
  demonstrated in the last section of this note.}
(Fig.\ref{fig:graphs})
\begin{equation}
  \begin{array}{lll}
  \mbox{Higgs-strahlung} \hskip-.75em &: & e^+e^-\to ZH \to \bar\nu\nu H
  \quad (e^+e^-H)
  \\[2mm]
  \mbox{$WW$ fusion}&: & e^+e^- \to \bar\nu_e\nu_e H
  \\[2mm]
  \mbox{$ZZ$ fusion}&: & e^+e^- \to e^+e^- H
  \end{array}
\end{equation}

For $\bar\nu_e\nu_e$ and $e^+e^-$ decays of the $Z$ bosons, the two
production amplitudes interfere.  It turns out that the interference
term is positive for $\bar\nu_e\nu_e$ and negative for $e^+e^-$
decays, respectively, in the cross-over region between the two
mechanisms.  The interference effect had been noticed
earlier~\cite{wwfi,boos}; however, we improve on these calculations by
deriving analytic results for the energy and polar angular
distribution of the Higgs particle $(E_H,\theta)$ in the final states
of $e^+e^- \to H + \mbox{neutrinos}$ and $e^+e^- \to He^+e^-$.  This
representation can comfortably serve as input for Monte Carlo
generators like PYTHIA/JETSET~\cite{PYTHIA} and HZHA~\cite{HZHA} which
include the leading QED bremsstrahlung corrections and the important
background processes.

\paragraph{Total cross sections.}
 The cross section for the Higgs-strahlung process can be written in
the following compact form:
\begin{equation}\label{eq:HZ}
  \sigma(e^{+}e^{-}\to ZH) =
  \frac{G_{F}^{2}m_{Z}^{4}}{96 \pi s}\,\left( v_{e}^{2}+a_{e}^{2}\right)
  \,\lambda^{\frac{1}{2}}\,
  \frac{\lambda + 12m_{Z}^{2}/s}{(1-m_{Z}^{2}/s)^{2}}
\end{equation}
where $\sqrt{s}$ is the center-of-mass energy, and $a_e=-1$,
$v_e=-1+4\sin^2\theta_W$ are the $Z$ charges of the electron; $\lambda
= \left(1-(m_H+m_Z)^2/s\right)\left(1-(m_H-m_Z)^2/s\right)$ is the usual
two-particle phase space function.  So long as the non-zero width of
the $Z$ boson%
\footnote{The results presented in this note are insensitive to
  non-zero width effects of the Higgs boson~\cite{nzw}.  For SM Higgs
  masses below $100$ GeV, $\Gamma_H$ is at least three orders of
  magnitude smaller than $\Gamma_Z$; for larger Higgs masses, $m_H$
  can be reinterpreted as the effective invariant mass of the Higgs
  decay products.}
is not taken into account, the cross section rises steeply at
threshold $\sim (s-(m_H+m_Z)^2)^{1/2}$.  After reaching a
maximum [about $25$ GeV above threshold in the LEP2 energy range], the
cross section falls off at high energies, according to the scaling law
$\sim g_W^4/s$ asymptotically.  Thus, Higgs-strahlung is the dominant
production process for moderate values of the energy.  The cross
section~(\ref{eq:HZ}) for Higgs-strahlung is reduced by a factor
$3\times {\rm BR}_\nu = 20\%$ if the final state of $Z$ decays is
restricted to neutrino pairs.

The total cross section for the $WW$ ($ZZ$) fusion of Higgs particles can be
cast into a similarly compact form%
\footnote{The variable $x$ is the invariant mass squared of $\nu_e$ plus $H$, 
 $(x-y)$ the 4-momentum transfer squared from $e^+$ to $\bar\nu_e$ 
 (all momenta in units of the total energy).} 
\cite{LC500}: 
\begin{equation}\label{WWF}
  \sigma
  = \frac{G_{F}^{3}m_{V}^{4}}{64\sqrt{2}\,\pi^{3}}\,\int_{x_{H}}^{1}\,dx\,
    \int_{x}^{1}\,\frac{dy}{[1+(y-x)/x_{V}]^{2}}
    \left[(v^2+a^2)^2 f(x,y) + 4v^2a^2 g(x,y)\right]
\end{equation}
where $V$ denotes either $W$ or $Z$, the charges are
$v=a=\sqrt{2}$ ($v=v_e$ and $a=a_e$) for $WW$ ($ZZ$) fusion,
respectively, and
\begin{eqnarray*}
  f(x,y)
  &=& \left( \frac{2x}{y^{3}}-\frac{1+2x}{y^{2}}
             +\frac{2+x}{2y}-\frac12\right)\,
    \left[\frac{z}{1+z}-\log(1+z)\right]\,
    +\,\frac{x}{y^{3}}\frac{z^{2}(1-y)}{1+z}
  \\
  g(x,y)
  &=& \left( -\frac{x}{y^2} + \frac{2+x}{2y} - \frac12 \right)
      \left[\frac{z}{1+z}-\log(1+z)\right]
\end{eqnarray*}
with $x_H=m_H^2/s$, $x_V=m_V^2/s$ and $z=y(x-x_H)/(xx_V)$.  For
moderate Higgs masses and energies, the cross section, being ${\cal
O}(g_W^6)$, is suppressed with respect to Higgs-strahlung by the
additional electroweak coupling.  The smaller value of the
$Z$-electron coupling suppresses the $ZZ$ fusion process by an
additional order of magnitude compared to $WW$ fusion.  At high
energies, the $WW$ fusion process becomes leading, nevertheless, since
the size of the cross section is determined by the $W$ mass, in
contrast to the scale-invariant Higgs-strahlung process,
\begin{eqnarray}
  \sigma(e^{+}e^{-}\to \bar\nu_e\nu_e H)
  &\approx&  \frac{G_{F}^{3}m_{W}^{4}}{4\sqrt{2}\,\pi^{3}}
     \left[\left(1+\frac{m_H^2}{s}\right) \log \frac{s}{m_H^2}
     - 2\left(1-\frac{m_H^2}{s}\right)\right] \nonumber\\[2mm]
  &\to& \frac{G_{F}^{3}m_{W}^{4}}{4\sqrt{2}\,\pi^{3}}
     \log\frac{s}{m_H^2}
\end{eqnarray}
The cross section rises logarithmically at high energies, as to be
anticipated for this $t$-channel exchange process.  

\begin{figure}[htbp]
\vspace*{16cm}
\caption{Total cross sections for the processes $e^+e^-\to
H\bar\nu\nu$ and $e^+e^-\to He^+e^-$ as a function of the Higgs mass.
The cross sections are broken down to the three components
Higgs-strahlung, vector boson fusion, and the interference term.
\emph{``thr''} denotes the maximum Higgs mass for on-shell $ZH$
production, \emph{``tot''} is the total cross section.  In $e^+e^-\to
H\bar\nu\nu$ (above) the interference term is negative for small Higgs
masses, for large Higgs masses positive.  In $e^+e^-\to He^+e^-$
(below), the interference term is of opposite sign.}
\label{fig:sigma500}
\end{figure}

\paragraph{Differential cross section and interference for $WW$ fusion.}
The compact form~(\ref{WWF}) for the fusion cross section cannot be
maintained once the interference term between vector boson fusion and
Higgs-strahlung is included.  Moreover, since in the case of $WW$
fusion the integration variables $x$ and $y$ do not correspond to
observable quantities, the formula is useful only for calculating the
total cross section without experimental cuts.  Nevertheless,
similarly compact expressions can be derived in this general case by
choosing the energy $E_H$ and the polar angle $\theta$ of the Higgs
particle as the basic variables in the $e^+e^-$ c.m.\ frame.  The
overall cross section that will be observed experimentally for the
process
\begin{displaymath}
  e^+e^- \to H + \bar\nu\nu
\end{displaymath}
receives contributions $3\times{\cal G}_S$ from Higgs-strahlung with
$Z$ decays into three types of neutrinos, ${\cal G}_W$ from $WW$
fusion, and ${\cal G}_I$ from the interference term between fusion and
Higgs-strahlung for $\bar\nu_e\nu_e$ final states.  We find%
\footnote{The analytic result for ${\cal G}_W$ had first been
  obtained in Ref.\cite{AMP}.}
for energies $\sqrt{s}$ above the $Z$ resonance:
\begin{equation}\label{totalWW}
  \frac{d\sigma(H\bar\nu\nu)}{dE_H\,d\cos\theta}
  = \frac{G_F^3 m_Z^8p}{\sqrt2\,\pi^3s}
  \left(3\,{\cal G}_S + {\cal G}_I + {\cal G}_W \right)
\end{equation}
with
\begin{eqnarray}
  {\cal G}_S &=& \frac{v_e^2+a_e^2}{96}\;
    \frac{ss_\nu + s_1s_2}{\left(s-m_Z^2\right)^2
               \left[(s_\nu-m_Z^2)^2 + m_Z^2\Gamma_Z^2\right]}\\
  {\cal G}_I &=& \frac{(v_e+a_e)\cos^4\theta_W}{8}\;
    \frac{s_\nu-m_Z^2}{\left(s-m_Z^2\right)
                \left[(s_\nu-m_Z^2)^2 + m_Z^2\Gamma_Z^2\right]}
    \nonumber\\
    &&\times
    \left[ 2 - (h_1+1)\log\frac{h_1+1}{h_1-1}
             - (h_2+1)\log\frac{h_2+1}{h_2-1}
           +\, (h_1+1)(h_2+1)\frac{{\cal L}}{\sqrt{r}}\right]
    \\
  {\cal G}_W &=& \frac{\cos^8\theta_W}{s_1 s_2 r}\,
    \Bigg\{(h_1+1)(h_2+1)
        \left[
        \frac{2}{h_1^2-1} + \frac{2}{h_2^2-1} - \frac{6s_\chi^2}{r}
        + \left(\frac{3t_1t_2}{r}-c_\chi\right)
          \frac{{\cal L}}{\sqrt{r}}\right]
    \nonumber\\
    &&\qquad\qquad
    - \left[\frac{2t_1}{h_2-1} + \frac{2t_2}{h_1-1}
      + \left(t_1+t_2+s_\chi^2\right)
        \frac{{\cal L}}{\sqrt{r}}\right]
    \Bigg\}
\end{eqnarray}
The cross section is written explicitly in terms of the Higgs momentum
$p= \sqrt{E_H^2-m_H^2}$, and the energy $\epsilon_\nu=\sqrt{s}-E_H$
and invariant mass squared $s_\nu=\epsilon_\nu^2-p^2$ of the neutrino
pair.  In addition, the following abbreviations have been adopted from
Ref.\cite{AMP},
\begin{displaymath}
  \begin{array}{r@{\;=\;}l}
  s_{1,2} & \sqrt{s}(\epsilon_\nu\pm p\cos\theta)\\[1mm]
  h_{1,2} & 1 + 2m_W^2/s_{1,2}\\[1mm]
  c_\chi & 1 - 2 s s_\nu/(s_1s_2)\\[1mm]
  s_\chi^2 & 1 - c_\chi^2
  \end{array}
  \qquad
  \begin{array}{r@{\;=\;}l}
  t_{1,2} & h_{1,2} + c_\chi h_{2,1}\\[1mm]
  r & h_1^2 + h_2^2 + 2c_\chi h_1h_2 - s_\chi^2\\[1mm]
  {\cal L} & {\displaystyle \log\frac{h_1h_2 + c_\chi + \sqrt{r}}
                        {h_1h_2 + c_\chi - \sqrt{r}}}
  \end{array}
\end{displaymath}

To derive the total cross section $\sigma(e^+e^-\to H\bar\nu\nu)$, the
differential cross section must be integrated over the region
\begin{equation}
  -1<\cos\theta<1 \quad\mbox{and}\quad
  m_H < E_H < \frac{\sqrt{s}}{2}\left(1+\frac{m_H^2}{s}\right)
\end{equation}

\paragraph{Differential cross section and interference for $ZZ$ fusion.}
Similarly, the overall cross section for the process
\begin{displaymath}
  e^+e^- \to H + e^+e^-
\end{displaymath}
receives contributions ${\cal G}_S$ from Higgs-strahlung with $Z$
decays into electron-positron pairs, ${\cal G}_Z$ from $ZZ$ fusion,
and ${\cal G}_I$ from the interference term between fusion and
Higgs-strahlung:
\begin{equation}\label{totalZZ}
  \frac{d\sigma(He^+e^-)}{dE_H\,d\cos\theta}
  = \frac{G_F^3 m_Z^8p}{\sqrt2\,\pi^3s}
  \left({\cal G}_S + {\cal G}_I + {\cal G}_{Z1} + {\cal G}_{Z2} \right)
\end{equation}
with
\begin{eqnarray}
  {\cal G}_S &=& \frac{\left(v_e^2+a_e^2\right)^2}{192}\;
    \frac{ss_e + s_1s_2}{\left(s-m_Z^2\right)^2
               \left[(s_e-m_Z^2)^2 + m_Z^2\Gamma_Z^2\right]}\\
  {\cal G}_I &=& \frac{\left(v_e^2+a_e^2\right)^2+4v_e^2a_e^2}{64}\;
    \frac{s_e-m_Z^2}{\left(s-m_Z^2\right)
                \left[(s_e-m_Z^2)^2 + m_Z^2\Gamma_Z^2\right]}
    \nonumber\\
    &&\times
    \left[ 2 - (h_1+1)\log\frac{h_1+1}{h_1-1}
             - (h_2+1)\log\frac{h_2+1}{h_2-1}
           +\, (h_1+1)(h_2+1)\frac{{\cal L}}{\sqrt{r}}\right]
    \\
  {\cal G}_{Z1} &=& 
    \frac{\left(v_e^2+a_e^2\right)^2+4v_e^2a_e^2}{32\, s_1 s_2 r}\,
    \Bigg\{
    (h_1+1)(h_2+1)\left[
        \frac{2}{h_1^2-1} + \frac{2}{h_2^2-1} - \frac{6s_\chi^2}{r}
        + \left(\frac{3t_1t_2}{r}-c_\chi\right)
          \frac{{\cal L}}{\sqrt{r}}\right]
    \nonumber\\
    && \qquad{}
        - \left[\frac{2t_1}{h_2-1} + \frac{2t_2}{h_1-1}
                + \left(t_1+t_2+s_\chi^2\right)
                \frac{{\cal L}}{\sqrt{r}}\right]
    \Bigg\} \\
  {\cal G}_{Z2} &=& \frac{\left(v_e^2-a_e^2\right)^2}{16\, s_1 s_2 r}\,
        (1 - c_\chi)
        \left[
        \frac{2}{h_1^2-1} + \frac{2}{h_2^2-1} - \frac{6s_\chi^2}{r}
        + \left(\frac{3t_1t_2}{r}-c_\chi\right)
          \frac{{\cal L}}{\sqrt{r}}\right]
\end{eqnarray}
where the same abbreviations as in the formulae following
Eq.(\ref{totalWW}), with the appropriate replacements
$\nu\to e$ and $W\to Z$, have been used.

\paragraph{}
To interpret the results, we display the three components of the total
cross sections $\sigma(e^+e^-\to H\bar\nu\nu)$ and
$\sigma(e^+e^-\to He^+e^-)$ in
Fig.\ref{fig:sigma500} for the linear collider energy
$\sqrt{s}=500$ GeV in the cross-over region.%
\footnote{Note that Higgs-strahlung dominates $WW$ fusion at $500$ GeV
  for moderate Higgs masses only if the total $ZH$ cross section is
  considered.}

While the energy distribution of the Higgs particle peaks at $E_H\sim
(s+m_H^2-m_Z^2)/2\sqrt{s}$ for Higgs-strahlung, it is nearly flat for
$WW$ fusion~(Fig.\ref{fig:e}, left).  Only with rising total energy the
lower part of the Higgs spectrum becomes more pronounced.  The angular
distribution for Higgs-strahlung is almost isotropic at threshold
while the standard $\sin^2\theta$ law is approached, in accordance
with the equivalence principle, at asymptotic
energies~(Fig.\ref{fig:e}, right).  The angular distribution peaks, by
contrast, in the $WW$ fusion process at $\theta\to 0$ and $\pi$ for
high energies as expected for $t$-channel exchange processes.

\paragraph{Polarized beams.}
At linear colliders the incoming electron and positron beams can be
polarized longitudinally.  Higgs-strahlung and $WW$ fusion both
require opposite helicities of the electrons and positrons.  If
$\sigma_{U,LR,RL}$ denote the cross sections in $e^+e^-\to H\bar\nu\nu$
for unpolarized electrons/positrons, left-handed
electrons/right-handed positrons, and right-handed
electrons/left-handed positrons, respectively, we can easily derive,
in the notation of Eq.(\ref{totalWW}):
\begin{eqnarray}
  \sigma_U &\propto&
     3\,{\cal G}_S + {\cal G}_I + {\cal G}_W\\
  \sigma_{LR} &\propto&
     6\,{\cal G}_S + 4\,{\cal G}_I + 4\,{\cal G}_W\\
  \sigma_{RL} &\propto&
     6\,{\cal G}_S
\end{eqnarray}
The cross section for $WW$ fusion of Higgs particles increases by a
factor four, compared with unpolarized beams, if left-handed electrons
and right-handed positrons are used.  By using right-handed electrons,
the $WW$ fusion mechanism is switched off.  [The interference term
cannot be separated from the $WW$ fusion cross section.]

For the process $e^+e^-\to He^+e^-$, the pattern is slightly more
complicated:
\begin{eqnarray}
  \sigma_U &\propto&
	{\cal G}_S + {\cal G}_I + {\cal G}_{Z1} + {\cal G}_{Z2}\\
  \sigma_{LR} &\propto&
	2\frac{(v_e+a_e)^2}{(v_e^2+a_e^2)}{\cal G}_S
	+ 2\frac{(v_e+a_e)^4}{(v_e^2+a_e^2)^2+4v_e^2a_e^2}
	\left({\cal G}_I + {\cal G}_{Z1}\right)\\
  \sigma_{RL} &\propto&
	2\frac{(v_e-a_e)^2}{(v_e^2+a_e^2)}{\cal G}_S
	+ 2\frac{(v_e-a_e)^4}{(v_e^2+a_e^2)^2+4v_e^2a_e^2}
	\left({\cal G}_I + {\cal G}_{Z1}\right)\\
  \sigma_{LL} = \sigma_{RR} &\propto&
	2\,{\cal G}_{Z2}.
\end{eqnarray}
However, since $v_e\ll a_e$, the difference between $\sigma_{RL}$ and
$\sigma_{LR}$ is suppressed.

\paragraph{Supersymmetric CP-even Higgs bosons.}
It is trivial to transfer all these results from the Standard
Model to the Minimal Supersymmetric Standard Model (MSSM).  Since the
couplings to $Z/W$ gauge bosons in the MSSM are shared~\cite{GH} by
the CP-even light and heavy scalar Higgs bosons, $h$ and $H$,
respectively, only the overall normalization of the cross sections is
modified with respect to the Standard Model:
\begin{eqnarray}
  \sigma(h)_{\rm MSSM} &=& \sin^2(\beta-\alpha)\times\sigma(H)_{\rm SM} \\
  \sigma(H)_{\rm MSSM} &=& \cos^2(\beta-\alpha)\times\sigma(H)_{\rm SM}
\end{eqnarray}
Higgs-strahlung, vector boson fusion, and the interference term are
affected in the same way.  [The angle $\alpha$ is the mixing angle in
the CP-even Higgs sector while the mixing angle $\beta$ is determined
by the ratio of the vacuum expectation values of the two neutral Higgs
fields in the MSSM.  A recent discussion of the size of the
coefficients $\sin^2/\cos^2(\beta-\alpha)$ may be found in
Ref.\cite{DKZ}.]



\newpage

\begin{figure}[htbp]
\unitlength 1cm
\vspace*{15cm}
\caption{Energy distribution (left) and angular distribution (right)
  of the Higgs bosons for the three components of the cross section
  [\/{\rm Hs} = Higgs-strahlung; $WW$ = fusion;\/ {\rm intf} =
  interference term].  The individual curves are normalized to the
  total cross section.  The\/ {\rm Hs} peak extends up to maximal
  values of\/ $0.22\;{\rm GeV}^{-1}$.  The total cross section is
  $69.4\;{\rm fb}$.}
\vspace*{-.6\baselineskip}
\label{fig:e}
\end{figure}

\newpage

\setcounter{footnote}{0}
\setcounter{figure}{0}
\setcounter{equation}{0} 

\newcommand{\nuclinst}{{\em Nucl.\ Instrum.\ Meth.\ }}
\newcommand{\annp}{{\em Ann.\ Phys.\ }}
\newcommand{\cms}{centre-of-mass\hspace*{.1cm}}
\newcommand{\epemt}{$e^{+} e^{-}\;$}
\newcommand{\epem}{e^{+} e^{-}\;}
\newcommand{\eennhht}{$e^{+} e^{-} \ra \nu_e \bar \nu_e HH\;$}
\newcommand{\eennhh}{e^{+} e^{-} \ra \nu_e \bar \nu_e HH\;}
\newcommand{\wwhh}{W^+W^-\ra HH\;}
\newcommand{\wwhht}{$W^+W^-\ra HH\;$}
\newcommand{\gamgamt}{$\gamma \gamma \;$}
\newcommand{\gamgam}{\gamma \gamma \;}
\newcommand{\ggwwht}{$\gamma \gamma \ra W^+ W^- H \;$}
\newcommand{\ggwwh}{\gamma \gamma \ra W^+ W^- H \;}
\newcommand{\ggwwhht}{$\gamma \gamma \ra W^+ W^- H H\;$}
\newcommand{\ggwwhh}{\gamma \gamma \ra W^+ W^- H H\;}
\newcommand{\cl}{{{\cal L}}}
\def\noi{\noindent}
\def\sm{${\cal{S}} {\cal{M}}\;$}

\begin{center}
{\large 
{\bf Associated Pair Production of the \sm Higgs}}

\vspace*{3mm}

{\large{\bf and the Probing of the Higgs Self-Coupling}}

\vspace*{0.5cm}

{\sc  E.~Chopin}

\vspace*{3mm} 

{\small Laboratoire de Physique Th\'eorique 
EN{\large S}{\Large L}{\large A}PP}
\footnote{ URA 14-36 du CNRS, associ\'ee \`a l'E.N.S de Lyon 
et \`a l'Universit\'e de Savoie.}\\ 
{\small B.P.110, 74941 Annecy-Le-Vieux Cedex, France} \\
{\small E-mail: chopin@lapphp0.in2p3.fr}
\end{center}

\subsection*{1 Associated Higgs Pair Production}

The interest in double Higgs production is the probing of the triple 
Higgs self coupling. It has been considered in \epemt sometime 
ago\cite{hh}. The most efficient means for double Higgs production 
is \eennhht (see fig.~\ref{eegghh100}). Double Higgs 
bremstrahlung ($\epem \ra ZHH$) is only competing at 
relatively low energies where the event sample is too low to 
be useful. The equivalent loop-induced double Higgs 
production in \epemt has been found to be much too small\cite{eehh} 
and is not sensitive to the $H^3$ coupling. However, the 
$\gamma\gamma$ mode can form a $J_Z=0$ state and  therefore 
$\gamma\gamma\ra HH$ is a candidate for testing the $H^3$ 
coupling\cite{Jikiahh}. It has been pointed out recently that 
another interesting process is \ggwwhht\cite{chopinhh} 
that is expected to compete with double Higgs production in \epemt. 
The reason is that in the TeV range, $W$ fusion processes are very 
much enhanced. The sub-process involved is \wwhht, where the 
dominant helicity amplitude is:
\begin{eqnarray}
\tilde{\cal M}_{LL} &=& \frac{g^2}{2} \left\{ 
\frac{1}{\beta_H\beta_W^3} \left( \frac{1}{x-x_0}
-\frac{1}{x+x_0}\right)(r\frac{M_H^2}{s}+\beta_H^2+\beta_W^4) 
\right. \nonumber \\
&+& \left.\frac{1}{\beta_W^2}(2-\beta_W^2-r)+\frac{3h_3r}{4}
\left(\frac{1+\beta_W^2}{1-M_H^2/s}\right)\right\}
\ra \frac{g^2}{4}r(3h_3-2)+\dots 
\end{eqnarray}
Where $h_3$ is the anomaly in the triple Higgs coupling 
$g$, {\em i.e.} $g=h_3g_{sm}$ where $g_{sm}$ is the minimal 
standard model coupling of $H^3$. We also denote 
$\beta_{W,H}=\sqrt{1-4M_{W,H}^2/s}$, $r=M_H^2/M_W^2$, 
$x_0=(1+\beta_H^2)/2\beta_W\beta_H$, $x=\cos\theta$. 

\begin{figure}
\vspace*{7cm}
\caption{\label{eegghh100}{\small Comparison of cross sections for 
double Higgs production at \epemt and $\gamma\gamma$ reactions 
for a light Higgs $M_H=100GeV$.}}
\end{figure}

Figure~\ref{mhggdep} shows that at 2TeV, the cross sections 
 drops precipitously with increasing Higgs mass.
One can also notice that in \ggwwhht the external outgoing 
$W$ (mainly transverse) are produced at small angle and take a large 
amount of energy. For the fusion diagrams of this process, the 
internal $W$ triggers \wwhht, which implies 
that these diagrams dominate for a heavy Higgs. 
When convoluting with the much advertised photon 
spectra of\cite{Laser} and for $M_H=100GeV$ the cross section 
of \ggwwhht drops by about at least a factor of 2 compared 
with the result without convolution. 
For $W$-fusion-like processes, the internal 
$W$'s are almost on shell and one may wonder if some structure 
functions could reproduce the exact results. 
\begin{figure}
\vspace*{7cm}
\caption{\label{mhggdep}{\small Higgs mass dependence of the 
\ggwwhht cross section at 2TeV. The contribution of the diagrams 
involving the triple Higgs vertex (Signal) and the rest
(Background) is shown separetely. Note the strong interference that 
occurs in the \sm especially for large $M_H$.}}
\end{figure}

\subsection*{2 The structure function approach}
There have been numerous derivations of the distribution (or 
structure function) of the $W$ inside
the light fermions (quarks and electrons)
\cite{EWA}. For the effective $W$ approximation, 
the most interesting aspect concerns the $W_L$ content, which 
 has been used to investigate manifestations of models of 
symmetry breaking and Higgs production. The $W_L$ distribution 
inside the photon has only very recently been studied \cite{Parisgg}. 
 For the case of the heavy Higgs the approximation is excellent, 
already at 2TeV. However, for a light Higgs, the approximation is 
not good and reproduce only the energy behaviour. If one 
makes the further approximation that the hard process cross 
section is independent of the energy, this additional 
``asymptotic'' approximation only reproduces the energy 
behaviour as well as the order of magnitude (even for a heavy Higgs, 
see figs.~\ref{eeapprox}).
\begin{figure}
\vspace*{11cm}
\caption{\label{eeapprox}{\small Comparing the result of the 
$W_L$ effective approximation ($\sigma^{EWA}$) to the 
exact result $\sigma^{exact}$ for \eennhht (left) and 
\ggwwhht (right) for a light Higgs 
and a heavy Higgs. Also shown is the asymptotic analytical 
cross section $\sigma^{EWA}_{\infty}$. $\sigma^{TT}$ is 
the cross section with both outgoing $W$'s transverse.}}
\end{figure}

\subsection*{3 Identifying and measuring the Higgs triple vertex}

There is a specific signature of the 
$H^3$ coupling in all processes that we have studied. Once we 
note that the two Higgses that originate from this vertex can 
be regarded as produced by a scalar $H^\star$ then in the centre 
of mass system of the pair, the angular distribution of the Higgses 
is flat. Therefore, we suggest to reconstruct the angle, 
$\theta^\star$, measured in the \cms of the pair, between the 
Higgs direction and the boost axis or the direction of the beam. 
For the ``signal''\footnote{We call 
``signal'' the part of the amplitude that 
include the $H^3$ vertex. The rest is called ``background''.} 
the distribution is flat, while the ``background'' 
is peaked in the forward/backward direction (see fig.~\ref{thetastar}).
\begin{figure}
\vspace*{11.4cm}
\caption{\label{thetastar}{\small The distribution in the reconstructed 
angle $\theta^{\star}$ for the signal, background and the 
interference in the case of \ggwwhht without convolution with photon 
spectra.}}
\end{figure}
We therefore consider the ratio $R$ of events that 
verify $|\cos(\theta^{\star})|<\cos(\theta^{\star}_0)$\footnote{
$\cos\theta^\star_0=0.5$ is taken in the following.}, 
over the number of events outside this region. 
Assuming a total integrated luminosity of $300 fb^{-1}$, and a 
$50\%$ efficiency for the reconstruction of the double Higgs 
events one obtain $\sim 68$ \eennhht events for
$M_H=100GeV$. Here the criterion for detection of an 
anomaly in $h_3$ is a $50\%$ deviation in the expected 
number of events, provided one has at least $30$ events. We 
conclude that with the total cross section one would only be able 
to claim New Physics if $\delta h_3<-0.75$ or $\delta h_3>2$. 
For $M_H=400GeV$, \sm values will not lead to a measurement, 
however if $|\delta h_3|>1$ a signal will be recorded 
(with more than 30 events) and would be a clear indication 
for an anomalous $h_3$ coupling. For $M_H=100GeV$ where one has 
enough events for a \sm value, the ratio $R$ is much more 
powerful in constraining the coupling. 
First the event sample within $|\cos \theta^\star|<0.5$ 
is about 7 out of 60 outside this region. Assuming that the ratio 
can be measured at $20\%$, we find $-.10<|\delta h_3|<.15$ 
(see fig.~\ref{Reegg}) which means a precision of about $10\%$ on $h_3$.
For \ggwwhht and considering the effective \gamgamt luminosity,  
for $M_H=100GeV$, one can hope to collect 15 events. 
In view of this number the criterion for detection of non-standard 
values is $100\%$ deviation in the number of events. 
However, for a Higgs mass 
of $400$GeV the effect of an anomalous $H^3$ coupling are 
dramatic and, by far, much more interesting than in \epemt.
Requiring observation of at least 15 events for $M_H=400$GeV 
(within the \sm one expects only 3) useful constraints 
on the coupling can be set: $-.7<\delta h_3 <0.5$. There is thus a 
complementarity between the \epemt and the \gamgamt depending on 
the Higgs mass in probing the Higgs triple vertex. 
As for the ratio R, taking $M_H=100GeV$ it is unlikely 
that with the number of total $WWHH$ events at \gamgamt one would 
be able to make such a measurement, nonetheless even 
if this ratio were measured with the same precision as in 
\epemt one would not constrain the couplings further than what 
is achieved in the classic \epemt mode. 
For $M_H >600GeV$, $\gamgam \ra HH$ is the only 
reaction where useful limits can be set. Thus, 
there is at a 2TeV collider a very nice coverage of the $h_3$ 
sensitivity by all three reactions.
\begin{figure}
\vspace*{8cm}
\caption{\label{Reegg}{\small Dependence of the ratio $R$ on $h_3$.}} 
\end{figure}

\subsection*{4 Conclusions}

We have seen that a 2TeV \epemt collider 
with the realistic luminosities expected for 
this machine one may hope to achieve a measurement of the 
tri-linear couplings at the level of $10\%$ (for a light Higgs). 
The results are also encouraging in the sense that the \epemt and 
the \gamgamt modes can cover different ranges of the Higgs mass. 
We find that for a light Higgs
(up to 250GeV) the best limits on the $H^3$ couplings come 
from $\eennhh$. However, for heavier Higgses up to mass of 
$500GeV$, the best channel is the associated double Higgs 
production in \gamgamt. For still heavier masses, the one-loop 
induced $\gamgam \ra HH$ is by far better. 
The variable R clearly helps in discriminating the triple 
Higgs vertex. As a by-product we have verified the validity 
of the distribution function describing the longitudinal 
$W$ content of the photon. We should also insist on 
the complementarity of the \epemt and the \gamgamt modes 
of the linear collider for these studies.

\newpage

\setcounter{footnote}{0}
\setcounter{figure}{0}
\setcounter{equation}{0}
\def\inpb{\mbox{$\hbox{pb}^{-1}$}}
\def\Gcs{\hbox{GeV}/\mbox{$c^2$}}

\def\thefootnote{\fnsymbol{footnote}}

\begin{center}
 {\large\bf
 Anomalous Couplings 
 in the Higgs-strahlung Process\\[\baselineskip]}
 {\sc W.~Kilian, M.~Kr\"amer, and P.M.~Zerwas}
 \\
 {\small Deutsches Elektronen-Synchrotron DESY, 
 D-22603 Hamburg/FRG}
\end{center}
\begin{abstract}
\nn The angular distributions in the Higgs-strahlung process $e^+e^-\to
  HZ\to H\bar f f$ are uniquely determined in the Standard Model.  We
  study how these predictions are modified if non-standard couplings
  are present in the $ZZH$ vertex, as well as lepton-boson contact
  terms.  We restrict ourselves to the set of operators which are
  singlets under standard $SU_3\times SU_2\times U_1$ transformations,
  CP conserving, dimension~6, helicity conserving, and custodial
  $SU_2$ conserving. 
\end{abstract}

\def\thefootnote{\arabic{footnote}}

\smallskip 

\paragraph{}
The Higgs-strahlung process~\cite{hst}
\begin{equation}\label{hst}
  e^+e^-\to HZ \to H\bar f f
\end{equation}
together with the $WW$ fusion process, are the most important
mechanisms for the production of Higgs bosons in $e^+e^-$
collisions~\cite{LEP2,FLC}.  Since the $ZZH$ vertex is uniquely
determined in the Standard Model (SM), the production cross section of
the Higgs-strahlung process, the angular distribution of the $HZ$
final state as well as the fermion distribution in the $Z$ decays can
be predicted if the mass of the Higgs boson is fixed~\cite{BZ}.  These
predictions may be modified when deviations from the pointlike
coupling are present, which can occur in models with non-pointlike
character of the Higgs boson itself or through interactions beyond the
SM at high energy scales.  Since the effective energy scale of the
Higgs-strahlung process is set by the c.m.\ energy $\sqrt{s}$, while
the fusion processes are essentially low-energy processes with an
effective scale of the order $M_W$, new interactions manifest
themselves more clearly in the total cross section and angular
distributions for the Higgs-strahlung process (see also~\cite{KKZa}).

\paragraph{Operator basis.}
Deviations from the pointlike coupling can occur in models with
non-pointlike character of the Higgs boson itself or through interactions
beyond the SM at high energy scales.  We need not specify the
underlying theory but instead we will adopt the usual assumption that
these effects can globally be parameterized by introducing a set of
dimension-6 operators
\begin{equation}\label{L-eff}
  {\cal L} = {\cal L}_{\rm SM} + \sum_i\frac{\alpha_i}{\Lambda^2}\op_i
\end{equation}
The coefficients are in general expected to be of the order
$1/\Lambda^2$, where $\Lambda$ denotes the energy scale of the new
interactions.  However, if the underlying theory is weakly
interacting, the $\alpha_i$ can be significantly smaller than unity,
in particular for loop-induced operators.  [It is assumed \emph{a
priori} that the ratio of the available c.m.\ energy to $\Lambda$ is
small enough for the expansion in powers of $1/\Lambda$ to be
meaningful.]

If we restrict ourselves to operators~\cite{BW85} which are singlets
under $SU_3\times SU_2\times U_1$ transformations of the SM gauge
group, CP conserving, and conserving the custodial $SU_2$ symmetry, the
following bosonic operators are relevant for the Higgs-strahlung
process:
\begin{eqnarray}
  \op_{\partial\varphi} 
   &=& \frac12|\partial_\mu(\varphi^\dagger\varphi)|^2\\
  \op_{\varphi W}
   &=& \frac12\varphi^\dagger \vec W_{\mu\nu}^2\varphi \\
  \op_{\varphi B}
   &=& \frac12\varphi^\dagger B_{\mu\nu}^2\varphi 
\end{eqnarray}
where the gauge fields $W^3,B$ are given by the $Z,\gamma$ fields.
This set of operators is particularly interesting because it does not
affect, at tree level, observables which do not involve the Higgs
particle explicitly.  [It is understood that the fields and parameters
are (re-)normalized in the Lagrangian ${\cal L}$ in such a way that
the particle masses and the electromagnetic coupling retain their
physical values.]

In addition, we consider the following helicity-conserving fermionic
operators which induce contact terms contributing to $e^+e^-\to ZH$:
\begin{eqnarray}
  \op_{L1}
   &=& (\varphi^\dagger iD_\mu\varphi)
       (\bar\ell_L\gamma^\mu\ell_L) + \hc \\
  \op_{L3}
   &=& (\varphi^\dagger \tau^a iD_\mu\varphi)
       (\bar\ell_L\tau^a\gamma^\mu\ell_L) + \hc \\
  \op_R
   &=& (\varphi^\dagger iD_\mu\varphi)
       (\bar e_R\gamma^\mu e_R) + \hc 
\end{eqnarray}
[$\ell_L$ and $e_R$ denote the left-handed lepton doublet and the
right-handed singlet, respectively.  The vacuum expectation value of
the Higgs field is given by $\langle\varphi\rangle = (0,v/\sqrt2)$
with $v=246\,{\rm GeV}$, and the covariant derivative acts on the
Higgs doublet as $D_\mu = \partial_\mu - \frac{i}{2}g\tau^a W^a_\mu +
\frac{i}{2}g'B_\mu$.]  Helicity-violating fermionic operators do not
interfere with the SM amplitude, so that their contribution to the
cross section is suppressed by another power of~$\Lambda^2$.  The
helicity-conserving fermionic operators modify the SM $Zee$ couplings
and are therefore constrained by the measurements at LEP1; however, it
is possible to improve on the existing limits by measuring the
Higgs-strahlung process at a high-energy $e^+e^-$ collider since the
impact on this process increases with energy~\cite{GW}.

\vspace*{2cm}
\begin{figure}[bth]
\caption{Anomalous $ZZH$/$\gamma ZH$ couplings and $e^+e^-ZH$
contact terms in the Higgs-strahlung process.}
\label{ZZH}
\end{figure}

The effective $ZZH$ and the induced $\gamma ZH$ interactions
(Fig.\ref{ZZH}, left diagram) may be written
\begin{eqnarray}\label{L-ZZH}
  {\cal L}_{ZZH} &=& g_ZM_Z\left(
        \frac{1+a_0}{2} Z_\mu Z^\mu H 
        + \frac{a_1}{4} Z_{\mu\nu}Z^{\mu\nu}H\right)\\
  {\cal L}_{\gamma ZH} &=&
        g_ZM_Z\, \frac{b_1}{2}  Z_{\mu\nu}A^{\mu\nu}H
\end{eqnarray}
where $g_Z=M_Z\,\sqrt{4\sqrt{2}G_F}$.  Additional operators $Z_\mu
Z^{\mu\nu}\partial_\nu H$ and $Z_\mu A^{\mu\nu}\partial_\nu H$ are
redundant in this basis: They may be eliminated in favor of the other
operators and the contact terms by applying the equations of motion.
The remaining coefficients are given by
\begin{eqnarray}
  \label{a0'}
  a_0 &=& -\frac{1}{2}\alpha_{\partial\varphi}\, v^2 / \Lambda^2\\
  a_1 &=& 4g_Z^{-2}
        \left( c_W^2\alpha_{\varphi W} + s_W^2\alpha_{\varphi B}\right) /
        \Lambda^2\\
  b_1 &=& 4g_Z^{-2}
        c_Ws_W\left( -\alpha_{\varphi W} + \alpha_{\varphi B}\right) /
        \Lambda^2
\end{eqnarray}
where $s_W$ and $c_W$ are the sine and cosine of the weak mixing
angle, respectively.

In the same way the $e\bar eHZ$ contact interactions (Fig.\ref{ZZH},
right diagram) can be defined for left/right-handed electrons and
right/left-handed positrons
\begin{equation}\label{L-contact}
  {\cal L}_{eeZH} 
  = g_ZM_Z\left[
   c_L \bar e_L\fmslash Z e_L H + c_R \bar e_R\fmslash Z e_R H
   \right]
\end{equation}
with
\begin{eqnarray}
  c_L &=& -2g_Z^{-1}  \left(\alpha_{L1}  + \alpha_{L3}\right)/
  \Lambda^2\\   
  c_R &=& -2g_Z^{-1}  \alpha_R / \Lambda^2
\end{eqnarray}

Some consequences of these operators for Higgs production in $e^+e^-$
collisions have been investigated in the past.  Most recently, the
effect of novel $ZZH$ vertex operators and $\ell\bar\ell ZH$ contact
terms on the total cross sections for Higgs production has been
studied in Ref.\cite{GW}.  The impact of vertex operators on angular
distributions has been analyzed in Refs.\cite{SH} and~\cite{GR}.  We
expand on these analyses by studying the angular distributions for the
more general case where both novel vertex operators and contact
interactions are present.  The analysis of angular distributions in
the Higgs-strahlung process~(\ref{hst}) allows us to discriminate
between various novel interactions.  In fact, the entire set of
parameters $a_0,a_1,b_1$ and $c_L,c_R$ can be determined by measuring
the polar and azimuthal angular distributions as a function of the
beam energy if the electron/positron beams are unpolarized. As
expected, the energy dependence of the polar angular distribution is
sufficient to provide a complete set of measurements if longitudinally
polarized electron beams are available\footnote{Since we can restrict
  ourselves to helicity-conserving couplings, as argued before,
  additional positron polarization need not be required.}.

\begin{figure}[bt]
\vspace*{7cm}
\caption{Polar and azimuthal angles in the Higgs-strahlung
process. [The polar angle $\theta_*$ is defined in the $Z$ rest frame.]}
\label{angles}
\end{figure}

\paragraph{Total cross section and polar angular distribution.}
Denoting the polar angle between the $Z$ boson and the $e^+e^-$ beam
axis by $\theta$, the differential cross section for the process
$e^+e^-_{L,R}\to ZH$ may be written as
\begin{equation}
  \frac{d\sigma^{L,R}}{d\cos\theta}
        =  \frac{G_F^2 M_Z^4}{96\pi s}
           \left(v_e\pm a_e\right)^2\lambda^{1/2}\,
           \frac{\frac34\lambda\,\sin^2\theta\,\left(1+\alpha^{L,R}\right)
                 + 6\left(1+\beta^{L,R}\right)M_Z^2/s}
                {(1-M_Z^2/s)^2}
\end{equation}
and the integrated cross section
\begin{equation}
  \sigma = \frac{G_F^2 M_Z^4}{96\pi s}
           \left(v_e\pm a_e\right)^2\lambda^{1/2}
           \frac{\lambda\left(1+\alpha^{L,R}\right)
                 + 12\left(1+\beta^{L,R}\right)M_Z^2/s}
                {(1-M_Z^2/s)^2}
\end{equation}
The $Z$ charges of the electron are defined as usual by $a_e=-1$ and
$v_e=-1+4s_W^2$.  $s$ is the c.m.\ energy squared, and $\lambda$ the
two-particle phase space coefficient $\lambda =
\left[1-(m_H+m_Z)^2/s\right]$ $\times\left[1-(m_H-m_Z)^2/s\right]$.  The 
coefficients $\alpha(s)^{L,R}$ and $\beta(s)^{L,R}$ can easily be 
determined for the interactions in Eqs.(\ref{L-ZZH}) and (\ref{L-contact}):
\begin{eqnarray}
  \alpha(s)^{L,R}
  &=& 2a_0 + (s-M_Z^2)\frac{8c_Ws_W}{v_e\pm a_e}c_{L,R}\\
  \beta(s)^{L,R}
  &=& \alpha(s)^{L,R} + 
      2\gamma\sqrt{s}\,M_Z
      \left[a_1 + \frac{4c_Ws_W}{v_e\pm a_e}
                  \left(1-\frac{M_Z^2}{s}\right)b_1\right]
\end{eqnarray}
where the boost of the $Z$ boson is given by $\gamma =
(s+M_Z^2-M_H^2)/2M_Z\sqrt{s}$.  

The modification of the cross section by the new interaction terms
has a simple structure.  The coefficient $a_0$ just renormalizes the
SM cross section.  By contrast, the contact interactions grow with
$s$.  [The ratio $s/\Lambda^2$ is assumed to be small enough for the
restriction to dimension-6 operators to be meaningful.]  The operators
$\op_{\varphi W}$, $\op_{\varphi B}$ affect the coefficient in the
cross section which is independent of $\theta$.  They damp the
fall-off of this term, changing the $1/s^2$ to a $1/s$ behavior;
however, these contributions remain subleading since they are
associated with transversely polarized $Z$ bosons which are suppressed
at high energies compared with the longitudinal components.  To
illustrate the size of the modifications $\alpha(s)^{L,R}$ and
$\beta(s)^{L,R}$, we have depicted these functions in
Fig.\ref{coeff}(a) for the special choice $\alpha_i = 1$.

\paragraph{Azimuthal distributions.}
The azimuthal angle $\phi_*$ of the fermion $f$ is defined as the
angle between the [$e^-,Z$\/] production
plane and the [$Z,f$\/] decay plane~(Fig.\ref{angles}).  It corresponds to the azimuthal angle of
$f$ in the $Z$ rest frame with respect to the [$e^-,Z$\/] plane.  On
general grounds, the $\phi_*$ distribution must be a linear function
of $\cos\phi_*$, $\cos 2\phi_*$, and $\sin\phi_*$, $\sin 2\phi_*$,
measuring the helicity components of the decaying spin-1 $Z$ state.
The coefficients of the sine terms vanish for CP invariant theories.
The $\cos\phi_*$ and $\cos 2\phi_*$ terms correspond to P-odd and
P-even combinations of the fermion currents.  The general azimuthal
distributions are quite involved~\cite{BZ,SH,GR}.  We therefore
restrict ourselves to the simplified case in which all polar angles
are integrated out, i.e., the polar angle $\theta$ of the $Z$ boson in
the laboratory frame and the polar angle $\theta_*$ of $f$ in the $Z$
rest frame. In this way we find for the azimuthal $\phi_*$ distribution:
\begin{equation}
  \frac{d\sigma^{L,R}}{d\phi_*} \sim 1 \mp\frac{9\pi^2}{32}\,
  \frac{2\,v_fa_f}{v_f^2+a_f^2}\,\frac{\gamma}{\gamma^2+2} 
  \left(1+f_{1}^{L,R}\right)\cos\phi_* + \frac{1}{2(\gamma^2+2)}
  \left(1+f_{2}^{L,R}\right)\cos 2\phi_*
\end{equation}
with 
\begin{eqnarray}
  f_1(s)^{L,R}
  &=& M_Z\sqrt{s}\,
         \frac{(\gamma^2-1)(\gamma^2-2)}{\gamma(\gamma^2+2)}
         \left[a_1 + \frac{4s_Wc_W}{v_e\pm a_e}
                     \left(1 - \frac{M_Z^2}{s}\right) b_1\right]
  \\
  f_2(s)^{L,R}
  &=& 2M_Z\sqrt{s}\,
         \frac{\gamma(\gamma^2-1)}{\gamma^2+2}
         \left[a_1 + \frac{4s_Wc_W}{v_e\pm a_e}
                     \left(1 - \frac{M_Z^2}{s}\right) b_1\right]
\end{eqnarray}
The cross section flattens with increasing c.m.\ energy in the
Standard Model, i.e.\ the coefficients of $\cos\phi_*$ and
$\cos2\phi_*$ decrease asymptotically proportional to $1/\sqrt{s}$ and
$1/s$, respectively.  The anomalous contributions modify this
behavior: The $\cos\phi_*$ term receives contributions which increase
proportional to $\sqrt{s}$ with respect to the total cross section,
while the $\cos2\phi_*$ term receive contributions from the
transversal couplings that approaches a constant value asymptotically.
The size of the new terms in $f_{1,2}^{L,R}$ is shown in
Fig.\ref{coeff}(b) as a function of the energy. [The special choice
$\alpha_i = 1$ we have adopted for illustration, implies $f_{1,2}^{L}
= f_{1,2}^{R}$.]

\paragraph{High-energy limit.} 
It is instructive to study the high-energy behavior of
the coefficients in the limit $M_Z^2\ll s\ll\Lambda^2$.  In this case
we obtain the simplified relations:
\begin{eqnarray}
  \alpha(s)^{L,R} 
  &\simeq& \mp\, s\cdot 8s_Wc_W\, c_{L,R} + {\cal O}(v_e) \label{25}\\
  \beta(s)^{L,R} 
  &\simeq& \alpha(s)^{L,R}
           + s\left(a_1 \mp 4s_Wc_W\, b_1\right)+ {\cal O}(v_e) \label{26}
\end{eqnarray}
and
\begin{eqnarray}
  f_1(s)^{L,R}
  &\simeq& 
  \frac{s}{2}\left(a_1 \mp 4s_Wc_W\, b_1\right) + {\cal O}(v_e) 
  \label{27}\\
  f_2(s)^{L,R}
  &\simeq& s\left(a_1 \mp 4s_Wc_W\, b_1\right) + {\cal O}(v_e) \label{28}
\end{eqnarray}
Terms which are proportional to $v_e=-1+4s_W^2$ are suppressed by an
order of magnitude. If longitudinally polarized electrons are
available, the asymptotic value of the coefficients $a_1,b_1,c_L$ and
$c_R$ can be determined by measuring the polar angular distribution
without varying the beam energy.  The analysis of the azimuthal
$\phi_*$ distribution provides two additional independent measurements
of the coefficients $a_1$ and $b_1$. On the other hand, the set of
measurements remains incomplete for fixed energy if only unpolarized
electron/positron beams are used at high energies; in this case the
coefficients cannot be disentangled completely without varying the
beam energy within the preasymptotic region.

\begin{figure}[htbp]
\vspace*{17cm}
\caption[dummy]{\label{coeff} Coefficients of the angular
  distributions as a function of the beam energy. Parameters are
  described in the text; in particular, $\alpha_i=1$ has ben chosen in
  the effective Lagrangian Eq.(\ref{L-eff}).  [The $L,R$ coefficients
  of the azimuthal distribution coincide for the special choice
  $\alpha_i = 1$.]}
\end{figure}


\newpage

\setcounter{footnote}{0}
\setcounter{figure}{0}
\setcounter{equation}{0}
\renewcommand{\thefootnote}{\fnsymbol{footnote}}

\begin {center}
{\large\sc {\bf  The hidden Higgs model at the NLC}}

\vspace{.4cm}

{\sc T. Binoth and  J. J. van der Bij} \\ 

\vspace{.5mm}

{\small  Albert--Ludwigs--Universit\"at Freiburg, 
Fakult\"at f\"ur Physik, \\
Hermann--Herder--Strasse 3, 79104 Freiburg, Germany}

\end{center}

\begin{abstract}
We investigate the influence of massless scalar singlets on Higgs signals
at the NLC. An exclusion bound is presented which restricts large regions of the  
parameter space but on the other hand implies that for strong 
interactions between the Higgs boson and the singlet 
fields of the hidden sector, detection of such a non standard Higgs 
signal can become impossible. 
\end{abstract}

\subsection*{1. Introduction}

Understanding of the electroweak symmetry breaking mechanism is one of the main
tasks in particle physics. The determination  of its nature would be 
a break-through in our knowledge about matter. So it is important to 
think about alternatives to the Standard Model Higgs sector. Various such extensions are 
available. Maybe the best motivated one is the supersymmetrized Standard Model with its
important phenomenological implication of a light Higgs boson and which allows 
a consistent frame for grand unified theories. Another well understood extension --
though in its minimal version disfavoured by the precision experiments at LEP -- are
technicolor theories. Though these theories avoid fundamental scalars, a rich
bosonic spectrum of techniquark condensates may exist. Thus in both theories, as 
long as they
do not occur in their minimal form, light bosonic matter could be present modifying
the standard Higgs signals we are looking for at present and future colliders.
If such bosons appear as singlets under the Standard Model gauge group, they do
not feel the color or electroweak forces,
but they can couple to the Higgs particle. As a consequence
radiative corrections to weak processes are not sensitive to the
presence of singlets in the theory, because no Feynman graphs containing
singlets  appear
at the one--loop level. Since effects at the two--loop level
are below the experimental precision,
the presence of a singlet sector is not ruled out by any 
of the LEP1 precision data. The only connection to such a hidden sector
is a possible Higgs--singlet coupling, leading to a nonstandard invisible
Higgs decay.
The invisible decay of the Higgs boson with a narrow width 
leads to relatively sharp missing energy signals, well known from discussions
on Majoron models \cite{valle}. However a strongly coupled hidden sector could lead to fast
Higgs decay and thereby to wide resonances. This would disturb the signal to background
ratio if necessary cuts are imposed.  

To check the influence of a hidden sector we will study the coupling
of a Higgs boson to an O(N) symmetric set of scalars, which  
is one of  the simplest possibilities, introducing only a few extra 
parameters in the theory. The effect of the extra scalars is practically
the presence of a possibly large invisible decay width of the Higgs particle.
When the coupling is large enough the Higgs resonance can become
wide even for a light Higgs boson. It was shown earlier that there
will be a range of parameters, where such a Higgs boson can be seen neither
at LEP nor at the LHC \cite{vladimir,valle}.  

In the next section we will introduce the 
model together with its theoretical constraints and in the last section
we will discuss exclusion limits at the NLC.  

\subsection*{2. The model}
The scalar sector of the model consists of the usual Higgs sector coupled 
to a real N--component vector $\vec\varphi$ of scalar fields, denoted by 
phions in the following. The Lagrangian density is given by,
\begin{displaymath}
\label{definition}
 {\cal L}  =
 - \partial_{\mu}\phi^+ \partial^{\mu}\phi -\lambda (\phi^+\phi - v^2/2)^2
   - 1/2\,\partial_{\mu} \vec\varphi \partial^{\mu}\vec\varphi
     -1/2 \, m^2 \,\vec\varphi^2 
     - \kappa/(8N) \, (\vec\varphi^2 )^2
    -\omega/(2\sqrt{N})\, \, \vec\varphi^2 \,\phi^+\phi 
\end{displaymath}
where $\phi$ is the standard Higgs doublet. 
Couplings to fermions and vector bosons are the same as in the Standard Model.
The ordinary
Higgs field acquires the vacuum expectation value $v/\sqrt{2}$. For positive $\omega$
the $\vec\varphi$--field acquires no vacuum expectation
value. After spontaneous
symmetry breaking one is left with the ordinary Higgs boson,
coupled to the phions into which it decays. Also the phions
receive an induced mass from the spontaneous symmetry breaking which is suppressed
by a factor $1/\sqrt{N}$.
If the factor N is taken
to be large,  the model can be analysed with $1/N$--expansion techniques.
By taking this limit the phion mass remains small, but as there
are many phions, the decay width of the Higgs boson can become large.
Therefore the main effect of the presence of the phions is to give
a large invisible decay rate to the Higgs boson. The 
invisible decay width is given by 
\beq \Gamma_H =\frac {\omega^2 v^2}{32 \pi M_H} = 
\frac {\omega^2 (\sin\theta_W\cos\theta_W M_Z)^2}{32 \pi^2 \alpha_{em} M_H}\quad .
\nonumber \eeq
The Higgs width is compared with the width in the Standard Model for various choices
of the coupling $\omega$ in Fig.~\ref{width}.
The model is different
from Majoron models \cite{valle}, since the width is not necessarily small.
The model is similar to the technicolor--like model of Ref.~\cite{chivukula}.
\begin{figure}[hbt]
\vspace*{8.5cm}
\caption{\it Higgs width in comparison with the Standard Model.}
\label{width}
\end{figure}

Consistency of the model requires two conditions.
One condition is the absence of a Landau pole below a certain scale
$\Lambda$. The other follows from the stability of the vacuum up to a certain
scale. An example of such limits is given in Fig.~\ref{stability},
where $\kappa=0$ was taken at the scale $2m_Z$, which allows for
the widest parameter range.
The regions of validity up to a given scale $\Lambda$ are sandwiched
between the lower--left and the upper--right contour lines in the figure. 
The first stem from instability of the 
vacuum, the second from the presence of a Landau pole at that scale.

\begin{figure}[htb]
\vspace*{9.1cm}
\caption{\it Theoretical limits on the parameters of the model
in the $\omega$ vs. $M_H$ plane. The contour lines correspond 
to the cutoff scales $\Lambda = 10^{19}$, $10^6$, $10^4$ and $10^3$ GeV.}
\label{stability}
\end{figure}

To search for the Higgs boson there are basically
two channels, one is the standard decay, which is reduced in branching
ratio due to the decay into phions.
The other is the invisible decay, which rapidly becomes dominant,
eventually making the Higgs resonance wide (see Fig.~\ref{width}). 
In order to give the bounds we 
neglect the coupling $\kappa$ as this is a small effect. We
also neglect the phion mass. For other values of the phion mass
the bounds can be found by rescaling the decay widths
with the appropriate phase space factor. Now we confront this
two dimensional parameter space with the experimental potential 
of the NLC.

\subsection*{3. NLC bounds}
At the NLC the upper limits on the couplings in the present model
come essentially from the invisible decay, as the branching ratio
into visible particles drops with increasing $\varphi$--Higgs
coupling ($\omega$), whereas for small $\omega$ one has to consider visible Higgs decays, too.
Since the main source for Higgs production, the $WW$--fusion process,
can not be used to look for invisible Higgs decay,
one is in principle left with the Higgsstrahlung und $ZZ$--fusion reaction. 
For energies up to 500 GeV the Higgsstrahlungs cross section is dominant and
is of comparable size to the $ZZ$--fusion process even if one is folding in 
the branching ratio $B(Z\rightarrow e^+e^-,\mu^+\mu^-)$. 
The possibility to tag an on--shell Z boson via a leptonic system which is extremely useful 
for the discrimination of possible backgrounds makes Higgsstrahlung to be the preferred production
mechanism. Thus we only have considered reactions 
containing an on shell Z boson with its decay into $e^+e^-$ or $\mu^+\mu^-$.  
One should be aware that a few events from the huge $WW$ background may survive \cite{eboli}, but
that the $Z\nu\nu$ background is dominant after imposing the cuts defined below.
Then the signal cross section is the well  known Higgsstrahlungs cross section modified
by the non standard Higgs width due to phion decay. With the invariant mass of the invisible
phion system, $s_I$, it has the form:  
\beq 
\sigma_{(e^+e^-\rightarrow Z+E\!\!\!/)} = 
\int ds_I \, \sigma_{(e^+e^-\rightarrow ZH)}(s_I) \,
\frac{\sqrt{s_I} \quad \Gamma(H\rightarrow E\!\!\!/)}
{\pi ((M_H^2-s_I)^2+s_I\,\Gamma(H\rightarrow \mbox{All})^2)}
\nonumber\eeq
We calculated the $Z\nu\nu$ background with the standard set of graphs for
Z production ($ZZ$--production, $WW$--fusion and Z initial, final state radiation) by a
Monte Carlo program (see Ref.~\cite{mele}). To reduce the background we used the fact
that the angular distribution of the Z--boson for the signal peaks for small values
of $|\cos\theta_Z|$ in contrast to the background. Thus we imposed the cut $|\cos\theta_Z|<0.7$.
Because we assume the reconstruction of the on-shell Z--boson we use the kinematical relation
$E_Z=(s+M_Z^2-s_I)/(2\sqrt{s})$
between the Z energy and the invariant mass of the invisible system
to define a second cut. Since the differential cross section $d\sigma/ds_I$ contains the 
Higgs resonance at $s_I=M_H^2$, we impose the following condition on the Z energy:
\beq \frac{s+M_Z^2-(M_H+\Delta_H)^2}
{2\sqrt{s}}<E_Z<\frac{s+M_Z^2-(M_H-\Delta_H)^2}{2\sqrt{s}}
\nonumber\eeq 
For the choice of $\Delta_H$ a comment is in order. As long as the Higgs width is small, one
is allowed to use small  $\Delta_H$, which reduces the background considerably keeping
most of the signal events. But in the case of large $\varphi$--Higgs coupling, $\omega$, one
looses valuable events. To compromise between both effects we took  $\Delta_H=30$ GeV.  
 
For the exclusion limits we assumed an integrated luminosity
of $20\,fb^{-1}$. To define the $95 \%$ confidence level we used 
Poisson statistics similar to the description of Ref. \cite{valle}.
The result is given  
in Fig.~\ref{exclusion}. One notices the somewhat reduced sensitivity
for $M_H\simeq M_Z$ due to a resonating $Z$ boson in the $ZZ$ background. For larger
values of $M_H$ the limit stems from the other $Z\nu\nu$ backgrounds
with $W$ bosons in the t--channel and kinematical constrains. 
For large $\omega$ the signal ceases
to dominate over the background because the Higgs peak is smeared out
to an almost flat distribution. 

\begin{figure}[htb]
\vspace*{9.1cm}
\caption{\it Exclusion limits at the NLC due to Higgs searches. The dashed
line corresponds to the invisible, the full line to all Higgs decay modes.}
\label{exclusion}
\end{figure}

We conclude from this analysis that the NLC can put further
limits on the parameter space of our invisible Higgs model. 
Note that within the kinematic range very strong limits on $\omega$ can be set. 
Again there is a range
where the Higgs boson will not be discovered, even if it does exist in this mass range.
This has already been shown for the Higgs search at LEP and also
holds true for the heavy Higgs search at LHC. 
We see that a sufficiently wide nonstandard Higgs resonance would make it very difficult
to test the mechanism of electroweak symmetry breaking at future colliders.

\newpage

\textheight 22.2cm
\setcounter{footnote}{0}
\setcounter{figure}{0}
\setcounter{equation}{0}
\def\thefootnote{\arabic{footnote}}
\begin{center}

{\large\sc {\bf  Heavy SUSY Higgs Bosons at $e^+ e^-$ Linear Colliders}}

\vspace{0.7cm}

{\sc A.~Djouadi$^{1,2}$, J. Kalinowski$^{3}$, P.~Ohmann$^{1,4}$ and
  P.M.~Zerwas$^1$ } 

\vspace{.5cm}

$^1$ Deutsches Elektronen--Synchrotron DESY, D-22603 Hamburg, FRG. \\
\vspace{0.3cm}

$^2$ Institut f\"ur Theoretische Physik, Universit\"at Karlsruhe,
D-76128 Karlsruhe, FRG. \\
\vspace{0.3cm}

$^3$ Institute of Theoretical Physics, Warsaw University, PL-00681 Warsaw,
Poland. \\
\vspace{0.3cm}

$^4$ Department of Theoretical Physics, Oxford University, OX1 3NP, 
Oxford, UK. 

\end{center}

\begin{abstract}
\normalsize
\noindent
\nn The production mechanisms and decay modes of the heavy neutral and
charged Higgs bosons in the Minimal  Supersymmetric Standard  Model are
investigated at future $e^+ e^-$ colliders in the TeV energy regime. We
generate supersymmetric particle spectra by requiring the MSSM Higgs
potential to produce correct radiative electroweak symmetry breaking,
and we assume a common scalar mass $m_0$, gaugino mass $m_{1/2}$
and trilinear coupling $A$, as well as gauge and Yukawa coupling
unification at the Grand Unification scale. Particular emphasis is put
on the low $\tb$ solution in this scenario where decays of the Higgs
bosons to Standard Model particles compete with decays to supersymmetric
charginos/neutralinos as well as sfermions. In the high $\tb$ case, the
supersymmetric spectrum is either too heavy or the supersymmetric decay
modes are suppressed, since the Higgs bosons decay almost exclusively
into $b$ and $\tau$ pairs. The main production mechanisms for the heavy
Higgs particles are the associated $AH$ production and $H^+H^-$ pair
production with cross sections of the order of a few fb. 
 
\end{abstract}

\setcounter{equation}{0}
\renewcommand{\theequation}{1.\arabic{equation}}

\subsection*{1. Introduction}

Supersymmetric theories \cite{R1,R2} are generally considered to be the
most natural extensions of the Standard Model (SM). This proposition is
based on several points. In these theories, fundamental scalar Higgs
bosons \cite{R3,R4} with low masses can be retained in the context of
high unification scales. Moreover, the prediction \cite{R5} of the
renormalized electroweak mixing angle $\sin^2 \theta_W = 0.2336 \pm
0.0017$, based on the spectrum of the Minimal Supersymmetric Standard
Model (MSSM) \cite{R6}, is in striking agreement with the electroweak
precision data which yield $\sin^2 \theta_W = 0.2314(3)$ \cite{R7}. An
additional attractive feature is provided by the opportunity to generate
the electroweak symmetry breaking radiatively \cite{R8}. If the top
quark mass is in the range between $\sim$ 150 and $\sim$ 200 GeV, the
universal squared Higgs mass parameter at the unification scale
decreases with decreasing energy and becomes negative at the electroweak
scale, thereby breaking the ${\rm SU(2)_L \times U(1)_Y}$ gauge symmetry
while leaving the U(1) electromagnetic and SU(3) color gauge symmetries
intact \cite{R8}. The analysis of the electroweak data prefers a light
Higgs mass \cite{R7,R9} as predicted in supersymmetric theories; however
since the radiative corrections depend only logarithmically on the Higgs
mass \cite{R10}, the dependence is weak and no firm conclusions can yet
be drawn. \s 

The more than doubling the spectrum of states in the MSSM gives rise to
a rather large proliferation of parameters. This number of parameters is
however reduced drastically by embedding the low--energy supersymmetric
theory into a grand unified (GUT) framework. This can be achieved in
supergravity models \cite{R8}, in which the effective low--energy
supersymmetric theory [including the interactions which break
supersymmetry] is described by the following parameters: the common
scalar mass $m_0$, the common gaugino mass $m_{1/2}$, the
trilinear coupling $A$, the bilinear coupling $B$, and the
Higgs--higgsino mass parameter $\mu$. In addition, two parameters are
needed to describe the Higgs sector: one Higgs mass parameter [in
general the mass of the pseudoscalar Higgs boson, $M_A$] and the ratio
of the vacuum expectation values, $\tb =v_2/v_1$, of the two Higgs
doublet fields which break the electroweak symmetry. \s

The number of parameters can be further  reduced by introducing additional
constraints which are based on physically rather natural assumptions: \s

$(i)$
Unification of the $b$ and $\tau$ Yukawa couplings at the GUT scale \cite{R11a}
leads to a correlation between the top quark mass and $\tb$ \cite{R11,R15,R19}.
Adopting the central value of the top mass as measured at the Tevatron
\cite{R12}, $\tb$ is restricted to two narrow ranges around $\tb \sim 1.7$ and
$50$, with the low $\tb$ solution theoretically somewhat favored 
\cite{R19}. \s

$(ii)$ If the electroweak symmetry is broken radiatively, then the
bilinear coupling $B$ and the Higgs--higgsino mass parameter $\mu$ are
determined up to the sign of $\mu$. [The sign of $\mu$ might be
determined by future precision measurements of the radiative
$b$ decay amplitude.] \s

$(iii)$ It turns out {\it a posteriori} that the physical observables are 
nearly independent of the GUT scale value of the trilinear coupling $A_G$, 
for $|A_G| \lsim 500$ GeV. \s

Mass spectra and couplings of all supersymmetric particles and Higgs
bosons are determined after these steps by just two mass parameters
along with the sign of $\mu$; we shall choose to express our results in
terms of the pseudoscalar Higgs boson $A$ mass $M_A$ and the common GUT
gaugino mass $m_{1/2}$.  \s

In this paper we focus on heavy Higgs particles $A$, $H$ and $H^{\pm}$
with masses of a few hundred GeV, and therefore close to the decoupling
limit \cite{R14}. The pattern of Higgs masses is quite regular in
this limit. While the upper limit on the mass of the lightest CP--even Higgs
boson $h$ is a function of $\tb$ \cite{R14a}, 
\begin{eqnarray}
M_h \lsim 100 \ {\rm to} \ 150 \ {\rm GeV \ \ [ for \ low \ to \ high 
\ \tb}] 
\end{eqnarray}
the heavy Higgs bosons are nearly mass degenerate [c.f. Fig.1]
\beq
M_A & \simeq & M_H  \ \simeq \ M_{H^\pm} 
\eeq
Moreover, the properties of the lightest CP--even Higgs boson $h$ become 
SM--like in this limit. The production of the heavy Higgs bosons becomes 
particularly simple in $e^+ e^-$ collisions; the heavy Higgs bosons can 
only be pair--produced, 
\begin{eqnarray}
e^+ e^- & \rightarrow & A \ H  \\
e^+ e^- & \rightarrow  & H^+ H^- 
\end{eqnarray}
Close to this decoupling limit, the cross section for $H$ Higgs--strahlung $\ee
\ra ZH$ is very small and the cross section for the $WW$ fusion mechanism $\ee
\ra \nu_e \bar{\nu}_e H$ is appreciable only for small values of $\tb$, $\tb
\sim 1$, and relatively small $H$ masses, $M_H \lsim 350$ GeV. The cross
section for $ZZ$ fusion of the $H$ is suppressed by an order of magnitude
compared to $WW$ fusion. The pseudoscalar $A$ particle does not couple to $W/Z$
boson pairs at the tree level. \s 

The decay pattern for heavy Higgs bosons is rather complicated in
general. For large $\tb$ the SM fermion decays prevail. For small $\tb$
this is true above the $t\bar{t}$ threshold of $M_{H,A} \gsim 350$ GeV
for the neutral Higgs bosons and above the $t\bar{b}$ threshold of
$M_{H^\pm} \gsim 180$ GeV for the charged Higgs particles. Below these
mass values many decay channels compete with each other: decays to SM
fermions $f\bar{f}$ [and for $H$ to gauge bosons $VV$], Higgs cascade
decays, chargino/neutralino $\chi_i \chi_j $ decays and decays to 
supersymmetric sfermions $\tilde{f} \tilde{\bar{f}}$
\begin{eqnarray}
H &\rightarrow& f \bar{f}  \ , \ VV \ , \ hh \ , \ \chi_{i} \chi_{j} 
\ , \ \tilde{f} \bar{\tilde{f}} \\
A &\rightarrow& f \bar{f} \ , \ hZ \ , \ \chi_{i} \chi_{j} 
\ , \  \tilde{f} \bar{\tilde{f}} \\ 
H^\pm &\rightarrow& f \bar{f}' \ , \ hW^\pm \ , \ \chi_{i} \chi_{j}
\ , \ \tilde{f} \bar{\tilde{f}} \; ' 
\end{eqnarray}

In this paper, we analyze in detail the decay modes of the heavy Higgs
particles and their production at $\ee$ linear colliders. The analysis will
focus on heavy particles for which machines in the TeV energy range are needed.
The paper is organized in the following way. In the next section we define the
physical set--up of our analysis in the framework of the MSSM embedded into a
minimal supergravity theory. In section 3, we discuss the production cross
sections of the heavy Higgs bosons. In the subsequent sections, we discuss the
widths of the various decay channels and the final Higgs decay products. 

\setcounter{equation}{0}
\renewcommand{\theequation}{2.\arabic{equation}}

\subsection*{2. The Physical Set--Up}

The Higgs sector of the Minimal Supersymmetric Standard Model is
described at tree-level by the following potential
\begin{eqnarray}
V_0 &=& (m_{H_1}^2+\mu ^2)|H_1|^2+(m_{H_2}^2+\mu
^2)|H_2|^2 - m_3^2(\epsilon_{ij}{H_1}^i{H_2}^j+{\rm h.c.})
\nonumber \\
&& +{1\over 8}(g^2+g^{\prime 2})\left [|H_1|^2-|H_2|^2\right ]^2
+{1\over 2}g^2|H_1^{i*}H_2^i|^2\; 
\end{eqnarray}
The quadratic Higgs terms associated with $\mu$ and the quartic
Higgs terms coming with the electroweak gauge couplings $g$ and $g'$ are
invariant under supersymmetric transformations. $m_{H_1}^{}$,
$m_{H_2}^{}$ and $m_3$ are soft--supersymmetry breaking parameters with
$m_3^2 = B \mu$. $\epsilon_{ij}$ [$i,j=1 , 2$ and $\epsilon_{12}=1$] is
the antisymmetric tensor in two dimensions and $H_1\equiv
(H_1^1,H_1^2)=(H_1^0,H_1^-)$, $H_2\equiv (H_2^1,H_2^2)=(H_2^+,H_2^0)$
are the two Higgs-doublet fields. After the symmetry breaking, three
out of the initially eight degrees of freedom will be absorbed to 
generate the $W^\pm$ and $Z$ masses, leaving a quintet of scalar Higgs
particles: two CP--even Higgs bosons $h$ and $H$, a CP--odd
[pseudoscalar] boson $A$ and two charged Higgs particles $H^\pm$. \s 

Retaining only the [leading] Yukawa couplings of the third generation
\beq
\lambda_t= \frac{\sqrt{2}m_t} {v \sin \beta} \ , \ \ 
\lambda_b= \frac{\sqrt{2}m_b} {v \cos \beta} \ \ {\rm and} \ \ 
\lambda_\tau= \frac{\sqrt{2}m_\tau} {v \cos \beta}
\eeq
where $\tb=v_2/v_1$ [with $v^2=v_1^2+v_2^2$ fixed by the $W$ mass, $v=
246$ GeV] is the ratio of the vacuum expectation values of the fields
$H_2^0$ and $H_1^0$, the superpotential is given in terms of the superfields
$Q=(t,b)$ and $L=(\tau, \nu_\tau)$ by\footnote{Note that 
our convention for the sign of $\mu$ is consistent with Ref.\cite{R2a},
which  is opposite to the one adopted in Ref.\cite{R16}.}
\beq
W= \epsilon_{ij} \left[ \lambda_t H_2^i Q^j t^c + \lambda_b H_1^i Q^j 
b^c + \lambda_\tau H_1^i L^j \tau^c - \mu H^i_1 H_2^j \right]
\eeq
Supersymmetry is broken by introducing the soft--supersymmetry breaking
bino $\tilde{B}$, wino $\tilde{W}^a$ $[a=$1--3] and gluino 
$\tilde{g}^a$ $[a=$1--8] mass 
terms,
\beq
\frac{1}{2} \,M_1 \,\overline{\tilde{B}}\,\tilde{B} \ + \
\frac{1}{2} \,M_2 \,\overline{\tilde{W}}^a \,\tilde{W}^a \ + \
\frac{1}{2} \,M_3 \,\overline{\tilde{g}}^a \,\tilde{g}^a \ , 
\eeq
soft--supersymmetry breaking trilinear couplings,
\beq
\epsilon_{ij} \left[ \lambda_t A_t H_2^i \tilde{Q}^j  \tilde{t}^c + 
\lambda_b A_b H_1^i \tilde{Q}^j \tilde{b}^c + \lambda_\tau A_\tau 
H_1^i \tilde{L}^j  \tilde{\tau}^c - \mu B H^i_1 H_2^j \right]
\eeq
and soft--supersymmetry breaking squark and slepton mass terms
\beq
M_Q^2 [\tilde{t}^*_L\tilde{t}_L +\tilde{b}^*_L\tilde{b}_L] +
M_U^2 \tilde{t}^*_R \tilde{t}_R  + M_D^2 \tilde{b}^*_R \tilde{b}_R+
M_L^2 [\tilde{\tau}^*_L\tilde{\tau}_L +\tilde{\nu_\tau}^*_L\tilde{\nu_\tau}_L] 
+ M_E^2 \tilde{\tau}^*_R \tilde{\tau}_R  \ + \ \cdots
\eeq
where the ellipses stand for the soft mass terms corresponding to the first 
and second generation sfermions. \s

The minimal SUSY--GUT model emerges by requiring at the GUT scale $M_G$: 
\s

\nn $(i)$ the unification of the U(1), SU(2) and SU(3) coupling constants 
$\alpha_i=g_i^2/4\pi$ $[i=1$--3],
\beq
\alpha_3 (M_{\rm G}) = \alpha_2 (M_{\rm G}) = \alpha_1 (M_{\rm G}) =\alpha_G
\eeq
$(ii)$ a common gaugino mass; the $M_i$ with $i=$1--3 at the
electroweak scale are then related through renormalization group equations
(RGEs) to the gauge couplings, 
\beq
M_i = \frac{\alpha_i(M_Z)}{\alpha_G} m_{1/2} \ \longrightarrow 
\ M_3(M_Z)=\frac{\alpha_3(M_Z)} {\alpha_2(M_Z)} M_2(M_Z)
          =\frac{\alpha_3(M_Z)} {\alpha_1(M_Z)} M_1(M_Z)
\eeq
$(iii)$ a universal trilinear coupling $A$
\beq
A_G = A_t (M_{\rm G}) = A_b (M_{\rm G}) = A_\tau (M_{\rm G})
\eeq
$(iii)$ a universal scalar mass $m_0$
\beq
m_0&=& M_Q=M_U=M_D=M_L=M_E \non \\
   &=& m_{H_1}(M_G)=m_{H_2}(M_G) 
\eeq
Besides the three parameters $m_{1/2}, A_G$ and $m_0$ the
supersymmetric sector is described at the GUT scale by the bilinear
coupling $B_G$ and the Higgs--higgsino mass parameter $\mu_G$. The
theoretically attractive assumption that the electroweak symmetry is
broken radiatively constrains the latter two parameters. Indeed,
radiative electroweak symmetry breaking results in two minimization
conditions [see Ref.\cite{R16} for details] of the Higgs
potential; at the low--energy scale in the tree approximation, they are
given by 
\begin{eqnarray}
{1\over 2}M_Z^2&=&{{m_{H_1}^2-m_{H_2}^2\tan ^2\beta }
\over {\tan ^2\beta -1}}-\mu ^2 \;  \\
B\mu &=&{1\over 2}(m_{H_1}^2+m_{H_2}^2+2\mu ^2)\sin 2\beta \;
\end{eqnarray}
For given values of the GUT parameters $m_{1/2}, m_0, A_G$ as well as 
$\tb$, the first minimization equation can be solved for $\mu$ [to within 
a sign]; the second equation can then be solved for $B$. Since $m_{H_1}^2$ 
and $m_{H_2}^2$ are related to $M_A$ through the RGEs, the solution for 
$\mu$ and $B$ can be approximately expressed as a function of $M_A$
and $\tb$.  The power of supergravity models with radiative
electroweak symmetry breaking becomes apparent when one includes the
one-loop contributions to the Higgs potential. It is through these
one--loop terms that most of the supersymmetric particle masses are
determined; the minimization conditions [which are also solved for
$\mu$ to within a sign and $B$] $\it{fix}$ the masses in order that the
electroweak symmetry is broken correctly, {\it i.e.} with the correct
value of $M_Z$. [U(1)$_{\rm EM}$ and SU(3) remain unbroken of course].
The one--loop contributions and the minimization equations are given in
Ref.\cite{R16} to which we refer for details. \s 

A heavy top quark is required to break the electroweak symmetry
radiatively,  since it is the large top Yukawa coupling which will drive
one of the Higgs mass parameters squared to a negative value. As
emphasized before, the additional condition of unification of the $b$--$\tau$
Yukawa couplings gives rise to stringent constraints on $\tb$. The
attractive idea of explaining the large top Yukawa coupling as a result
of a fixed point solution of the RGEs leads to a relationship between
$M_t$ and the angle $\beta$, $M_t \simeq (200~{\rm GeV})\sin \beta$ 
for $\tb \lsim 10$, giving a further constraint on the model. \s 

To limit the parameter space further, one could require that the SUGRA
model is not fine--tuned and the SUSY breaking scale should not be too
high, a constraint which can be particularly restrictive in the small
$\tb$ region. However, the degree of fine--tuning which can be
considered acceptable is largely a matter of taste, so we disregard this
issue in our analysis. \s 

We now detail the calculations of the supersymmetric particle spectrum more
precisely. We incorporate boundary conditions at both electroweak and GUT
scales, adopting the ambidextrous approach of Ref.\cite{R16}. We specify the
values of the gauge and Yukawa couplings at the electroweak scale, in
particular $M_t$, $\tb$ and $\alpha_s$. The gauge and Yukawa couplings are then
evolved to the GUT scale $M_G$ [defined to be the scale $\tilde{\mu}$ for which
$\alpha_1(\tilde{\mu}) = \alpha_2( \tilde{\mu})$] using the two--loop RGEs
\cite{R15}. At $M_G$ we specify the soft supersymmetry breaking parameters
$m_{1/2}$, $m_0$ and $A_G$. We then evolve parameters down to the electroweak
scale where we apply the one--loop minimization conditions derived from the
one--loop effective Higgs potential and solve for $\mu$ to within a sign and
$B$ [we then can integrate the RGEs back to $M_G$ and obtain $\mu_G$ and
$B_G$]. 
By this procedure, the supersymmetric spectrum is
completely specified; that is, we generate a unique spectrum corresponding to
particular values of $m_{1 /2}$, $m_0$, $A_G$, $\tb$ and the sign of $\mu$. It
turns out that the spectrum is nearly independent of $A_G$, for $|A_G| \lsim
500$ GeV. In most of our calculations, we substitute a particular value of
$M_A$ for $m_0$ in order to introduce a mass parameter which can be measured
directly. \s 

We discuss the SUSY spectrum and its phenomenological
implications for two representative points in the $M_t$--$\tan\beta$
plane\footnote{Our numerical analysis is consistent with the numbers
obtained in Ref.\cite{R35}, once their value of $A_\tau$ in 
Tab.2 is corrected. We thank W. de Boer for a 
discussion on this point.}. We choose $M_t^{\rm pole} = 175$ GeV, 
consistent with the most
recent Tevatron analyses \cite {R12} throughout our calculations, and 
values of $\tb = 1.75$ and 50, which are required (within
uncertainties) by $b$--$\tau$ unification at $M_G$. In particular, we
emphasize the low $\tb$ solutions; they are theoretically favored from
considerations such as $b \rightarrow s\gamma$ \cite{R13} and cosmological
constraints \cite{R17}. The low $\tb$ solutions generate much lighter
SUSY spectra, more likely to be seen at future $e^+ e^-$ colliders. In
both the low and high $\tb$ regions we take\footnote{This 
corresponds to the $\sin^2\theta_W$ value quoted and 
compared with the high--precision
electroweak analyses in the Introduction.} 
$\alpha_s(M_Z^{}) = 0.118$ \cite{R18} and $A_G = 0$, though the
qualitative behavior in each region should not depend greatly on these
parameters. 

\smallskip

\nn (a) \underline{Low $\tan \beta$}

\smallskip

\nn As a typical example of the low $\tan \beta$ region we consider the
point $M^{\rm pole}_t = 175$ GeV and $\tan \beta = 1.75$ for which
$\lambda_t(M_G)$ lies in its ``fixed-point'' region \cite{R11,R19}. If $M_A$
is fixed, the scalar mass parameter $m_0$ can be calculated as a
function of the common gaugino mass parameter $m_{1/2}$ so that
all Higgs and supersymmetric particle masses can in principle be
parameterized by $m_{1/2}$. The correlation between $m_0$ and
$m_{1/2}$ is shown in Fig.2 for three values of $M_A =
300,600$ and 900 GeV in the low $\tb$ region. \s 

Some of the parameter space is already eliminated by experimental bounds
on the light Higgs mass, the chargino/neutralino masses, the light stop
mass, the slepton masses and the squark/gluino masses from LEP1/1.5 and
the Tevatron \cite{R20}. The lower limits are indicated by the 
non--solid lines in Fig.2. Low values of $m_{1/2} \lsim 60$ GeV are excluded
by the lower bound on the gaugino masses. For $\mu>0$, the bound from
the negative search of charginos at LEP1.5 almost rules out completely
the scenario with $M_A \lsim 300$ GeV. If the $h$ boson is not
discovered at LEP2, i.e. if $M_h \gsim 95$ GeV, the whole $\mu<0$
scenario [for $m_{1/2},m_0 < 500$ GeV] can be excluded, 
while for $\mu > 0$ only the $m_{1/2}>200$
GeV range [which implies very large values of $M_A$] would survive. The
requirement that the lightest neutralino is the LSP, and therefore its
mass is larger than the lightest $\tilde{\tau}$ mass, excludes a small
edge of the parameter space [dotted line] at small $m_0$ with $m_{1/
2} > 200$ GeV in the $\mu < 0$ case. \s 

The masses of the Higgs bosons are shown in Fig.3a as a function of
$m_{1/2}$ for $\tb=1.75$, both signs of $\mu$ and for two
representative values of $m_0=100$ and 500 GeV. The lightest Higgs boson
has a rather small tree--level mass and $M_h$ comes mainly from
radiative corrections; the maximal values [for $m_{1/2} \sim 400$ GeV]
are $M_h^{\rm max} \sim 90$
GeV for $\mu<0$ and $\sim 100$ for $\mu>0$. Because the pseudoscalar
mass is approximately given by $M_A^2 \sim B\mu /\sin2\beta \sim
B\mu$ [at the tree--level] and since $B\mu $ turns out to be large in
this scenario, the pseudoscalar $A$ is rather heavy especially for large
values of $m_0$, and thus is almost mass degenerate with the heavy
CP--even and charged Higgs bosons, $M_A \sim M_H \simeq M_{H^\pm}$. Note
that $M_A$ is below the $t\bar{t}$ threshold, $M_A \lsim 350$ GeV, only
if $m_0$ and $m_{1/ 2}$ are both of ${\cal O}(100)$ GeV. \s 

The chargino/neutralino and sfermion masses are shown Fig.3b-d as a
function of $m_{1 /2}$ for the two values $M_A=300$ and 600 GeV and
for both signs of $\mu$. In the case of charginos and neutralinos, the 
masses are related through RGEs by the same ratios that
describe the gauge couplings at the electroweak scale. The LSP is almost
bino--like [with a mass $m_{\chi_1^0} \sim M_1$] while the
next--to--lightest neutralino and the lightest chargino are
wino--like [with masses $m_{\chi_2^0} \sim m_{\chi_1^+} \sim M_2
\sim 2 m_{\chi_1^0}$]. The heavier neutralinos and chargino are
primarily higgsinos with masses $m_{\chi_3^0} \sim m_{\chi_4^0} \sim
m_{\chi_2^+} \sim |\mu|$. 
Note that the masses approximately scale as $M_A$ and that 
the decay of the heavy scalar and pseudoscalar Higgs bosons into 
pairs of the heaviest charginos 
and neutralinos is kinematically not allowed. \s 

The left-- and right--handed charged sleptons and sneutrinos are
almost mass degenerate, the mass differences not exceeding ${\cal
O}(10)$ GeV; the mixing in the $\tau$ sector is rather small 
for small $\tb$, allowing one to
treat all three generations of sleptons on the same footing. In the case of
squarks, only the first two generations are degenerate, with left-- and
right--handed squarks having approximately the same mass. The mixing in
the stop as well as in the sbottom sector leads to a rather substantial
splitting between the two stop or sbottom mass eigenstates. Only for
small values of $M_A$ and for $\mu<0$ is $\tilde{b}_1$ the lightest
squark; otherwise $\tilde{t}_1$ is the lightest squark state. Note that
the squark masses increase with $m_{1 /2}$ and that they scale as $M_A$ 
i.e. as $|\mu|$. The slepton masses decrease with increasing $m_{1/2}$:
this is due to the fact that when $m_{1/2}$ increases and $M_A$
is held constant, $m_0$ decreases 
[see Fig.2], and the dependence of the slepton masses on $m_0$ is 
stronger [for fixed $m_0$, the slepton masses would increase
with increasing $m_{1/2}$]. 

\smallskip

\nn (b) \underline{High $\tan \beta$}

\smallskip

In this region we take $\tb = 50$ as a representative example, a value
consistent with the unification of the $t$, $b$ and $\tau$ Yukawa 
couplings. The set of possible solutions in the parameter space [$m_{1
/2}$, $m_0$] for $M_A = 300$ and 600 GeV is shown in Fig.4. 
At $\tb = 50$ and $M^{\rm pole}_t = 175$ GeV, we find solutions only for
$\mu <0 $; this is a result of the large one--loop contribution
to $M_A$, the sign of which depends on $\mu$ \cite{R36}. The boundary
contours given in the figure correspond to tachyonic solutions in the
parameter space: $m_{\tilde{\tau_1}}^2 < 0$, $M_A^2 < 0$ or $M^2_h <0$
at the tree--level. The latter constraint is important for algorithmic
reasons: $M^2_h$ at the tree--level enters into the minimization
equations in the form of a logarithm \cite{R16}. Also the
requirement of the lightest neutralino to be the LSP excludes a small
edge of the parameter space at small values of $m_0$; this explains
why the curves for $M_A=300$ and 600 GeV do not terminate for low
$m_0$ values. \s

\begin{small}

\vskip 0.3in
{\center \begin{tabular}{|c|c|c|c|c|}
\hline 
\multicolumn{1}{|c|}{Particle}
&\multicolumn{1}{|c|}{Mass (GeV)}
&\multicolumn{1}{|c|}{Mass (GeV)}
&\multicolumn{1}{|c|}{Mass (GeV)}
&\multicolumn{1}{|c|}{Mass (GeV)}
\\[0.2cm] \hline 
\multicolumn{1}{|c|}{$M_A$}
&\multicolumn{1}{|c|}{300}
&\multicolumn{1}{|c|}{300}
&\multicolumn{1}{|c|}{600}
&\multicolumn{1}{|c|}{600}
\\[0.2cm]  \hline 
\multicolumn{1}{|c|}{($m_{1 /2}$, $m_0$)}
&\multicolumn{1}{|c|}{(364,250)}
&\multicolumn{1}{|c|}{(352,800)}
&\multicolumn{1}{|c|}{(603,300)}
&\multicolumn{1}{|c|}{(590,800)}
\\[0.2cm]  \hline \hline 
\multicolumn{1}{|c|}{$\tilde{g}$}
&\multicolumn{1}{|c|}{940}
&\multicolumn{1}{|c|}{910}
&\multicolumn{1}{|c|}{1557}
&\multicolumn{1}{|c|}{1524}
\\[0.2cm]  \hline 
\multicolumn{1}{|c|}{$\tilde{t_1}$,$\tilde{t_2}$}
&\multicolumn{1}{|c|}{662,817}
&\multicolumn{1}{|c|}{753,896}
&\multicolumn{1}{|c|}{1115,1285}
&\multicolumn{1}{|c|}{1156,1325}
\\[0.2cm]  \hline 
\multicolumn{1}{|c|}{$\tilde{b_1}$,$\tilde{b_2}$}
&\multicolumn{1}{|c|}{689,787}
&\multicolumn{1}{|c|}{804,894}
&\multicolumn{1}{|c|}{1159,1260}
&\multicolumn{1}{|c|}{1220,1312}
\\[0.2cm]  \hline 
\multicolumn{1}{|c|}{$\tilde{u_1}$,$\tilde{u_2}$}
&\multicolumn{1}{|c|}{881,909}
&\multicolumn{1}{|c|}{1144,1164}
&\multicolumn{1}{|c|}{1431,1479}
&\multicolumn{1}{|c|}{1586,1628}
\\[0.2cm]  \hline 
\multicolumn{1}{|c|}{$\tilde{d_1}$,$\tilde{d_2}$}
&\multicolumn{1}{|c|}{878,912}
&\multicolumn{1}{|c|}{1142,1167}
&\multicolumn{1}{|c|}{1425,1481}
&\multicolumn{1}{|c|}{1582,1630}
\\[0.2cm]  \hline 
\multicolumn{1}{|c|}{$\tilde{\tau_1}$,
$\tilde{\tau_2}$; $\tilde{\nu_{\tau}}$}
&\multicolumn{1}{|c|}{165,365; 325}
&\multicolumn{1}{|c|}{567,740; 729}
&\multicolumn{1}{|c|}{255,517; 485}
&\multicolumn{1}{|c|}{586,812; 799}
\\[0.2cm]  \hline 
\multicolumn{1}{|c|}{$\tilde{e_1}$, $\tilde{e_2}$; $\tilde{\nu_{e}}$}
&\multicolumn{1}{|c|}{290,360; 351}
&\multicolumn{1}{|c|}{813,838; 835}
&\multicolumn{1}{|c|}{381,519; 513}
&\multicolumn{1}{|c|}{833,901; 898}
\\[0.2cm]  \hline 
\multicolumn{1}{|c|}{$\chi^{\pm}_i$}
&\multicolumn{1}{|c|}{268,498}
&\multicolumn{1}{|c|}{261,536}
&\multicolumn{1}{|c|}{452,764}
&\multicolumn{1}{|c|}{443,779}
\\[0.2cm]  \hline 
\multicolumn{1}{|c|}{$\chi^0_i$}
&\multicolumn{1}{|c|}{144,268,485,496}
&\multicolumn{1}{|c|}{139,261,526,534}
&\multicolumn{1}{|c|}{239,452,754,763}
&\multicolumn{1}{|c|}{234,443,771,778}
\\[0.2cm]  \hline 
\multicolumn{1}{|c|}{$M_A$,$M_{H^{\pm}}$,$M_H$,$M_h$}
&\multicolumn{1}{|c|}{300,315,300,124}
&\multicolumn{1}{|c|}{300,315,300,124}
&\multicolumn{1}{|c|}{600,608,600,131}
&\multicolumn{1}{|c|}{600,608,600,131}
\\[0.2cm]  \hline 
\end{tabular}
\vskip .1in }
\begin{center}
\nn {\bf Tab.1:}  Particle spectra for $M_t^{\rm pole} =$ 175
GeV, $\tan\beta=50$ for selected $M_A, m_{1/2}$ and $m_0$ values.   \\
\end{center}
\vskip .2in

\end{small}

\smallskip

The sparticle spectra for $M_A =$ 300 and 600 GeV and two sets of $m_{1
/2}$ and (extreme) $m_0$ values are shown in Table 1. In all these cases, the
particle spectrum is very heavy; hence most of the SUSY decay channels
of the Higgs particles are shut for large $\tb$. The only allowed decay
channels are $H,A \rightarrow \tilde{\tau_1} \tilde{\tau_1}, \chi^0_1
\chi^0_1$ and $H^\pm \rightarrow \tilde{\tau_1} \tilde{\nu}$ [for large
$M_A$ values]. 
However, the branching ratios of these decay channels are suppressed by
large $b \bar{b}$ and $t \bar{b}$ widths of the Higgs particles for
large $\tb$: while the supersymmetric decay widths are of the order 
${\cal O}(0.1$ GeV), the decays involving $b$ quarks have widths 
${\cal O}(10$ GeV) and dominate by 2 orders of magnitude. 

\bigskip 

\nn (c) \underline{Additional Constraints}

\bigskip

\nn There are additional experimental constraints on the 
parameter space for both high and low $\tb$; the most important
of these are the $b \rightarrow s \gamma$, $Z \rightarrow 
b \overline{b}$, and dark matter [relic LSP abundance] constraints.
These constraints are much more restrictive in the high $\tb$ case.
\s

Recent studies \cite{R13} have indicated that the combination of $b
\rightarrow s \gamma$, dark matter and $m_b$ constraints 
disfavor the high $\tb$ solution for which the $t$, $b$ and $\tau$
Yukawa couplings are equal, in particular the minimal SUSY--SO(10) model
with universal soft-supersymmetry breaking terms at $M_G$. This model
can, however, be saved if the soft terms are not universal [implying a
higgsino--like lightest neutralino], and there exist theoretical motivations
for non--universal soft terms at $M_G$ \cite{R22}. The presently favored
$Z \ra b\bar{b}$ decay width would favor a very low $A$ mass for large 
$\tb$. \s

For low $\tb$, these additional constraints do not endanger the model,
yet they can significantly reduce the available parameter space. In
particular the $Z \rightarrow b \overline{b}$ constraint
favors a light chargino and light stop for small to moderate
values of $\tb$ \cite{R23,R24} so that they could be detected
at LEP2 \cite{R24}. The dark matter constraint essentially places an
upper limit on $m_0$ and $m_{1/2}$ \cite{R26}.
The $b \rightarrow s \gamma$ constraint \cite{R32}, on the other 
hand, is
plagued with large theoretical uncertainties mainly stemming from the
unknown next-to-leading QCD corrections and uncertainties
in the measurement of $\alpha_s(M_Z)$. However, it is consistent with
the low $\tb$ solution and may in the future be useful in determining
the sign of $\mu$ \cite{R33}. 


\setcounter{equation}{0}
\renewcommand{\theequation}{3.\arabic{equation}}

\subsection*{3. Production Mechanisms}

The main production mechanisms of neutral Higgs bosons at $\ee$ 
colliders are the Higgs--strahlung process and pair production,
\begin{eqnarray}
(a) \ \ {\rm Higgs\mbox{-}strahlung} \hspace{1cm} \ee & 
           \ra &  (Z) \ra Z+h/H 
\hspace{5cm} \non \\
(b) \ \ {\rm pair \ production} \hspace{1cm} \ee & \ra & (Z) \ra A+h/H  
\non
\end{eqnarray}
as well as the $WW$ and $ZZ$ fusion processes, 
\begin{eqnarray}
(c) \ \ {\rm fusion \ processes} \hspace{0.8cm} \ \ee & \ra &  \bar{\nu} 
\nu \ (WW) \ra \bar{\nu} \nu \ + h/H \hspace{3.5cm} \non \\
\ee & \ra &  \ee (ZZ) \ra \ee + h/H  \non
\end{eqnarray}
[The ${\cal CP}$--odd Higgs boson $A$ cannot be produced in the
Higgs--strahlung and fusion processes to leading order since it does not
couple to $VV$ pairs.] 
The charged Higgs particle can be pair produced through virtual photon and
$Z$ boson exchange, 
\begin{eqnarray}
(d) \ \ {\rm charged \ Higgs } \hspace{0.8cm} \ \ee & \ra &  \ (\gamma , 
Z^* ) \ \ra \ H^+ H^- \hspace*{3.93cm} \nonumber 
\end{eqnarray}
[For masses smaller than $\sim 170$ GeV, the charged Higgs boson is 
also accessible in top decays, $\ee \ra t\bar{t}$ with $t \ra H^+b$.] \s

The production cross sections\footnote{The complete analytical expressions 
of the cross sections can be found, e.g., in Ref.\cite{R25}. Note that 
in this 
paper there are a few typos that we correct here: in eq.(20), the factor 
92 should replaced by 96; in the argument of the $\lambda$ function of the 
denominator in eq.(21), the parameter $M_A^2$ should be replaced by 
$M_Z^2$; finally, the minus
sign in the interference term in eq.(25) should be replaced by a plus sign.} 
for the neutral 
Higgs bosons are suppressed by mixing angle factors compared to the SM 
Higgs production,
\begin{eqnarray}
\sigma(\ee \ra Zh) \ , \ \sigma(VV \ra h) \ , \ \sigma(\ee \ra AH) \ \ 
\sim \ \sin^2(\beta-\alpha) \\
\sigma(\ee \ra ZH) \ , \ \sigma(VV \ra H) \ , \ \sigma(\ee \ra Ah) \ \ 
\sim \ \cos^2(\beta-\alpha) 
\end{eqnarray}
while the cross section for the charged Higgs particle does not depend 
on any parameter other than $M_{H^\pm}$. \s

In the decoupling limit, $M_A \gg M_Z$, the $HVV$ coupling vanishes, 
while the $hVV$ coupling approaches the SM Higgs value 
\beq
g_{HVV} & = & \cos(\beta-\alpha) \ra  M_Z^2\sin4\beta /2 M_A^2  \ra 0 \\
g_{hVV} & = & \sin(\beta-\alpha)  \ra 1- {\cal O}(M_Z^4/M_A^4) \ \ \ra 1 
\eeq
Hence, the only relevant mechanisms for the production of the heavy Higgs 
bosons in this limit will be the associated pair production (b) and the pair 
production of the charged Higgs particles (d). The cross sections, in the 
decoupling limit and for $\sqrt{s} \gg M_Z$, are given by [we use $M_H \sim 
M_A$]
\beq
\sigma (\ee \ra AH) &=& \frac{G_F^2 M_Z^4}{96 \pi s} (v_e^2+a_e^2)
\beta_A^3 \\
\sigma (\ee \ra H^+H^-) &=& \frac{2G_F^2 M_W^4 s_W^4 }{3 \pi s } \left[
1+ \frac{v_e v_H}{8 s_W^2 c_W^2} + \frac{(a_e^2+ v_e^2)v_H^2}{
256 c_W^4 s_W^4} \right] \beta_{H^\pm}^3
\eeq
where  $\beta_j=(1-4M_j^2/s)^{1/2}$ is the velocity of Higgs bosons, 
 the $Z$ couplings to electrons are given by $a_e=-1, v_e=-1+4\sin^2
\theta_W$, and to the charged Higgs boson by $v_H=-2+4\sin^2\theta_W$. The
cross sections for $hA$ and $HZ$ production vanish in the decoupling limit
since they are proportional to $\cos^2 (\beta- \alpha)$. \s

The cross section for the fusion process, $\ee \ra \bar{\nu}_e \nu_e H$, is
enhanced at high energies since it scales like $M_W^{-2}\log s/M_H^2$. 
This mechanism
provides therefore a useful channel for $H$ production in the mass range of a
few hundred GeV below the decoupling limit and small values of $\tb$, where
$\cos^2(\beta-\alpha)$ is not prohibitively small; the cross section, though,
becomes gradually less important for increasing $M_H$ and vanishes in the
decoupling limit. In the high energy regime, the $WW\ra H$ 
fusion cross section
is well approximated by the expression 
\beq
\sigma( \ee \ra \bar{\nu}_e \nu_e H ) = \frac{G_F^3 M_W^4}{4 \sqrt{2}\pi^3} 
\left[ \left(1+\frac{M_H^2}{s} \right) \log \frac{s}{M_H^2} -2 \left(1-
\frac{M_H^2}{s} \right) \right] \cos^2(\beta-\alpha) 
\eeq
obtained in the effective longitudinal $W$ approximation. Since the NC
couplings are small compared to the CC couplings, the cross section for the
$ZZ$ fusion process is $\sim 16\cos^4 \theta_W$, {\it i.e.} one order of
magnitude smaller than for $WW$ fusion. \s 

Numerical results for the cross sections are shown in Fig.5 at high--energy 
$\ee$ colliders as a function of $\sqrt{s}$ TeV for the two values
$\tb=1.75$ and 50, and for pseudoscalar masses $M_A=300$, 600 and 900 
GeV [note that $M_H \simeq M_{H^\pm} \simeq M_A$ as evident from Figs. 1 
and 3a]. For a luminosity of
$\int {\cal L}=200$ fb$^{-1}$, typically a sample of about $1000$ $HA$
and $H^+H^-$ pairs are predicted for heavy Higgs masses of $\sim 500$
GeV at $\sqrt{s}=1.5$ TeV. For small $\tb$ values, $\tb \lsim 2$, a few 
hundred events are predicted in the $WW \ra H$ fusion process for $H$ 
masses $\sim 300$ GeV. The cross sections for the $hA$ and $HZ$ processes 
are too low, less than $\sim 0.1$ fb, to be useful for $M_H \gsim 300$ GeV; 
Fig.5b. \s 

Note that the cross sections for the production of the lightest Higgs
boson $h$ in the decoupling limit and for $\sqrt{s} \gg M_Z, M_h$ are 
simply given by
\beq
\sigma (\ee \ra ZZ) &\simeq& \frac{G_F^2 M_Z^4}{96 \pi s} (v_e^2+a_e^2) \\
\sigma( \ee \ra \bar{\nu}_e \nu_e h ) &\simeq& \frac{G_F^3 M_W^4}{4 \sqrt{2}
\pi^3} \log \frac{s}{M_h^2} 
\eeq
The cross sections are the same as for the SM Higgs particle and 
are very large $\sim 100$ fb, 
especially for the $WW$ fusion mechanism.

\setcounter{equation}{0}
\renewcommand{\theequation}{4.\arabic{equation}}

\subsection*{4. Decay Modes}

\noindent {\bf 4.1 Decays to standard  particles }

\smallskip

For large $\tb$ the Higgs couplings to down--type fermions dominate over
all other couplings. As a result, the decay pattern is in general very
simple. The neutral Higgs bosons will decay into $b\bar{b}$ and
$\tau^+\tau^-$ pairs for which the branching ratios are close to $\sim
90$~\% and $\sim 10$~\%, respectively. The charged Higgs particles decay
into $\tau\nu_{\tau}$ pairs below and into $tb$ pairs above the
top--bottom threshold. \s 

The partial decay widths of the neutral Higgs bosons\footnote{We refrain
from a discussion of the $h$ decays which become SM--like in the
decoupling limit. In addition, we discuss only the dominant two--body
decay modes of the heavy Higgs bosons; for an updated and more detailed
discussion, including also three--body decays, see Ref.\cite{R28}.}, $\Phi=H$
and $A$, to fermions are given by \cite{R4} 
\beq
\Gamma(\Phi \ra \bar{f}f) = N_c \frac{G_F M_\Phi }{4 \sqrt{2} \pi} 
g^2_{\Phi ff}m_f^2 \beta_f^p
\eeq
with $p=3(1)$ for the CP--even (odd) Higgs bosons;  $\beta_f=(1-4m_f^2/
M_\Phi^2) ^{1/2}$ is the velocity of the fermions in the final state,
$N_c$ the color factor. For the decay widths to quark  pairs, the QCD 
radiative corrections are large and must be included; for a recent 
update and a more detailed discussion, see Ref.\cite{R27}. \s

\textheight 22.2cm

The couplings of the MSSM neutral Higgs bosons [normalized to the SM Higgs 
coupling $g_{H_{\rm SM}ff} = \left[ \sqrt{2} G_F \right]^{1/2} m_f$
and $g_{H_{\rm SM}VV} = 2 \left[ \sqrt{2} G_F \right]^{1/2} M_V^2$] are given 
in Table 2. 
\vspace*{3mm}

\begin{center}
\begin{tabular}{|c||c|c||c|} \hline
& & & \s
$\hspace{1cm} \Phi \hspace{1cm} $ &$ g_{ \Phi \bar{u} u} $ & $
g_{\Phi \bar{d} d} $ & $ g_{\Phi VV} $ \\
& & & \\ \hline \hline
& & & \\ 
$h$  & \ $\; \cos\alpha/\sin\beta       \; $ \ & \ $ \; -\sin\alpha/
\cos\beta \; $ & $ \sin(\beta-\alpha) $ \\
$H$  & \       $\; \sin\alpha/\sin\beta \; $ \ & \ $ \; \cos\alpha/
\cos\beta \; $ & $ \cos(\beta-\alpha) $ \\
$A$  & \ $\; 1/ \tb \; $        \ & \ $ \; \tb \; $ & $0$ \\[0.3cm] \hline
\end{tabular}
\end{center}

\vspace*{3mm}

\nn {\small {\bf Tab.~2}: Higgs boson couplings in the MSSM to fermions 
and gauge bosons relative to the SM Higgs  couplings.}

\vspace*{4mm}

In the decoupling limit, $M_A \gg M_Z$, we have 
\beq
\cos\alpha & \sim &
\sin\beta -\cos\beta \frac{M_Z^2}{2M_A^2}\sin4\beta 
\ra \sin\beta \\
\sin\alpha& \sim&
{} -\cos\beta+\sin\beta \frac{M_Z^2}{2M_A^2}\sin4\beta 
\ra {}-\cos\beta
\eeq
Therefore the $hff$ couplings reduce to the SM 
Higgs couplings, while the $Hff$ couplings become equal
to those of the pseudoscalar boson $A$,
\beq
\cos\alpha/ \sin\beta & \ra &  \ 1 \non \\
-\sin\alpha/ \cos\beta & \ra & \ 1 \non \\
-\sin\alpha/\sin\beta & \ra & \ 1/\tb \non \\
\cos\alpha/\cos\beta &\ra &  \ \tb 
\eeq

The partial width of the decay mode $H^+\ra u \bar{d}$ is given by
\beq
\Gamma(H^+\ra u\bar{d}) &=& \frac{N_c G_F}{4\sqrt{2} \pi} \frac{\lambda^{1/2}
_{ud,H^{\pm} } } {M_{H^\pm}} \, |V_{ud}|^2 \times \non \\
& & \left[ (M_{H^\pm}^{2} -m_{u}^{2}-m_{d}^{2}) \left( m_{d}^{2} {\tg}^2
\beta + m_u^2{\rm ctg}^2 \beta \right) -4m_u^2m_{d}^2 \right]
\end{eqnarray}
with  $V_{ud}$ the CKM--type matrix element for quarks and
$\lambda$ is the two--body phase space function defined by
\beq
\lambda_{ij,k} = (1-M_i^2/M_k^2- M_j^2/M_k^2)^2-4M_i^2 M_j^2/M_k^4 
\eeq
For decays into quark pairs, the QCD corrections must be also included. \s

Below the $\bar{t}t$ threshold, a variety of channels is open
for the decays of the heavy CP--even Higgs bosons, the most important
being the cascade decays $H \ra \Phi \Phi$ with $\Phi=h$ or $A$, with
a partial width [for real light Higgs bosons]
\begin{eqnarray}
\Gamma(H \ra \Phi \Phi) = \frac{G_F}{16\sqrt{2} \pi} \frac{M_Z^4}{M_H}
 g^2_{H\Phi \Phi} \beta_{\Phi}
\label{HAA}
\end{eqnarray}
where $\beta_{\Phi}=(1-4 M_{\Phi}^2/M_H^2)^{1/2}$ and 
the radiatively corrected three--boson self--couplings [to leading 
order], in units of $g_Z'=(\sqrt{2}G_F)^{1/2}M_Z^2$, are given by 
\begin{eqnarray}
g_{Hhh} &=& 2 \sin 2\alpha \sin (\beta+\alpha) -\cos 2\alpha \cos(\beta
+ \alpha) + 3 \frac{\epsilon}{M_Z^2} \frac{\sin \alpha \cos^2\alpha }{\sin
\beta}
\label{3h2}\\
g_{HAA} &=& - \cos 2\beta \cos(\beta+ \alpha)+ \frac{\epsilon}{M_Z^2}
\frac{\sin \alpha \cos^2\beta }{\sin \beta} \label{3h3} \non 
\end{eqnarray}
In contrast to the previous couplings, the leading $m_t^4$ radiative
corrections cannot be absorbed entirely in the redefinition of the
mixing angle $\alpha$, but they are renormalized by an explicit term
depending on the parameter $\epsilon$ given by [$M_S$ is the common
squark mass at the electroweak scale]
\begin{eqnarray}
\epsilon = \frac{3 G_F}{\sqrt{2} \pi^2} \frac{m_t^4}{ \sin^2\beta} \, \log 
\left( 1+ \frac{M_S^2}{m_t^2} \right)
\end{eqnarray}
In the decoupling limit, these couplings approach the values
\beq
g_{Hhh} & \ra & \frac{3}{2}\sin4\beta \non \\
g_{HAA} & \ra & -\frac{1}{2}\sin4\beta
\eeq

In the mass range above the $WW$ and $ZZ$ thresholds, where the $HVV$ 
couplings
are not strongly suppressed for small values of $\tb$, the partial widths of
the $H$ particle into massive gauge bosons can also be substantial; they are
given by 
\begin{eqnarray}
\Gamma (H \ra VV) =  \frac{ \sqrt{2} G_F \cos^2 (\alpha-\beta)} {32 \pi}
M_H^3 (1-4\kappa_V+12\kappa_V^2) (1-4\kappa_V)^{1/2} \ \delta_V'
\end{eqnarray}
with $\kappa_V=M_V^2/M_H^2$ and $\delta_V'=2(1)$ for  $V=W(Z)$. \s

For small values of $\tb$ and below the $\bar{t}t$ and the $t\bar{b}$ 
thresholds, the pseudoscalar and charged Higgs bosons can decay
into the lightest Higgs boson $h$ and a gauge boson; however these
decays are suppressed by $\cos^2(\beta-\alpha)$ and therefore are 
very rare for large $A$ masses. The partial decay widths are given by
\begin{eqnarray} 
\Gamma(A \ra hZ) &=& \frac{G_F \cos^2 (\beta-\alpha) }{8\sqrt{2} \pi} 
\ \frac{M_Z^4}{M_A} \lambda^{1/2}_{Zh,A} \lambda_{Ah,Z} \non \\
\Gamma(H^{+} \rightarrow hW^+) &=& 
\frac{G_F \cos^2 (\beta-\alpha) }{8\sqrt{2}\pi} \frac{M_W^4}{M_{H^\pm}} 
\lambda^{\frac{1}{2}}_{Wh,H^{\pm}} \lambda_{H^{\pm}h,W} 
\hspace*{0.2cm}
\end{eqnarray}

In the decoupling limit, the partial widths of all decays of the
heavy CP--even, CP--odd and charged Higgs bosons involving gauge bosons
vanish since $\cos^2(\beta-\alpha)\ra0$. In addition, the $H \ra hh$  
decay width is very small
since it is inversely proportional to $M_H$, and $H\ra AA$ is not 
allowed kinematically. Therefore, the only
relevant channels are the decays into $\bar{b}b/\bar{t}t$ for the
neutral and $t\bar{b}$ for the charged Higgs bosons. The total decay widths of
the three bosons $H,A$ and $H^\pm$, into standard particles can be 
approximated in this limit by 
\beq
\Gamma(H_k \ra {\rm all}) = \frac{3 G_F}{4\sqrt{2} \pi} M_{H_k} 
\left[ m_{b}^{2} {\tg}^2\beta + m_t^2{\rm ctg}^2 \beta \right]
\eeq 
[We have neglected the small contribution of the decays into $\tau$ leptons 
for large $\tb$.] 

\vspace{0.3in}

\noindent {\bf 4.2 Decays to charginos and neutralinos}

\vspace{0.3in}

The decay widths of the Higgs bosons $H_k$  [$k=(1,2,3,4)$ correspond
to $ (H,h,A,H^\pm)$] into neutralino and chargino pairs are given by 
\cite{R29}
\begin{eqnarray}
\Gamma (H_k \ra \chi_i \chi_j) = \frac{G_F M_W^2}{2 \sqrt{2} \pi}
\frac{ M_{H_k} \lambda_{ij,k}^{1/2} }{1+\delta_{ij}} \hspace*{-3mm}
&& \left[ (F_{ijk}^2 + F_{jik}^2) \left(1- \frac{ m_{\chi_i}^2}{M_{H_k}^2}
        - \frac{ m_{\chi_j}^2}{M_{H_k}^2} \right) \right. \non \\
&& \left. -4\eta_k \epsilon_i \epsilon_j F_{ijk} F_{jik} \frac{ m_{\chi_i} 
m_{\chi_j}} {M_{H_k}^2} \right]
\end{eqnarray}
where $\eta_{1,2,4}=+1$, $\eta_3=-1$ and $\delta_{ij}=0$ unless the
final state consists of two identical (Majorana) neutralinos in which
case $\delta_{ii}=1$; $\epsilon_i =\pm 1$ stands for the sign of the 
$i$'th eigenvalue of the neutralino mass matrix [the matrix $Z$ is defined 
in the convention of Ref.\cite{R2a}, and the eigenvalues of the mass matrix 
can be either positive or negative] while  $\epsilon_i=1$ for charginos;
$\lambda_{ij,k}$ is the usual two--body phase 
space function given in eq.(4.4).\s 

In the case of neutral Higgs boson decays, the coefficients $F_{ijk}$ are
related to the elements of the matrices $U,V$ for charginos and
$Z$ for neutralinos,
\begin{eqnarray}
H_k \ra \chi_i^+ \chi_j^- \ &:& F_{ijk}= \frac{1}{\sqrt{2}} \left[
e_k V_{i1}U_{j2} -d_k V_{i2}U_{j1} \right] \\
H_k \ra \chi_i^0 \chi_j^0 \ \ &:& F_{ijk}= \frac{1}{2}
\left( Z_{j2}- \tan\theta_W
Z_{j1} \right) \left(e_k Z_{i3} + d_kZ_{i4} \right) \ + \ i \leftrightarrow j
\end{eqnarray}
with the coefficients $e_k$ and $d_k$ given by
\begin{eqnarray}
e_1/d_1=\cos\alpha/-\sin \alpha \ , \
e_2/d_2=\sin\alpha/\cos \alpha \ ,  \
e_3/d_3=-\sin\beta/\cos \beta
\end{eqnarray}
For the charged Higgs boson, the coupling to neutralino/chargino
pairs are  given by  
\begin{eqnarray}
F_{ij4} &  = & \cos\beta \left[ Z_{j4} V_{i1} + \frac{1}{\sqrt{2}}
\left( Z_{j2} + \tan \theta_W Z_{j1} \right) V_{i2} \right] \non \\
F_{ji4} & = & \sin \beta \left[ Z_{j3} U_{i1} - \frac{1}{\sqrt{2}}
\left( Z_{j2} + \tan \theta_W Z_{j1} \right) U_{i2} \right]
\end{eqnarray}
The matrices $U,V$ for charginos and $Z$ for neutralinos
can be found in Ref.\cite{R2a}.

Since in most of the parameter space discussed in Section 2, the
Higgs--higgsino mass parameter $|\mu|$ turned out to be very large, $|\mu| \gg
M_1 , M_2 , M_Z$, it is worth discussing the Higgs decay widths into charginos
and neutralinos in this limit. First, the decays of the neutral Higgs bosons
into pairs of [identical] neutralinos and charginos $H_k \ra \chi_i \chi_i$
will be suppressed by powers of $M_Z^2/\mu^2$. This is due to the fact that 
neutral Higgs bosons mainly couple to {\it mixtures} of higgsino and gaugino 
components, and in the large $\mu$ limit, neutralinos and charginos are either
pure higgsino-- or pure gaugino--like. For the same reason, decays $H^+ \ra
\chi_{1,2}^0 \chi_1^+$ and $\chi_{3,4}^0 \chi_2^+ $ of the charged Higgs 
bosons
will be suppressed. Furthermore, since in this case 
$M_A$ is of the same order as $|\mu|$, decays into pairs of heavy charginos 
and neutralinos will be kinematically forbidden. Therefore, the channels 
\beq
H,A & \ra & \chi_1^0 \, \chi_{3,4}^0  \ , \ \chi_2^0 \, \chi_{3,4}^0  \ \ 
{\rm and} \ \chi_1^\pm \, \chi_2^\mp \non \\
H^+ & \ra & \chi_1^+ \, \chi_{3,4}^0  \ \ {\rm and} 
\ \chi_2^+ \chi_{1,2}^0 
\eeq
will be the dominant decay channels of the heavy Higgs particles. Up to 
the phase space suppression [i.e. for $M_{A}$ sufficiently larger 
than $|\mu|$], the partial widths of these decay channels, in units of 
$G_F M_W^2 M_{H_k}/( 4 \sqrt{2} \pi)$, are given by \cite{R29}
\beq
\Gamma( H \ra \chi_1^0 \chi_{3,4}^0)& =& 
\frac{1}{2} {\rm tan}^2 \theta_W ( 1 \pm \sin 2\beta) \non \\
\Gamma( H \ra \chi_2^0 \chi_{3,4}^0)&=& 
 \frac{1}{2} (1\pm \sin 2\beta) \non \\ 
\Gamma( H \ra \chi_1^\pm \, \chi_{2}^\mp) & =& 1 \\
\Gamma( A \ra \chi_1^0 \chi_{4,3}^0)  &=&
\frac{1}{2} {\rm tan}^2 \theta_W ( 1 \pm \sin 2\beta) \non \\
\Gamma( A \ra \chi_2^0 \chi_{4,3}^0)  &=& 
\frac{1}{2} (1\pm \sin 2\beta) \non \\
\Gamma( A \ra \chi_1^\pm \, \chi_{2}^\mp)  &=& 1 \\
\Gamma( H^+ \ra \chi_1^+ \, \chi_{3,4}^0) & =& 1 \non \\
\Gamma( H^+ \ra \chi_2^+ \, \chi_{1}^0) &=&  1  \non \\
\Gamma( H^+ \ra \chi_2^+ \, \chi_{2}^0)  &=& {\rm tan}^2 \theta_W 
\hspace{1.3cm}
\eeq
[We have used the fact that in the decoupling limit $\sin 2\alpha=
-\sin 2\beta$.] If all these channels are kinematically allowed, the total
decay widths of the heavy Higgs bosons to chargino and neutralino
pairs will be given by the expression
\begin{eqnarray}
\Gamma (H_k \ra \sum \chi_i \chi_j) = \frac{3 G_F M_W^2}{4 \sqrt{2} \pi}
M_{H_k} \left( 1+\frac{1}{3} \tan^2 \theta_W \right)
\end{eqnarray}
which holds universally for all the three Higgs bosons $H,A,H^\pm$. 

\bigskip 

\noindent {\bf 4.3 Decays to squarks and sleptons}

\bigskip

\nn Decays of the neutral and charged Higgs bosons, $H_k=h,H,A,H^\pm$,
to sfermion pairs can be written as 
\begin{eqnarray}
\Gamma (H_k \ra \tilde{f}_i \tilde{f}_j ) = \frac{N_C G_F }{2 \sqrt{2} 
\pi M_{H_k} } \, \lambda^{1/2}_{ \tilde{f}_i \tilde{f}_j, H_k} \, 
g_{H_k \tilde{f}_i \tilde{f}_j}^2
\end{eqnarray}
$\tilde{f}_{i}$ with $i=1,2$ are the sfermion mass eigenstates
which are related to the current eigenstates $\tilde{f}_{L}, \tilde{f}_{R}$ 
by
\beq
\tilde{f}_1 &=& \ \tilde{f}_L \cos\theta_f + \tilde{f}_R \sin \theta_f 
\non \\
\tilde{f}_2 &=& -\tilde{f}_L \sin\theta_f + \tilde{f}_R \cos \theta_f 
\eeq
The mixing angles $\theta_f$ are proportional to the masses of the partner 
fermions and they are important only in the case of third generation 
sfermions. The couplings $g_{H_k \tilde{f}_i \tilde{f}'_j}$ of the neutral 
and charged Higgs bosons $H_k$ to sfermion mass eigenstates are 
superpositions of the couplings of the current eigenstates, 
\beq
g_{H_k \tilde{f}_i \tilde{f}'_j} = \sum_{\alpha, \beta=L,R} T_{ij \alpha 
\beta} \ g_{\Phi \tilde{f}_\alpha \tilde{f}'_\beta} 
\eeq
The elements of the $4 \times 4$ matrix $T$ are given in Tab.3a. 
The couplings $g_{H_k \tilde{f}_\alpha \tilde{f}'_\beta}$, in the 
current eigenstate basis $\tilde{f}_{\alpha,\beta}=
\tilde{f}_{L,R}$ [normalized to $2 (\sqrt{2} G_F)^{1/2}$] 
may be written as \cite{R4,R29}
\beq
g_{H_k \tilde{f}_L \tilde{f}_L} &=& m_f^2 g^{\Phi }_1 + M_Z^2 (T_3^f 
-e_f s_W^2) g_2^{\Phi } \non \\
g_{H_k \tilde{f}_R \tilde{f}_R} &=& m_f^2 g^{\Phi 
}_1 + M_Z^2 e_f s_W^2 g_2^{\Phi } \non \\
g_{H_k \tilde{f}_L \tilde{f}_R} &=& - \frac{1}{2} m_f 
\left[ \mu g^{\Phi }_3 - A_f g_4^{\Phi } \right] 
\eeq
for the neutral Higgs bosons, $H_k=h,H,A$. $T_3=\pm 1/2$ is the isospin 
of the [left--handed] sfermion and $e_f$ its 
electric charge. The coefficients 
$g_i^{\Phi}$ are given in Tab.3b; in the decoupling limit, the coefficients
$g_2^{\Phi}$ reduce to 
\beq
\cos(\beta+\alpha) & \ra & \sin2\beta \non \\
\sin(\beta+\alpha) & \ra & -\cos2\beta 
\eeq
[for the other coefficients, see eqs.(4.2)].
For the charged Higgs bosons, the couplings [also normalized to $2 (\sqrt{2}
G_F )^{1/2}$] are 
\beq
g_{H^+ \tilde{u}_\alpha \tilde{d}_\beta} &=& - \frac{1}{\sqrt{2}}
\left[  g_1^{\alpha \beta} + M_W^2 g_2^{\alpha \beta} \right]
\eeq
with the coefficients $g_{1/2}^{\alpha \beta}$ with $\alpha, \beta=L,R$ 
listed in Table 3c. \s

\bigskip

\begin{center}
\begin{tabular}{|c|cccc|} \hline
&&&& \\
$i,j \ /  \ \alpha, \beta$ & $ \mbox{\small LL} $ & $
\mbox{\small RR}$ &  $\mbox{\small LR} $ & $\mbox{\small RL}$ \\ 
&&&& \\ \hline 
&&&&  \\
11 & $ \cos \theta_f \cos \theta_{f'} $
   & $ \sin \theta_f \sin \theta_{f'} $
   & $ \cos \theta_f \sin \theta_{f'} $
   & $ \sin \theta_f \cos \theta_{f'} $ \\ 
12 & $ -\cos \theta_f \sin \theta_{f'}$
   & $\sin \theta_f \cos \theta_{f'}$
   & $\cos \theta_f \cos \theta_{f'}$
   & $-\sin \theta_f \sin \theta_{f'}$ \\ 
21 & $-\sin \theta_f \cos \theta_{f'}$
   & $\cos \theta_f \sin \theta_{f'}$
   & $-\sin \theta_f \sin \theta_{f'}$
   & $\cos \theta_f \cos \theta_{f'} $ \\ 
22 & $\sin \theta_f \sin \theta_{f'}$
   & $\cos \theta_f \cos \theta_{f'}$
   & $-\sin \theta_f \cos \theta_{f'} $
   & $-\cos \theta_f \sin \theta_{f'} $\\ 
&&&& \\ \hline
\end{tabular}
\end{center}

\nn {\small {\bf Tab.~3a}: Transformation matrix for the Higgs couplings to 
sfermions in the presence of mixing.}

\bigskip

\begin{center}
\begin{tabular}{|c|c|c|c|c|c|} \hline
& & & & & \\
$ \ \ \tilde{f} \ \ $ & $\ \ \Phi \ \ $ & $\ \ g_1^{\Phi } \ \ $ 
& $g_2^{\Phi }$ & $g_3^{\Phi }$ & $g_4^{\Phi } $ \\ 
&&&&& \\ \hline  &&&&& \\
&$h$ & $\cos\alpha/ \sin \beta $ & $-\sin(\alpha+\beta)$ 
               & $-\sin\alpha/\sin\beta$ & $\cos \alpha/ \sin\beta$ \\
$ \tilde{u}$& $H$ & $\sin\alpha/\sin\beta$ & $\cos(\alpha+\beta)$ & 
    $\cos \alpha /\sin \beta$ & $\sin \alpha/ \sin\beta$ \\ 
& $A$ & $0$ & $0$ & $1$ & $-1/\tb$ \\ 
&&&&& \\
& $h$ & $-\sin\alpha/ \cos \beta $ & $-\sin(\alpha+\beta)$ 
               & $\cos\alpha/\cos\beta$ & $-\sin \alpha/ \cos\beta$ \\
$\tilde{d}$ & $H$ & $\cos\alpha/\cos\beta$ & $\cos(\alpha+\beta)$ & 
    $\sin \alpha /\cos \beta$ & $\cos \alpha/ \cos\beta$ \\ 
&    $A$ & $0$ & $0$ & $1$ & $-\tb$ \\ 
&&&&& \\ \hline
\end{tabular}
\end{center}

\nn {\small {\bf Tab.~3b}: Coefficients in the couplings of 
neutral Higgs bosons to sfermion pairs. } 

\bigskip

\begin{small}
\begin{center}
\begin{tabular}{|c|c|c|c|} \hline
&&& \\
$g^{LL}_{1/2} $ & $g^{RR}_{1/2} $  & $g^{LR}_{1/2} $ & $g^{RL}_{1/2} $ \s
&&& \\ \hline
&&& \\
$ m_u^2/{\rm tan} \beta + m_d^2 {\rm tan} \beta$ &
$ m_u m_d (\tb + 1/ \tb )$ & $m_d (\mu + A_d \tb )$ &
$ m_u( \mu +  A_u / \tb )$ \\
$- \sin 2\beta $ & 0 &0 &0 \\
&&& \\ \hline
\end{tabular}
\end{center}
\end{small}

\nn {\small {\bf Tab.~3c}: Coefficients in the couplings of 
charged Higgs bosons to sfermion pairs. } 

\bigskip

Mixing between sfermions occurs only in the third--generation sector. 
For the first two generations the decay pattern is rather simple. In 
the limit of massless fermions, the pseudoscalar Higgs boson does 
not decay into sfermions since by virtue of CP--invariance it couples 
only to pairs of left-- and right--handed sfermions with the coupling 
proportional to $m_f$. In the asymptotic regime, where the masses 
$M_{H,H^\pm}$ are 
large, the decay widths of the heavy CP--even and charged \cite{Bartl} 
Higgs bosons 
into sfermions are proportional to 
\beq
\Gamma (H,H^+ \ra \tilde{f} \tilde{f} ) \sim \frac{G_F M_W^4}{M_{H}}
\, \sin^2 2 \beta
\eeq
These decay modes can be significant only for low values
$\tb$ [which implies $\sin^2 2 \beta \sim 1$]. However, in this regime 
the decay widths are inversely proportional to $M_H$, and thus cannot 
compete with the decay widths into charginos/neutralinos and ordinary fermions
which increase with increasing Higgs mass. Therefore, the decays
into first and second generations are unlikely to be important. \s

In the case of the third generation squarks, the Higgs decay widths can be
larger by more than an order of magnitude. For instance the decay widths 
of the heavy neutral Higgs boson into top squarks of the same helicity 
is proportional to 
\beq
\Gamma (H \ra \tilde{t} \tilde{t} ) \sim \frac{G_F m_t^4 }{M_{H} {\rm 
tan}^2 \beta } 
\eeq
in the asymptotic region, and it will be enhanced by large coefficients 
[for small $\tb$] compared to first/second generation 
squarks. Conversely, the decay widths into sbottom quarks can be very large 
for large $\tb$. Furthermore, the decays of heavy neutral CP--even 
and CP--odd Higgs bosons to top squarks of different helicities will 
be proportional in the asymptotic region [and for the CP--even, up to 
the suppression by mixing angle] to 
\beq
\Gamma (H,A \ra \tilde{t} \tilde{t} ) \sim \frac{G_F m_t^2 }{M_{H}} \left[ 
\mu + A_t/ {\rm tan} \beta \right]^2 
\eeq
For $\mu$ and $A_t$ values of the order of the Higgs boson masses, these 
decay widths will be competitive with the chargino/neutralino and standard
fermion decays. Therefore, if kinematically allowed, these decay modes
have to be taken into account. 

\bigskip

\noindent {\bf 4.4 Numerical results}

\bigskip

The decay widths of the $H,A$ and $H^\pm$ Higgs bosons into the sum of
charginos and neutralinos, squark or slepton final states, as well as the
standard and the total decay widths are shown in Figs.6a, 7a and 8a 
as a function of
$m_{1/2}$ for two values of the pseudoscalar Higgs boson mass $M_A=300$
and $600$ GeV, and for positive and negative $\mu$ values; $\tb$ is
fixed to $1.75$. \s 

Fig.6a shows the various decay widths for the heavy CP--even Higgs
boson. For $M_A=300$ GeV, the $H\ra t\bar{t}$ channel is still closed
and the decay width into standard particles is rather small, being of
${\cal O}(250)$ MeV. In this case, the decays into the lightest stop
squarks which are kinematically allowed for small values of $m_{1/2}$ 
will be by far the dominant decay channels. This occurs in most of the
$m_{1/2}$ range if $\mu>0$, but if $\mu<0$ only for $m_{1/2} \lsim 50$ GeV 
which is already ruled out by CDF and LEP1.5 data. \s 

The decays into charginos and neutralinos, although one order of magnitude
smaller than stop decays when allowed kinematically, are also very important.
They exceed the standard decays in most of the $m_{1/2}$ range, except for
large values of $m_{1/2}$ and $\mu <0$  where no more phase space 
is available for the Higgs boson to decay into combinations of the heavy 
and light chargino/neutralino states. For small $m_{1/2}$ values, chargino and 
neutralino decays can be larger than the standard decays by up to an 
order of magnitude. \s

As expected, the decay widths into sleptons are rather small and they never
exceed the widths into standard particles, except for large values of 
$m_{1/2}$.
Note that due to the isospin and charge assignments, the coupling of the $H$
boson to sneutrinos is approximately a factor of two 
larger than the coupling to
the charged sleptons. Since all the sleptons of the three generations are
approximately mass degenerate [the mixing in the $\tilde{\tau}$ sector is very
small for low values of $\tb$], the small decay widths into sleptons 
are given by the approximate relation: 
$\Gamma (H \ra \tilde{\nu} \tilde{\nu}) \simeq 4 \Gamma (H \ra \tilde{l}_L
\tilde{l}_L) \simeq 4\Gamma( H \ra \tilde{l}_R \tilde{l}_R)$. \s 

For larger values of $M_H$, $M_H \gsim 350$ GeV, the decay widths into
supersymmetric particles have practically the same size as discussed
previously. However, since the $H\ra t\bar{t}$ channel opens up, the decay
width into standard model particles becomes rather large, ${\cal O}(10$ GeV),
and the supersymmetric decays are no longer dominant. For $M_H\simeq 600$
GeV, Fig.6a, only the $H \ra \tilde{q} \tilde{q}$ decay width can be larger
than the decay width to standard particles; this occurs in the lower range of
the $m_{1/2}$ values. The chargino/neutralino decays have a branching ratio of
$\sim 20\%$, while the branching ratios of the decays into sleptons are below
the $1\%$ level. \s

Fig.6b and 6c show the individual decay widths of the heavy $H$ boson with a
mass $M_H \simeq 600$ GeV into charginos, neutralinos, stop quarks
and sleptons for the set of parameters introduced previously. For decays into
squarks, only the channels $H\ra \tilde{t}_1 \tilde{t}_1, \tilde{t}_1
\tilde{t}_2$, and in a very small range of $m_{1/2}$ values the channel
$H\ra \tilde{b}_1 \tilde{b}_1$, are allowed kinematically [see Fig.3c].
The decay into two different stop states is suppressed by the [small]
mixing angle, and due to the larger phase space the decay
$H \ra \tilde{t}_1 \tilde{t}_1$ is always dominating. \s 

For the decays into chargino and neutralinos, the dominant channels are decays
into mixtures of light and heavy neutralinos and charginos, in particular $H
\ra \chi_1^+ \chi_2^-$ and $H \ra \chi_1^0 \chi_3^0$ or $\chi_2^0 \chi_3^0$.
This can be qualitatively explained, up to phase space suppression factors,  
by recalling the approximate values of the relative branching ratios in 
the large $|\mu|$ limit given in eqs.(4.18--20):
$\Gamma( H \ra \chi_1^\pm \chi_2^\mp) \sim 1$, while $ \Gamma(H \ra \chi_2^0
\chi_3^0) \sim 1$ and $ \Gamma(H \ra \chi_1^0 \chi_3^0) \sim \tan^2 \theta_W$
because $\sin 2\beta$ is close to one.  The mixed decays involving $\chi_4^0$
are suppressed since they are proportional to $(1-\sin 2\beta)$, and all 
other decay channels are suppressed by powers of $M_Z^2/\mu^2$ for large
$|\mu|$ values. \s

The decay widths for the pseudoscalar Higgs boson are shown in Fig.7a. There
are no decays into sleptons, since the only decay allowed by CP--invariance, $A
\ra \tilde{\tau}_1 \tilde{\tau}_2$, is strongly suppressed by the very small
$\tilde{\tau}$ mixing angle. For $M_A=300$ GeV, the decay into the two stop
squark eigenstates, $A\ra \tilde{t}_1 \bar{\tilde{t}}_2$, is not allowed
kinematically and the only possible supersymmetric decays are the decays into
charginos and neutralinos. The sum of the decay widths into these states can be
two orders of magnitude larger than the decay width into standard particles. \s

For values of $M_A$ above the $t\bar{t}$ threshold, the decay width into
charginos and neutralinos is still of the same order as for low $M_A$, but
because of the opening of the $A \ra t\bar{t}$ mode, the total decay width
increases dramatically and the chargino/neutralino decay branching ratio drops
to the level of $20\%$. As in the case of the heavy CP--even Higgs boson $H$,
the relative decay widths of the pseudoscalar boson into charginos and
neutralinos, Fig.7b, are larger for the channels involving mixtures of light
and heavy neutralinos or charginos; the dominant decay modes are, roughly,
$A\ra \chi_1^+ \chi_2^-$ and $A \ra \chi_1^0 \chi_4^0$ or $\chi_2^0 \chi_4^0$.
Again, this can be qualitatively explained, up to phase space suppression
factors,  by recalling the approximate formulae of eqs.(4.18--19), since the
situation is the same as for $H$, with the two neutralino states $\chi_3^0$ and
$\chi_4^0$ being interchanged. \s 

For small values of the common gaugino mass, $m_{1/2} \lsim 100$ GeV, the decay
mode of the pseudoscalar Higgs boson into stop squarks, $A\ra \tilde{t}_1
\bar{\tilde{t}}_2$, is phase space allowed. In this case, it is competitive
with the top--antitop decay mode. As discussed previously, the $1/M_A^2$
suppression [and to a lesser extent the suppression due to the mixing angle]
of the $A\ra \tilde{t}_1 \bar{\tilde{t}}_2$ decay width compared to 
$\Gamma (A \ra t\bar{t})$ will be compensated by the enhancement of the 
$A \tilde{t}_1 \bar{\tilde{t}}_2$ coupling for large values of $\mu$ and 
$A_t$. \s

Fig.8a shows the decay widths for the charged Higgs boson. Since the dominant
decay channel $H^+ \ra t\bar{b}$ is already open for $M_{H^\pm} \simeq 300$ 
GeV [although still slightly suppressed by phase space], the charged 
Higgs decay width into standard particles is rather large and it increases
only by a factor of $\sim 4$ when increasing the pseudoscalar mass to 
$M_A=600$ GeV. 
The situation for the supersymmetric decays is quite similar for the two 
masses: the chargino/neutralinos decay modes have branching ratios of the 
order of a few ten percent, while the branching ratios for the decays 
into sleptons, when kinematically allowed, do not exceed the level of a few 
percent, as expected. Only the decay $H^+ \ra \tilde{t}_1 \tilde{b}_1$, 
the only squark decay mode allowed by phase space [see Fig.3c] for relatively 
low values of $m_{1/2}$, is competitive with the $t\bar{b}$  decay mode. \s

The decay widths of the charged Higgs into the various combinations of
charginos and neutralinos are shown in Fig.8b for $M_{H^\pm} \sim 600$ GeV. 
The dominant channels are again decays into mixtures of gauginos and 
higgsinos, since $|\mu|$ is large. The pattern follows approximately the 
rules of eq.(4.22), modulo phase suppression. \s

As discussed in section 2, since the chargino, neutralino and sfermion
masses scale as $M_A$, the situation for even larger values of the
pseudoscalar Higgs boson mass, $M_A \sim 1$ TeV, will be qualitatively
similar to what has been discussed for $M_A \sim 600$ GeV. The only
exception is that there will be slightly more phase space available for
the supersymmetric decays to occur. 


\setcounter{equation}{0}
\renewcommand{\theequation}{5.\arabic{equation}}

\subsection*{5. Final Decay Products of the Higgs Bosons}

In this section, we will qualitatively describe the final decay products
of the produced Higgs bosons. Assuming that $M_A$ is large, $M_A \gsim
500$ GeV, the decays into standard particles [and more precisely, the
$t\bar{t}$ for the neutral and the $t\bar{b}$ decays for the charged
Higgs bosons] always have substantial branching ratios, even for the
value $\tb=1.75$ which will be chosen for the discussion. Therefore, to
investigate decays into SUSY particles in the main production processes,
$\ee \ra HA$ and $H^+H^-$, one has to look for final states where one of
the Higgs bosons decays into standard modes while the other Higgs boson
decays into charginos, neutralinos or stop squarks. As discussed
previously, the decays into the other squarks are disfavored by phase space,
while the branching ratios into sleptons are always small
and can be neglected. \s 

Let us first discuss the case where one of the Higgs bosons decays into
chargino and neutralino pairs, 
\beq
\ee & \ra & H \ A \ \ra [\,t\bar{t}\,]\,[\,\chi^+ \chi^-\,] \ \ 
\mbox{and} \ \ 
[\,t\bar{t}\,]\,[\,\chi^0 \chi^0\,] 
\ \ \non \\
\ee & \ra & H^+ H^- \ra [\,tb\,]\,[\,\chi^\pm \chi^0\,] 
\eeq
The lightest chargino $\chi_1^+$ and next--to--lightest neutralino 
$\chi_{2}^0$ decay into [possibly virtual] $W,Z$ 
and the lightest Higgs boson $h$, assuming that decays into sleptons and 
squarks are kinematically disfavored. In the limit of large $|\mu|$, the 
decay widths [in the decoupling limit] are proportional to 
\cite{R30}
\beq
\Gamma( \chi_1^+ \ra  \chi_1^0 W^+) & \sim & \sin^2 2\beta \\
\Gamma( \chi_2^0 \ra  \chi_1^0 Z )  & \sim & \cos^2 2\beta 
\left[ (M_2-M_1)/2\mu  \right]^2 \non \\
\Gamma( \chi_2^0 \ra  \chi_1^0 h ) & \sim & \sin^2 2 \beta 
\eeq
In most of the parameter space, the $W/Z/h$ are virtual [in addition to the 
three--body phase space factors, the decay widths are suppressed by powers 
of $M_2 M_Z/\mu^2$] except near the upper values of $m_{1 /2}$.
In the case of $\chi_2^0$, the channel $ \chi_2^0 \ra \chi_1^0 Z$ mode is 
always dominant although suppressed by additional powers of $M_2^2/\mu^2$ 
compared to the $ \chi_2^0 \ra h \chi_1^0$ mode, since both $h$ and $Z$ are 
off--shell, and the
$Z$ boson width is much larger than the width of the $h$ boson for small
values of $\tb$. The radiative decay $\chi_2^0 \ra \chi_1^0 \gamma$
should play a marginal role except for very small values of $m_{1 /
2}$ where the difference between the $\chi_2^0$ and $Z$ boson masses
becomes too large. \s 

For large values of $m_{1 /2}$, the sleptons become rather light 
compared to the gauginos
and the decays of the light chargino and neutralino into
leptons+sleptons are kinematically possible. In this case, these cascade
decays become dominant since the partial widths for large $|\mu|$ are
given by 
\beq
\sum_{l} \Gamma(\chi_2^0 \ra l \tilde{l}) =
\sum_{l} 2 \Gamma(\chi_1^\pm \ra l \tilde{\nu}) = \frac{3 G_F^2 M_W^2}{\sqrt{2}
\pi} M_2 
\eeq
and therefore not suppressed by powers of $M_Z M_2/\mu^2$, unlike the
previous decay modes [we assume of course that there is no suppression by
phase--space]. The sleptons will then decay into the LSP and massless
leptons, leading to multi--lepton final states. \s 

The heavier chargino, in the absence of squark and slepton decay modes, will
decay preferentially into the lightest chargino and neutralinos plus 
gauge or light Higgs bosons. The decay widths, in units of 
$G_FM_W^2 |\mu| /(8\sqrt{2}\pi)$ may be approximated in the 
decoupling limit by \cite{R30}
\beq
\chi_2^+ &  \ra &   \  \chi_1^+ Z \  \ : \ \Gamma = 1  \non \\
         & \ra &  \  \chi_1^+ h \  \ : \ \Gamma = 1 \non \\
         & \ra & \  \chi_1^0 W^+ \  \ : \ \Gamma = \tan ^2 \theta_W \non \\
         & \ra & \  \chi_2^0 W^+ \  \ : \ \Gamma = 1 
\eeq
The branching ratios for the various final states are roughly equal. Since 
$\chi_2^+$ is almost higgsino--like, the decay  widths into sleptons and 
partners of the light quarks, when kinematically allowed, are extremely 
small since they are suppressed by powers of $m_f^2/M_Z^2$. Because
of the large $m_t$ value, only the decays into stop squarks and bottom 
quarks will be very important. This decay is allowed in most of the parameter 
space for $M_A \gsim 600$ GeV and, up to suppression by 
mixing angles, it is enhanced by a power $m_t^2$ \cite{R30}
\beq
\frac{\Gamma( \chi_2^+ \ra \tilde{t} b )} 
{\Gamma( \chi_2^+ \ra W,Z,h) } \sim \frac{3m_t^2}{M_W^2} \frac{1}{\sin^2\beta
(3+\tan^2\theta_W)} \sim 4
\eeq
compared to the other decays. Therefore, when kinematically possible, 
this decay will be the dominant decay mode of the heavy charginos. \s

For the heavier neutralinos, $\chi_{3,4}^0$, the decay
widths into $W/Z/h$ bosons, again in units of $G_FM_W^2|\mu| 
/(8\sqrt{2}\pi)$ may be be written in the decoupling limit as \cite{R30}
\beq
\chi_{3/4}^0 & \ra & \  \chi_1^0 Z \ \  \ : \ \Gamma = \frac{1}{2}
\tan^2 \theta_W (1\pm \sin 2 \beta)   \non \\
             & \ra & \  \chi_1^0 h \ \  \ : \ \Gamma = \frac{1}{2}
\tan^2 \theta_W (1\mp \sin 2 \beta)   \non \\
             & \ra &  \  \chi_2^0 Z \ \  \ : \ \Gamma = \frac{1}{2} (1\pm 
\sin 2 \beta)   \non \\
             & \ra & \ \chi_2^0 h \ \  \ : \ \Gamma = \frac{1}{2} (1 \mp 
\sin 2 \beta)   \non \\
             &  \ra & \  \chi_1^+ W^- \ \  \ : \ \Gamma = 2 
\eeq
The dominant mode is the charged decay, $\chi^0_{3,4} \ra \chi_1^+W^- $,
followed by the modes involving the $h(Z)$ boson for $\chi^0_4
(\chi^0_3)$. Because $\sin 2\beta \sim 1$, only one of the $h$ or $Z$
decay channels is important. Here again, because of the higgsino nature
of the two heavy neutralinos, the decay widths into sleptons and the
scalar partners of the light quarks are negligible; the only important
decays are the stop decays, $\chi_{3,4}^0 \ra t \tilde{t}_1$,  when they
are allowed kinematically [i.e. for not too large values of $m_{1 /
2}]$. The ratio between stop and $W/Z/h$ decay widths, up to suppression 
by mixing angles, is also given by eq.(5.6), and the stop decays will 
therefore dominate. 

\bigskip

We now turn to the case where one of the produced Higgs particles decays 
into stop squarks
\beq
\ee & \ra & \ H \ A \ \ra [\,t\bar{t}\,]\,[\,\tilde{t}_1 \tilde{t}_1\,] \ \  
\mbox{and} \ \ [\,t\bar{t}\,]\,[\,\tilde{t}_1 \tilde{t_2}\,] 
\ \ \non \\
\ee & \ra & H^+ H^- \ra [tb][\tilde{t}_1 \tilde{b}_1] 
\eeq
{}From the squark mass plots, Fig.~3c, the only decay modes of the lightest 
stop squark allowed by phase space are
\beq
\tilde{t}_1 \ra t\chi_1^0 \ \ , \ \ t\chi_2^0 \ \ , \ \ b \chi_1^+
\eeq
Only the last decay mode occurs for relatively small values of $m_{1 
/2}$, since $m_{\tilde{t}_1} < m_t + m_{\chi^0_{1,2}}$ in this case. 
For larger values of 
$m_{1 /2}$, $\tilde{t}_1$ is heavy enough to decay into top quarks plus the 
lightest neutralinos. For these $m_{1 /2}$ values, the three decay
modes of eq.(5.9) will have approximately the same magnitude since the chargino 
and the neutralinos are gaugino--like and there is no enhancement due to 
the top mass for the $\tilde{t}_1 \ra t \chi^0$ decays. \s

The heavier stop squark, in addition to the previous modes, has 
decay channels with $\tilde{t}_1$ and $Z/h$ bosons in the
final state 
\beq
\tilde{t}_2 \ra \tilde{t}_1 Z \ \ , \ \ \tilde{t}_1 h 
\eeq
These decays, in particular the decay into the lightest Higgs boson
$h$, will be dominant in the large $|\mu|$ limit, since they will be 
enhanced by powers of $\mu^2$.

\setcounter{equation}{0}
\renewcommand{\theequation}{A\arabic{equation}}

\subsection*{Appendix A: Chargino and Neutralino Masses and Couplings}

In this Appendix we collect the analytical expressions of the chargino 
and neutralino masses and couplings, and we discuss the limit 
in which the  Higgs--higgsino mass parameter $|\mu|$ is large. \s

The general chargino mass matrix \cite{R2a},
\begin{eqnarray}
{\cal M}_C = \left[ \begin{array}{cc} M_2 & \sqrt{2}M_W \sin \beta
\\ \sqrt{2}M_W \cos \beta & \mu \end{array} \right]
\end{eqnarray}
is diagonalized by two real matrices $U$ and $V$, 
\begin{eqnarray}
U^* {\cal M}_C V^{-1} \ \ \ra \ \ U={\cal O}_- \ {\rm and} \ \ V = 
\left\{
\begin{array}{cc} {\cal O}_+ \ \ \ & {\rm if \ det}{\cal M}_C >0  \\
            \sigma  {\cal O}_+ \ \ \ & {\rm if \ det}{\cal M}_C <0  
\end{array}
\right. 
\end{eqnarray}
where $\sigma$ is the matrix 
\begin{eqnarray}
{\sigma} = \left[ \begin{array}{cc} \pm 1 & 0
\\ 0 & \pm 1 \end{array} \right] 
\end{eqnarray}
with the appropriate signs depending upon the values of $M_2$, $\mu$, and
$\tan\beta$ in the chargino mass matrix.
${\cal O}_\pm$ is given by:
\begin{eqnarray}
{\cal O}_\pm = \left[ \begin{array}{cc} \cos \theta_\pm & \sin \theta_\pm
\\ -\sin \theta_\pm & \cos \theta_\pm \end{array} \right] 
\end{eqnarray}
with
\begin{eqnarray}
\tan 2 \theta_- &= & \frac{ 2\sqrt{2}M_W(M_2 \cos \beta
+\mu \sin \beta)}{ M_2^2-\mu^2-2M_W^2 \cos \beta} \non \\
\tan 2 \theta_+ & = & \frac{ 2\sqrt{2}M_W(M_2 \sin \beta
+\mu \cos \beta)}{M_2^2-\mu^2 +2M_W^2 \cos \beta} 
\end{eqnarray}
This leads to the two chargino masses, the 
$\chi_{1,2}^+$ masses
\begin{eqnarray}
m_{\chi_{1,2}^+} = && \frac{1}{\sqrt{2}} \left[ M_2^2+\mu^2+2M_W^2
\right. \non \\
&& \left. \mp \left\{ (M_2^2-\mu^2)^2+4 M_W^4 \cos^2 2\beta+4M_W^2 (M^2_2+\mu^2
+2M_2\mu \sin 2\beta) \right\}^{\frac{1}{2}} \right]^{\frac{1}{2}}
\end{eqnarray}
In the limit $|\mu| \gg M_2, M_Z$, the masses of the two charginos reduce to
\begin{eqnarray}
m_{\chi_{1}^+} & \simeq &  M_2 - \frac{M_W^2}{\mu^2} 
\left( M_2 +\mu \sin 2 \beta
\right) \non \\
m_{\chi_{2}^+} & \simeq & |\mu| + 
\frac{M_W^2}{\mu^2} \epsilon_\mu \left( M_2 \sin 
2 \beta +\mu \right) 
\end{eqnarray}
where $\epsilon_\mu$ is for the sign of $\mu$. For $|\mu| \ra \infty$,
the lightest chargino corresponds to a pure wino state with mass 
$m_{\chi_{1}^+} \simeq M_2$, while the heavier chargino corresponds to a 
pure higgsino state with a mass $m_{\chi_{1}^+} = |\mu|$.  

\bigskip 

In the case of the neutralinos, the four-dimensional neutralino mass matrix 
depends on the same two mass parameters $\mu$ and $M_2$, if the GUT 
relation
$M_1=\frac{5}{3} \tan^2 \theta_W$ $ M_2 \simeq \frac{1}{2} M_2$ 
\cite{R2a} is used. In the $(-i\tilde{B}, -i\tilde{W}_3, \tilde{H}^0_1,$ 
$\tilde{H}^0_2)$ basis, it has the form  
\begin{eqnarray}
{\cal M}_N = \left[ \begin{array}{cccc}
M_1 & 0 & -M_Z s_W \cos\beta & M_Z  s_W \sin\beta \\
0   & M_2 & M_Z c_W \cos\beta & -M_Z  c_W \sin\beta \\
-M_Z s_W \cos\beta & M_Z  c_W \cos\beta & 0 & -\mu \\
M_Z s_W \sin \beta & -M_Z  c_W \sin\beta & -\mu & 0
\end{array} \right]
\end{eqnarray}

\smallskip

It can be diagonalized analytically \cite{R31} by a single real matrix $Z$;
the [positive] masses of the neutralino states $m_{\chi_i^0}$ are given by
\beq
\epsilon_1 m_{\chi_1^0} &=& C_1 -\left( \frac{1}{2} a- \frac{1}{6}C_2
\right)^{1/2} + \left[ - \frac{1}{2} a- \frac{1}{3}C_2 + \frac{C_3}
{(8a-8C_2/3)^{1/2}} \right]^{1/2} \non \\
\epsilon_2 m_{\chi_2^0} &=& C_1 +\left( \frac{1}{2} a- \frac{1}{6}C_2
\right)^{1/2} - \left[ - \frac{1}{2} a- \frac{1}{3}C_2 - \frac{C_3}
{(8a-8C_2/3)^{1/2}} \right]^{1/2} \non \\
\epsilon_3 m_{\chi_3^0} &=& C_1 -\left( \frac{1}{2} a- \frac{1}{6}C_2
\right)^{1/2} - \left[ - \frac{1}{2} a- \frac{1}{3}C_2 + \frac{C_3} 
{(8a-8C_2/3)^{1/2}} \right]^{1/2} \non \\
\epsilon_4 m_{\chi_4^0} &=& C_1 +\left( \frac{1}{2} a- \frac{1}{6}C_2
\right)^{1/2} + \left[ - \frac{1}{2} a- \frac{1}{3}C_2 - \frac{C_3}
{(8a-8C_2/3)^{1/2}} \right]^{1/2}
\eeq
where $\epsilon_i = \pm 1$; the coefficients $C_i$ and $a$ are given by
\beq
C_1 &=& (M_1+M_2)/4 \non \\
C_2 &=& M_1 M_2 - M_Z^2 -\mu^2 -6 C_1^ 2 \non \\
C_3 &=& 2 C_1 \left[ C_2 + 2 C_1^2 +2 \mu^2 \right]+
M_Z^2 (M_1 c_W^2 + M_2 s_W^2) - \mu M_Z^2 \sin 2 \beta \non \\
C_4 &=& C_1 C_3- C_1^2 C_2 -C_1^4 -M_1 M_2 \mu^2 +(M_1 c_W^2 + M_2 s_W^2)
M_Z^2 \mu \sin 2\beta
\eeq
and
\beq
a = \frac{1} {2^{1/3}} {\rm Re} \left[ S+ i \left( \frac{D}{27} 
\right)^{1/2} \right]^{1/3}
\eeq
with
\beq
S &=& C_3^2+\frac{2}{27} C_2^3 -\frac{8}{3} C_2 C_4 \non \\
D &=& \frac{4}{27} (C_2^2 +12 C_4)^3 -27 S^2 
\eeq

In the limit of large $|\mu|$ values, the masses of the neutralino states 
simplify to 
\begin{eqnarray}
m_{\chi_{1}^0} &\simeq& M_1 - \frac{M_Z^2}{\mu^2} \left( M_1 +\mu \sin 2 \beta
\right) s_W^2 \non \\
m_{\chi_{2}^0} &\simeq& M_2 - \frac{M_Z^2}{\mu^2} \left( M_2 +\mu \sin 2 \beta
\right) c_W^2 \non \\
m_{\chi_{3}^0} &\simeq& |\mu| + \frac{1}{2}\frac{M_Z^2}{\mu^2} \epsilon_\mu 
(1-\sin 2\beta) \left( \mu + M_2 s_W^2+M_1 c_W^2 \right) \non \\
m_{\chi_{4}^0} &\simeq& |\mu| + \frac{1}{2}\frac{M_Z^2}{\mu^2} \epsilon_\mu 
(1+\sin 2\beta) \left( \mu - M_2 s_W^2 - M_1 c_W^2 \right) 
\end{eqnarray}
Again, for $|\mu| \ra \infty$, two neutralinos are pure gaugino states 
with masses $m_{\chi_{1}^0} \simeq M_1$ , $m_{\chi_{2}^0} =M_2$, while
the two others are pure higgsino states, with masses 
$m_{\chi_{3}^0} \simeq m_{\chi_{4}^0} \simeq |\mu|$. 

\bigskip

The matrix elements of the diagonalizing matrix, $Z_{ij}$ with $i,j=1,..4$,
are given by
\begin{eqnarray}
Z_{i1} &=& \left[ 1+ \left(\frac{Z_{i2}}{Z_{i1}}\right)^2+\left(\frac{Z_{i3}}
{Z_{i1}}\right)^2+\left(\frac{Z_{i4}}{Z_{i1}}\right)^2 \right]^{-1/2} \\
\frac{Z_{i2}}{Z_{i1}} &=&  -\frac{1}{\tan \theta_W} \frac{ M_1-
\epsilon_i  m_{\chi_{i}^0} } {M_2 -\epsilon_i m_{\chi_{i}^0} } \non \\
\frac{Z_{i3}}{Z_{i1}} &=&  \frac{
\mu (M_1-\epsilon_i  m_{\chi_{i}^0} )(M_2 -\epsilon_i m_{\chi_{i}^0} )
-M_Z^2 \sin \beta \cos \beta [(M_1-M_2)c_W^2+M_2 -\epsilon_i m_{\chi_{i}^0}] } 
{M_Z (M_2 -\epsilon_i m_{\chi_{i}^0} )s_W [\mu\cos \beta 
+\epsilon_i m_{\chi_{i}^0} \sin \beta) } \non \\
\frac{Z_{i4}}{Z_{i1}} &=&  \frac{
-\epsilon_i m_{\chi_{i}^0} (M_1-\epsilon_i m_{\chi_{i}^0} )(M_2 -\epsilon_i 
m_{\chi_{i}^0} ) -M_Z^2 \cos^2 \beta [(M_1-M_2)c_W^2+M_2 
-\epsilon_i  m_{\chi_{i}^0}] } 
{M_Z (M_2 -\epsilon_i m_{\chi_{i}^0} )s_W [\mu\cos \beta
+\epsilon_i m_{\chi_{i}^0} \sin \beta) } \non 
\end{eqnarray}
where $\epsilon_i$ is the sign of the $i$th eigenvalue of the neutralino 
mass matrix, which in the large $|\mu|$ limit are: $\epsilon_1
=\epsilon_2=1$ and $\epsilon_4=-\epsilon_3=\epsilon_\mu$. 

\setcounter{equation}{0}
\renewcommand{\theequation}{B\arabic{equation}}

\subsection*{Appendix B: Sfermion Masses and Mixing}

We now present the explicit expressions of the squark and slepton masses. 
We will assume a universal scalar mass $m_0$ and gaugino mass $m_{1/2}$ at 
the GUT scale, and we will neglect the Yukawa couplings in the RGE's 
[see Appendix C]. For third generation squarks this is a poor approximation 
since these couplings can be large; these have been taken into account in 
the numerical analysis. \s

By performing the RGE evolution to the electroweak scale, one obtains for the
left-- and right--handed sfermion masses at one--loop order [we include the 
full two--loop evolution of the masses in the numerical analysis] 
\begin{eqnarray}
m_{\tilde{f}_{L,R}}^2 = m_0^2 + \sum_{i=1}^{3} F_i (f) 
m^2_{1/ 2} \pm ( T_{3 {f} } - e_{f} s_W^2 ) M_Z^2 
\cos 2 \beta
\end{eqnarray}
$T_{ 3 {f} }$ and $e_{{f}}$ are the weak isospin and the 
electric charge of the corresponding fermion ${f}$, and $F_i$ are the RGE 
coefficients for the three gauge couplings at 
the scale $Q \sim M_Z$, given by
\begin{eqnarray}
F_i = \frac{c_i(f)}{b_i} \left[1- \left( 1 - \frac{\alpha_G}{4\pi} 
b_i \log \frac{Q^2}{M_G^2} \right)^{-2} \right] 
\end{eqnarray}
The coefficients $b_i$, assuming that all the MSSM particle spectrum 
contributes to the evolution from $Q$ to the GUT scale $M_G$, are
\begin{eqnarray}
b_1=33/5 \ \  , \ \ b_2=1 \ \ , \ \ b_3=-3
\end{eqnarray}
The coefficients $c(\tilde{f})=(c_1,c_2,c_3) (\tilde{f})$ depend on the 
hypercharge and color of the sfermions  [$F_L=L_L$ or $Q_L$ is the 
slepton or squark doublet]
\begin{eqnarray}
c(\tilde{L}_L) = \left( \begin{array}{c} 3/ 10 \\ 3/2\\ 0 \end{array} \right) 
\ , \ 
c(\tilde{E}_R)= \left( \begin{array}{c} 6/5 \\ 0 \\ 0 
\end{array} \right) \non 
\end{eqnarray}
\begin{eqnarray}
c(\tilde{Q}_L)= \left( \begin{array}{c} 1/30 \\ 3/2 \\ 
8/3 \end{array} \right) \ , \ 
c(\tilde{U}_R)= \left( \begin{array}{c} 8/15 \\ 0 \\ 8/3 
\end{array} \right) \ , \ 
c(\tilde{D}_R)= \left( \begin{array}{c} 2/15 \\ 0 \\ 8/3
 \end{array} \right) 
\end{eqnarray}

With the input gauge coupling constants at the scale of the $Z$ boson mass
\beq
\alpha_1 (M_Z) \simeq 0.01 \ \ , \ \ 
\alpha_2 (M_Z) \simeq 0.033 \ \ ,  \ \ 
\alpha_3 (M_Z) \simeq 0.118 
\eeq
one obtains for the GUT scale $M_G$ and for the coupling constant $\alpha_G$
\beq
M_G \sim 1.9 \times 10^{16} \ {\rm GeV} \ \ \ {\rm and} \ \ 
\alpha_G = 0.041
\eeq
Using these values, and including only gauge loops in the one--loop 
RGE's, one obtains for the left-- and right--handed sfermion masses
\cite{DN}
\beq
m^2_{\tilde{u}_L} &=& m_0^2 +6.28 m^2_{1/2} +0.35 M_Z^2 \cos(2\beta) \non \\
m^2_{\tilde{d}_L} &=& m_0^2 +6.28 m^2_{1/2} -0.42 M_Z^2 \cos(2\beta) \non \\
m^2_{\tilde{u}_R} &=& m_0^2 +5.87 m^2_{1/2} +0.16 M_Z^2 \cos(2\beta) \non \\
m^2_{\tilde{d}_R} &=& m_0^2 +5.82 m^2_{1/2} -0.08 M_Z^2 \cos(2\beta) \non \\
m^2_{\tilde{\nu}_L} &=& m_0^2 +0.52 m^2_{1/2} +0.50 M_Z^2 \cos(2\beta) \non \\
m^2_{\tilde{e}_L} &=& m_0^2 +0.52 m^2_{1/2} -0.27 M_Z^2 \cos(2\beta) \non \\
m^2_{\tilde{e}_R} &=& m_0^2 +0.15 m^2_{1/2} -0.23 M_Z^2 \cos(2\beta) 
\eeq

In the case of the third generation sparticles, left-- and right--handed 
sfermions will mix; for a given sfermion $\tilde{f} = \tilde{t}, \tilde{b}$
and $\tilde{\tau }$, the mass matrices which determine the mixing are 
\begin{eqnarray}
\left[ \begin{array}{cc} m_{\tilde{f}_L}^2 + m_f^2 & m_f (A_f - \mu r_f) 
\\ m_f (A_f - \mu r_f)  & m_{\tilde{f}_R}^2 + m_f^2 \end{array} \right]
\end{eqnarray}
where the sfermion masses $m_{\tilde{f}_{L,R}}$ are given above, $m_f$
are the masses of the partner fermions and $r_{b} = r_\tau =1/r_t= \tb$. 
These matrices are diagonalized by orthogonal matrices with mixing angles
$\theta_f$ defined by
\begin{eqnarray}
\sin 2\theta_f = \frac{2 m_f (A_f -\mu r_f)} { m_{\tilde{f}_1}^2
-m_{\tilde{f}_2}^2 } \ \ , \ \ 
\cos 2\theta_f = \frac{m_{\tilde{f}_L}^2 -m_{\tilde{f}_R}^2}
{ m_{\tilde{f}_1}^2 -m_{\tilde{f}_2}^2 } 
\end{eqnarray}
and the masses of the squark eigenstates given by
\begin{eqnarray}
m_{\tilde{f}_{1,2}}^2 = m_f^2 + \frac{1}{2} \left[ 
m_{ \tilde{f}_L}^2 + m_{\tilde{f}_R}^2 \mp \sqrt{
(m_{\tilde{f}_L}^2 - m_{\tilde{f}_R}^2)^2 + 4m_f^2 (A_f -\mu r_f)^2 } 
\right].
\end{eqnarray}

\setcounter{equation}{0}
\renewcommand{\theequation}{C\arabic{equation}}

\subsection*{Appendix C: Renormalization Group Equations}

Finally, we collect for completeness the renormalization group equations for 
the soft--SUSY breaking parameters [the trilinear couplings, scalar masses 
as well as for $\mu$ and $B$], including the dependence on $A_t, A_b$ and 
$A_\tau$. We restrict ourselves to the one--loop RGE's and we keep only the leading 
terms in the mass hierarchy in the MSSM with three fermion generations. 
The complete expressions for the RGE's can be found 
in Refs.\cite{R15,R16}. \s

For the trilinear couplings of the third generation sfermions, the RGE's
are given by
\begin{eqnarray}
{{dA_t}\over {dt}}&=&{2\over {16\pi ^2}}\Big (\sum c_ig_i^2M_i
+6\lambda _t^2A_t+\lambda _b^2A_b\Big )\; \non \\
{{dA_b}\over {dt}}&=&{2\over {16\pi ^2}}\Big (\sum c_i^{\prime }g_i^2M_i
+6\lambda _b^2A_b+\lambda _t^2A_t+\lambda _{\tau}^2A_{\tau}\Big )\; \non \\
{{dA_{\tau }}\over {dt}}&=&{2\over {16\pi ^2}}\Big (\sum c_i^{\prime \prime }
g_i^2M_i+3\lambda _b^2A_b+4\lambda _{\tau}^2A_{\tau}\Big )\; 
\end{eqnarray}
while for the scalar masses of the third generation sfermions, one has
\begin{eqnarray}
{{dM_{Q_L}^2}\over {dt}}&=&{2 \over {16\pi ^2}}
\Big (-{1\over 15}g_1^2M_1^2-3g_2^2M_2^2-{16\over 3}g_3^2M_3^2
+\lambda _t^2X_t+\lambda _b^2X_b\Big )\; \non \\
{{dM_{t_R}^2}\over {dt}}&=&{2 \over {16\pi ^2}}
\Big (-{16\over 15}g_1^2M_1^2-{16\over 3}g_3^2M_3^2
+2\lambda _t^2X_t\Big )\; \non \\
{{dM_{b_R}^2}\over {dt}}&=&{2 \over {16\pi ^2}}
\Big (-{4\over 15}g_1^2M_1^2-{16\over 3}g_3^2M_3^2
+2\lambda _b^2X_b\Big )\; \non \\
{{dM_{L_L}^2}\over {dt}}&=&{2 \over {16\pi ^2}}
\Big (-{3\over 5}g_1^2M_1^2-3g_2^2M_2^2
+\lambda _{\tau}^2X_{\tau }\Big )\; \non \\
{{dM_{\tau _R}^2}\over {dt}}&=&{2 \over {16\pi ^2}}
\Big (-{12\over 5}g_1^2M_1^2
+2\lambda _{\tau}^2X_{\tau }\Big )\;
\end{eqnarray}
The evolution parameter is defined by $t=\log(Q/M_G)$,  
\begin{eqnarray}
b_i&=& (\, 33/5 \, , \, 1 \, , \, -3 \, ) \; \non \\
c_i&=&(\, 13/ 15 \, , \, 3 \, , \, 16/ 3 \, ) \; \non \\
c_i^{\prime}&=&(\, 7/ 15 \, , \, 3 \, , \, 16/ 3 \, ) \; \non \\
c_i^{\prime \prime}&=&(\, 9/ 5 \, , \, 3 \, , \, 0 \, ) \;
\end{eqnarray}
and
\begin{eqnarray} 
X_t      &=&  M_{Q_L}^2+M_{t _R}^2+M_{H_2}^2+A_t^2\; \non \\
X_b      &=&  M_{Q_L}^2+M_{b _R}^2+M_{H_1}^2+A_b^2\; \non \\
X_{\tau }&=&  M_{L_L}^2+M_{\tau _R}^2+M_{H_1}^2+A_{\tau }^2\;
\end{eqnarray}
For the first and second generation sfermions, these expressions 
reduce to
\begin{eqnarray}
{{dA_u}\over {dt}}&=&{2\over {16\pi ^2}}\Big (\sum c_ig_i^2M_i
+\lambda _t^2A_t\Big )\; \non \\
{{dA_d}\over {dt}}&=&{2\over {16\pi ^2}}\Big (\sum c_i^{\prime }g_i^2M_i
+\lambda _b^2A_b+{1\over 3}\lambda _{\tau}^2A_{\tau}\Big )\; \non \\
{{dA_e}\over {dt}}&=&{2\over {16\pi ^2}}\Big (\sum c_i^{\prime \prime }
g_i^2M_i+\lambda _b^2A_b+{1\over 3}\lambda _{\tau}^2A_{\tau}\Big )\;  
\end{eqnarray}
and
\begin{eqnarray}
{{dM_{q_L}^2}\over {dt}}&=&{2 \over {16\pi ^2}}
\Big (-{1\over 15}g_1^2M_1^2-3g_2^2M_2^2-{16\over 3}g_3^2M_3^2
\Big )\; \non \\
{{dM_{u_R}^2}\over {dt}}&=&{2 \over {16\pi ^2}}
\Big (-{16\over 15}g_1^2M_1^2-{16\over 3}g_3^2M_3^2
\Big )\; \non \\
{{dM_{d_R}^2}\over {dt}}&=&{2 \over {16\pi ^2}}
\Big (-{4\over 15}g_1^2M_1^2-{16\over 3}g_3^2M_3^2
\Big )\; \non \\
{{dM_{l_L}^2}\over {dt}}&=&{2 \over {16\pi ^2}}
\Big (-{3\over 5}g_1^2M_1^2-3g_2^2M_2^2
\Big )\; \non \\
{{dM_{e_R}^2}\over {dt}}&=&{2 \over {16\pi ^2}}
\Big (-{12\over 5}g_1^2M_1^2
\Big )\; 
\end{eqnarray}
For the gauge coupling constants and the other soft--SUSY breaking parameters, 
the RGE's are given by
\begin{eqnarray}
{{dg_i}\over {dt}}&=&{1\over {16\pi ^2}}b_ig_i^3 \;  \\
{{dM_i}\over {dt}}&=&{2\over {16\pi ^2}}b_ig_i^2  M_i\;  \\
{{dB}\over {dt}}&=&{2\over {16\pi ^2}}\Big ({3\over 5}g_1^2M_1+3g_2^2M_2
+3\lambda _b^2A_b+3\lambda _t^2A_t+\lambda _{\tau}^2A_{\tau}\Big )\; \\
{{d\mu }\over {dt}}&=&{\mu \over {16\pi ^2}}\Big (-{3\over 5}g_1^2-3g_2^2
+3\lambda _t^2+3\lambda _b^2+\lambda _{\tau}^2\Big )\; \\
{{dm_{H_1}^2}\over {dt}}&=&{2 \over {16\pi ^2}}
\Big (-{3\over 5}g_1^2M_1^2-3g_2^2M_2^2
+3\lambda _b^2X_b+\lambda _{\tau}^2X_{\tau }\Big )\; \\
{{dm_{H_2}^2}\over {dt}}&=&{2 \over {16\pi ^2}}
\Big (-{3\over 5}g_1^2M_1^2-3g_2^2M_2^2
+3\lambda _t^2X_t\Big )\;
\end{eqnarray}

\newpage

\vspace*{18.5cm}
\nn {\bf Fig.~1:}  Masses of the CP--even Higgs bosons $h,H$ and of the 
charged Higgs particles $H^\pm$ as a function of $M_A$ for two values of 
$\tb =1.75$ and 50; the common squark mass $M_S$ at the weak scale is fixed 
to $M_S =1$ TeV and we take $\mu=A_t=0$.
\newpage

\vspace*{18.5cm}
\nn {\bf Fig.~2:}  The correlation between $m_0$ and $m_{1/2}$ for
$\tb=1.75$ and three values of $M_A=300, 600$ and 900 GeV. The non-solid 
lines show the boundaries which can be excluded by including the 
experimental bounds from LEP1.5 and Tevatron. 
\newpage

\vspace*{20cm}
\nn {\bf Fig.~3a:}  The masses of the Higgs bosons as a function of $m_{1/2}$ 
for $\tb =1.75$, for the two values $m_0=100$ and $500$ GeV and both
signs of $\mu$.
\newpage

\vspace*{20cm}
\nn {\bf Fig.~3b:}  The masses of the two charginos (dashed lines) and the 
four neutralinos (solid lines) as a function of $m_{1/2}$ for $\tb 
=1.75$,  $M_A=300$ and $600$ GeV and for both signs of 
$\mu$. The chargins/neutralinos are ordered with increasing masses. 
\newpage

\vspace*{20cm}
\nn {\bf Fig.~3c:} The masses of the two stop (solid lines), sbottom 
(dotted lines) and first/second generation squark (dashed lines) 
eigenstates as a function of $m_{1/2}$ for $\tb =1.75$, 
$M_A=300$ and $600$ GeV and for both signs of $\mu$. 
\newpage 

\vspace*{20cm}
\nn {\bf Fig.~3d:} The masses of the charged sleptons (solid and dotted
lines) and the sneutrino (dashed lines) of the three generations as a
function of $m_{1/2}$ for $\tb =1.75$,  $M_A=300$ and $600$
GeV and for both signs of $\mu$. 
\newpage

\vspace*{20cm}
\nn {\bf Fig.~4:}  The correlation between $m_0$ and $m_{1/2}$ for
$\tb \simeq 50$, $\mu <0$, and two values of $M_A=300$ and $600$ GeV. The boundary
contours correspond to tachyonic solutions, $m_{\tilde{\tau}}^2 <0$, $M_A^2<0$
and $M_h^2 <0$ at the tree--level. 
\newpage

\vspace*{20cm}
\nn {\bf Fig.~5a:} Cross sections for the pair production processes
$\ee \ra HA$ and $\ee \ra H^+H^-$ as a function of $\sqrt{s}$ for $\tb=1.75$
(solid lines) and $\tb=50$ (dashed lines) and three values of $M_A=300,
600$ and $900$ GeV. 
\newpage

\vspace*{20cm}
\nn {\bf Fig.~5b:} Cross sections for the production processes $\ee \ra
HZ$, $\ee \ra hA$ and $\ee \ra H \nu \bar{\nu}$ as a function of 
$\sqrt{s}$ for $\tb=1.75$ and the values $M_A=300$ and $600$ GeV. 
\newpage

\vspace*{18cm}
\nn {\bf Fig.~6a:}  Decay widths (in GeV) of the heavy CP--even Higgs 
boson $H$ into charginos and neutralinos 
(dotted lines), squarks (dashed lines), 
sleptons (dash--dotted lines), standard particles (dott--long--dashed 
lines) 
and the total decay widths (solid lines) as a function of $m_{1/2}$ for 
$\tb =1.75$,  $M_A=300$ and $600$ GeV and for both signs of $\mu$. 
\newpage

\vspace*{18cm}
\nn {\bf Fig.~6b:}  Partial decay widths (in GeV) of the heavy CP--even Higgs
boson $H$ into all combinations of chargino and neutralino pairs 
[$ij \equiv \chi_i \chi_j$]
as a function of $m_{1/2}$ for $\tb =1.75$, $M_A=600$ GeV 
and for both signs of $\mu$. 
\newpage

\vspace*{18cm}
\nn {\bf Fig.~6c:}  Partial decay widths (in GeV) of the heavy CP--even Higgs
boson $H$ into stop and sbottom squarks and into slepton pairs
as a function of $m_{1/2}$ for $\tb =1.75$,
$M_A=600$ GeV and for both signs of $\mu$. 

\newpage
\vspace*{18cm}
\nn {\bf Fig.~7a:}  Decay widths (in GeV) of the pseudoscalar Higgs 
boson $A$ into charginos and neutralinos (dotted lines), stop squarks (dashed 
lines), standard particles (dott--long--dashed lines) and the total decay 
widths (solid lines) as a function of $m_{1/2}$ for $\tb =1.75$, 
$M_A=300$ and $600$ GeV and for both signs of $\mu$. 

\newpage

\vspace*{18cm}
\nn {\bf Fig.~7b:}  Partial decay widths (in GeV) of the pseudoscalar Higgs
boson $A$ into all combinations of chargino and neutralino pairs
[$ij \equiv \chi_i \chi_j$]
as a function of $m_{1/2}$ for $\tb =1.75$, $M_A=$  
$600$ GeV and for both signs of $\mu$. 
\newpage 

\vspace*{18cm}
\nn {\bf Fig.~8a:}  
Decay widths (in GeV) of the charged Higgs bosons into charginos and
neutralinos (dotted lines), squarks (dashed lines), sleptons (dash--dotted
lines), standard particles (dott--long--dashed lines) and the total decay
widths (solid lines) as a function of $m_{1/2}$ for $\tb =1.75$, 
 $M_A=300$ and $600$ GeV and for both signs of $\mu$. 

\newpage

\vspace*{20cm}
\nn {\bf Fig.~8b:}  Partial decay widths (in GeV) of the charged Higgs 
boson $H^\pm$ into all combinations of charginos and neutralinos 
[$ij \equiv \chi_i^+ \chi_j^0$]
as a function of $m_{1/2}$ for $\tb =1.75$,  $M_A= 
600$ GeV and for both signs of $\mu$. 

\newpage

\textheight 22.cm
\setcounter{footnote}{0}
\setcounter{figure}{0}
\setcounter{equation}{0}
\def\thefootnote{\arabic{footnote}}
\renewcommand{\theequation}{\arabic{equation}}

\def\epsi{``$\epsilon$-app\-ro\-xi\-ma\-tion''~}
\def\Hb{\mbox{$H^0$~}}
\def\hb{\mbox{$h^0$~}}
\def\Ab{\mbox{$A^0$~}}
\def\MH{\mbox{$M_H$~}}
\def\Mh{\mbox{$M_h$~}}
\def\MA{\mbox{$M_A$~}}
\def\mH{\mbox{$m_H$~}}
\def\mh{\mbox{$m_h$~}}
\def\mA{\mbox{$m_A$~}}
\def\aeff{\mbox{$\alpha_{eff}$~}}
\def\tb{\mbox{$\tan\beta$~}}
\def\ra{\rightarrow}
\def\dmh{\mbox{$\delta M_h^{EPA}$}}
\def\dmH{\mbox{$\delta M_H^{EPA}$}}
\def\dmhe{\mbox{$\delta M_h^{\epsilon}$}}
\def\dszh{\mbox{$\delta\sigma_{Zh}^{EPA}$}}
\def\dsah{\mbox{$\delta\sigma_{Ah}^{EPA}$}}
\def\dszH{\mbox{$\delta\sigma_{ZH}^{EPA}$}}
\def\dsaH{\mbox{$\delta\sigma_{AH}^{EPA}$}}
\def\sqrts{\mbox{$\surd{s}$~}}
\def\admh{\mbox{$|\delta M_h^{EPA}|$}}
\def\admH{\mbox{$|\delta M_H^{EPA}|$}}
\def\admhe{\mbox{$|\delta M_h^{\epsilon}|$}}
\def\adszh{\mbox{$|\delta\sigma_{Zh}^{EPA}|$}}
\def\adsah{\mbox{$|\delta\sigma_{Ah}^{EPA}|$}}
\def\adszH{\mbox{$|\delta\sigma_{ZH}^{EPA}|$}}
\def\adsaH{\mbox{$|\delta\sigma_{AH}^{EPA}|$}}

\renewenvironment{thebibliography}[1]           
 {\normalsize\rm
   \begin{list}{\arabic{enumi}.}                
    {\usecounter{enumi} \setlength{\parsep}{0pt}
     \setlength{\itemsep}{3.5pt} \settowidth{\labelwidth}{#1}    
     \sloppy    
 }}{\end{list}} 

\renewenvironment{itemize}      
{\begin{list}{--}               
    {\setlength{\parsep}{0pt}   
     \setlength{\itemsep}{2pt}  
     \setlength{\topsep}{2pt}   
     \sloppy    
 }}{\end{list}} 

\newcounter{secno}
\newcommand{\jrsec}[1]{
\addtocounter{secno}{1}
\vskip 0.4 cm 
{\normalsize \bf \noindent \arabic{secno}. #1}
\vskip 0.3 cm
}
\baselineskip = 0.98\baselineskip

\def\thefootnote{\fnsymbol{footnote}} 

\begin{center}
{\large{\bf 
Production of Heavy Neutral MSSM Higgs Bosons}}

\vspace*{3mm}

{\large{\bf 
a complete 1-loop calculation.}}
\vskip 5mm

{\sc V. Driesen, W. Hollik and J. Rosiek}\footnote{Supported 
in part by  the Alexander von Humboldt Stiftung  and by the 
Polish Committee for Scientific Research.}
\vspace*{3mm}

{\small Institut f\"ur Theoretische Physik, Universit\"at Karlsruhe, 
D-76128 Karlsruhe, Germany}
\end{center}

\begin{abstract} 
The complete 1-loop diagrammatic calculations of the cross sections for the neutral 
Higgs production processes 
$e^+e^-\ra Z^0H^0(Z^0h^0)$ and $e^+e^-\ra A^0H^0(A^0h^0)$ in the MSSM
are presented and compared the with the corresponding 
results of the simpler and compact effective potential approximation.
\end{abstract}

\jrsec{Introduction}
\label{sec:intro}

In order to experimentally detect possible signals of the neutral MSSM
Higgs bosons, 
detailed studies for the decay and production processes of Higgs boson are 
required.
As has been discovered several years ago~[1-3], 
radiative corrections in the MSSM Higgs sector are large and have to be 
taken into account for phenomenological studies. Three main approaches
have been developed to calculate the 1-loop 
radiative corrections to the MSSM Higgs
boson masses, production and decay rates:
\begin{itemize}
\item[a)] The Effective Potential Approach (EPA)~\cite{EPA}.
\item[b)] The Renormalization Group approach (RGE)~\cite{RGE}. 
\item[c)] The diagrammatic calculation in the on-shell renormalization 
scheme (Feynman Diagram Calculation, 
FDC)~\cite{MYNPB,BRI}: The masses are calculated from the pole positions of 
the Higgs propagators, and the cross sections are obtained from the full set
of 1-loop diagrams contributing to the amplitudes, including~\cite{MYNPB}:
\begin{itemize}
\item[--] the most general form of the MSSM lagrangian with soft 
breaking terms,
\item[--] the virtual contributions from all the particles of the MSSM 
spectrum,
\item[--] all 2-, 3- and 4-point Green's functions for a given process with 
Higgs particles,
\item[--] the momentum dependence of the Green's functions,
\item[--] the leading reducible diagrams of higher orders corrections.
\end{itemize}
\end{itemize}

The experimental searches for Higgs bosons at LEP1~\cite{lepexp} and studies 
for the future searches at higher energies~\cite{highen} conventionally make 
use of the most compact 
effective potential approximation. 
We present the complete 1-loop diagrammatic results for 
the cross sections for the neutral Higgs production processes 
$e^+e^-\ra Z^0H^0(Z^0h^0)$ and $e^+e^-\ra A^0H^0(A^0h^0)$,  compare them 
with the corresponding 
ones of the simpler and compact EPA approximation and discuss 
the typical size of the differences.

\newpage

\jrsec{Outline of the calculations}
\label{sec:outline}

The tree level potential for the neutral MSSM Higgs bosons can be
written as:
\begin{eqnarray}
V^{(0)} = m_1^2 H_1^2 +  m_2^2 H_2^2 
+ \epsilon_{ij}(m_{12}^2 H_1^i H_2^j + H.c.)
+ {g^2 + {g'}^2\over 8}(H_1^2-H_2^2)^2 + {g^2\over 4}(H_1 H_2)^2
\label{eq:pot0}
\end{eqnarray}
Diagonalization of the mass matrices following from the
potential~(\ref{eq:pot0}) leads to three physical particles: two
CP-even Higgs bosons \Hb, \hb and one CP-odd Higgs boson \Ab, and
defines their tree-level masses \mH, \mh and \mA, with
$\mH>\mh$,  and the mixing angles $\alpha$, $\beta$.
The way of calculating the radiative corrections in the EPA and FDC
methods is briefly described as follows:

In the EPA, the tree level potential $V^{(0)}$ is improved by adding the
1-loop terms~\cite{EPA}:
\begin{eqnarray}
V^{(1)}(Q^2) = V^{(0)}(Q^2) +
 {1\over 64 \pi^2} \sum_{\stackrel{quarks}{squarks}} 
{\mathrm Str} {\cal M}^4 \left(\log{{\cal M}^2\over Q^2} - {3\over 2}\right) 
\label{eq:pot}
\end{eqnarray}
\vskip -3mm
\begin{figure}[htbp]         
\vspace*{3cm}
\caption{Classes of diagrams contributing to the 
$e^+e^-\ra Z^0h^0(H^0)$ process in the FDC approach. 
\label{fig:feyn}} 
\end{figure}
\noindent 
where $V^{(0)}(Q^2)$ is the tree level potential evaluated with couplings
renormalized at the scale $Q^2$, and $\mathrm{Str}$ denotes the supertrace
over the third generation of quark and squark fields contributing to the
generalized mass matrix ${\cal M}^2$. 
The 1-loop potential $V^{(1)}$ is rediagonalized yielding the 1-loop
corrected physical masses \MH, \Mh and the effective mixing angle
\aeff (for explicit formulae see~\cite{EPA}).

\indent In the FDC the 1-loop physical Higgs boson masses are
obtained as the pole positions  of the dressed scalar propagators. 
$M_H^2$ and $M_h^2$ are given by the solution of the 
equation~(\ref{eq:FDCmass}).
For the calculations of the cross sections we need the full set of 
2-, 3- and 4-point functions. In Fig.~\ref{fig:feyn} the 
diagrams contributing
to the $e^+e^-\ra Z^0h^0(H^0)$ process are collected. The  diagrams 
contributing to the $e^+e^-\ra A^0h^0(H^0)$ process can be obtained by
changing $Z^0$ into $A^0$ on the external line and 
skipping the diagrams i), j).
\vskip -3mm
\begin{eqnarray}
{\mathrm Re}\left[\left(p^2 - m_h^2 - \Sigma_{hh}(p^2)\right)
\left(p^2 - m_H^2 - \Sigma_{HH}(p^2)\right)
- \Sigma_{hH}^2(p^2) \right]=0
\label{eq:FDCmass}
\end{eqnarray}
\vskip -2mm
The formulae for the cross sections obtained in the FDC differ from the
Born expressions, because not only the effective  masses are
corrected but also new form factors and momentum dependent effects are
considered (see~\cite{MYNPB} for a detailed description).

\jrsec{Results on production cross sections}
\label{sec:results}

In this section we present the results for $Z^0H^0(Z^0h^0)$ and 
$A^0H^0(A^0h^0)$ production from the FDC and discuss the
quality of simpler EPA approximation.
In all figures we use as an example the set of parameters listed
in Table~\ref{tab:par}.
$\mu$ is the parameter describing the Higgs 
doublet mixing in the MSSM superpotential. $M_2$ denotes the SU(2) gaugino 
mass parameter. For the U(1) gaugino mass we use the value 
$M_1=\frac{5}{3} \tan^2\theta_W M_2$, suggested by GUT constraints.  
$M_{sq},M_{sl},A_t$ and $A_b$ are the parameters entering the sfermion mass 
matrices (for the detailed expressions see e.g.~\cite{PRD41}). 
For simplicity we assume a common value $M_{sq}$ for all generations 
of squarks, and a common $M_{sl}$ for sleptons. 
\vskip -4mm
\begin{table}[htbp]
\begin{center}
\begin{tabular}{|c|c|c|c|c|c|c|c|}
\hline
Parameter & $m_t$ & \MA & $M_{sq}$ & $M_{sl}$ & $M_2$ & $\mu$ & $A_t=A_b$\\
\hline
Value (GeV) & 175 & 200 & 1000 & 300 & 1000 & 500 & 1000\\
\hline
\end{tabular}
\vskip -.2mm
\caption{
\baselineskip 12pt \small
Parameters used for the numerical analysis.
\label{tab:par}}
\end{center}
\vskip -4mm
\end{table}

From the theoretical point of view, the most convenient parameters for the 
Higgs sector are the mass \MA of the CP-odd Higgs boson and the ratio
$\tan\beta=\frac{v_2}{v_1}$. From the experimental point of view it is more 
natural to use, depending on the process considered,  the masses \Mh
or \MH of the CP-even Higgs instead of the formal quantity
$\tan\beta$. 

As a first step, we apply the conventional  \MA, $\tan\beta$ parameterization. 
Figs.~\ref{fig:crl500} and~\ref{fig:cr500} show the production cross sections 
for the processes $\sigma(e^+e^-\ra Z^0h^0, A^0h^0)$ and 
$\sigma(e^+e^-\ra Z^0H^0, A^0H^0)$ for $\sqrts = 500$ GeV.
For the chosen set of parameters the numerical differences can reach 30\% at
$\sqrts=500$ GeV.
\begin{figure}[htbp]
\vskip -7mm
\begin{tabular}{p{0.48\linewidth}p{0.48\linewidth}}             
\vspace*{8.5cm}
\end{tabular}   
\vskip -4mm
\caption{
\small
\baselineskip=12pt
Comparison of the cross sections \mbox{$\sigma(e^+e^-\ra Z^0h^0,
A^0h^0)$}
obtained in the EPA and FDC. Parameters as given in
\protect{Table~\ref{tab:par}}, \sqrts = 500 GeV.
\label{fig:crl500}
}               
\vskip -3mm
\end{figure}
\begin{figure}[htbp]
\vspace*{7cm}
\vskip -5mm
\begin{tabular}{p{0.48\linewidth}p{0.48\linewidth}}             
\end{tabular}   
\vskip -6mm
\caption{
\small
\baselineskip=12pt
Comparison of the cross sections \mbox{$\sigma(e^+e^-\ra Z^0H^0,
A^0H^0)$}
obtained in the  EPA and FDC. Parameters as given in
\protect{Table~\ref{tab:par}}, \sqrts = 500 GeV.
\label{fig:cr500}
}               
\vskip -7mm
\end{figure}
They become more important with increasing energies, 
exceeding 40\% at 1 TeV. Note, however, that in the region of large
cross sections the EPA accuracy is better (20\% at 500 GeV). 
More detailed discussion of the 
lighter CP-even Higgs boson production can be found in ref.~\cite{DHR}. 

Fig.~\ref{fig:esig} shows the production cross sections for the
processes $\sigma(e^+e^-\ra Z^0H^0, A^0H^0)$ as a function of \sqrts. 
The effect of the additional form factors included in the FDC grows
when center-of-mass energy increases. For $\sqrts=1.5$ TeV the
differences between FDC and EPA can reach 50\% for the
$\sigma(e^+e^-\ra Z^0H^0)$ production channel. In addition, the
angular dependence of the cross section given by the FDC is
modified in compare to the effective Born approximation.

We now turn to the more physical parameterization of the cross sections in 
terms of the two Higgs boson masses \MA and \Mh or \MH. 
This parameterization is more
clumsy in the calculations, but it has the advantage of physically well defined
input quantities avoiding possible confusions from different renormalization 
schemes. Varying \MH (\MA and other input quantities fixed)
we obtain $\tan\beta$ and $\sigma_{ZH}$, $\sigma_{AH}$ as functions of
\MH. For the parameter values given in Table~\ref{tab:par}, 
the differences between
the \tb values obtained in the EPA and FDC can reach 10\% (up to 20\% for 
smaller $\MA\approx 100$ GeV).
Also significant differences can occur for the cross sections, as displayed 
in Fig.~\ref{fig:tbcr} where the predictions of EPA and FDC for the
$\sigma_{ZH}$ and $\sigma_{AH}$ are plotted as functions of \MH. 
The typical size of differences between the methods is 
10-20\% for $\sqrts=500$ GeV, but they may became as large as  60\% in case 
of the process $\sigma(e^+e^- \ra Z^0 H^0)$. This particularly large
deviation occurs for large \MH values, corresponding to small $\tb\leq 1$ 
(compare Fig.~\ref{fig:crl500}).

We have analyzed also the dependence of the differences between the EPA and
the FDC predictions on the SUSY parameters: sfermion and gaugino masses, $\mu$
parameter and sfermion mixing parameters.
In most cases the variation of those parameters does not
have a large effect on the size of the differences between the EPA and FDC
(a more detailed discussion can be found in ref.~\cite{JRAS}).

To give a more global impression of the typical size of 
the differences between the EPA and FDC results, 
we have chosen 1000 random points (for each \sqrts value in 
Table~\ref{tab:avdif}) 
from the hypercube in the MSSM parameters space with the following bounds:
\begin{center}
\begin{tabular}{lcl}
$0.5<\tb<50$   & \mbox{\hskip 1cm }  & 50 GeV $<\MA<$ 250 GeV\\
-500 GeV $<\mu<$ 500 GeV &  & 200 GeV $<M_2<$ 1000 GeV\\
200 GeV $<M_{sq}=2M_{sl}<$ 1000 GeV & & $-M_{sq}<A_t=A_b<M_{sq}$\\
\end{tabular}
\end{center}
We define the relative differences for the masses and cross sections 
as follows:
\begin{eqnarray}
\delta X^{EPA} = {X^{FDC}-X^{EPA}\over X^{FDC}}.
\label{eq:diffmet}
\end{eqnarray}
where $X$ can be chosen as \Mh, \MH, $\sigma_{ZH}$, $\sigma_{Zh}$,
$\sigma_{AH}$ or $\sigma_{Ah}$.

We calculated the quantities \dmh, \dmH, \dszh, \dsah, \dszH and \dsaH
and averaged them (and also their absolute values) 
arithmetically over all generated points of the
parameter space. The average mass differences are small and
equal $\admh=2\%$ and $\admH=1\%$.
The results for the cross sections are summarized in Table~\ref{tab:avdif}. 
It shows that the predictions of both methods deviate in particular 
for $\sigma_{ZH}$.
\vskip -5mm
\begin{table}[htbp]
\begin{center}
\begin{tabular}{|c|c|c|c|c|c|c|c|c|}
\hline
\sqrts (GeV)& \dszh & \dsah &\dszH &\dsaH & \adszh & \adsah &\adszH &\adsaH \\ 
\hline
500 & 16.4\% & -2.4\% & 57\% & 4.4\% & 21\% & 31\% & 62\% & 14\% \\
\hline
1000 & 10.3\% & 1.1\% & 56\% & -3.0\% & 15\% & 31\% & 62\% & 14\% \\
\hline
1500 & 4.2\% & 4.9\% & 53\% & -9.0\% & 17\% & 32\% & 63\% & 18\% \\
\hline
\end{tabular}
\vskip -2mm
\caption{\baselineskip 12pt \small
Differences between the EPA and FDC predictions averaged over a 
random sample of parameters.\label{tab:avdif}}
\end{center}
\end{table}
\vskip -7mm

Summarizing, comparisons between the FDC predictions with the simpler
EPA approximation have shown that at $\sqrts=500$ GeV the EPA has an
accuracy of typically 10-20\% in the parameter regions where the cross
sections are large.  The differences become larger with increasing
energy, where also modifications of the Born-like angular
distributions are more visible.  The use of the physical input
variables $M_A,$ $M_h$ or $M_A,$ $M_H$ avoids ambiguities from the
definition of $\tan\beta$ in higher order, but the observed
differences remain of the same size.  For a better accuracy, the full
FDC would be required.

Recently some papers on the leading 2-loop corrections
to the CP-even MSSM Higgs boson masses have been published~\cite{twoloop}. 
The main conclusion is that 2-loop corrections are also significant and 
tend to compensate partially the effects of 1-loop corrections. 
The calculations are based on the EPA and RG methods. Since the main 
emphasis of this study is to figure out the difference between complete
and approximate results in a given order, we have not implemented the 2-loop
terms. They would improve the 1-loop FDC results in the same way as the 
approximations and thus do not influence the remaining differences which can 
only be obtained by an explicit diagrammatic calculation.

\bigskip
The library of  FORTRAN codes for the calculation of the 1-loop radiative
corrections in the 
on-shell renormalization scheme to the MSSM neutral Higgs
production and decay rates~\cite{MYNPB} can be found at the URL address:\\
{\sl http://itpaxp1.physik.uni-karlsruhe.de/$\sim$rosiek/neutral\_higgs.html}

\newpage

\begin{figure}[htbp] 
\vspace*{8cm}
\vskip -7mm
\begin{tabular}{p{0.48\linewidth}p{0.48\linewidth}}             
\end{tabular}   
\vskip -7mm
\caption{
\small
\baselineskip=12pt
Comparison of the cross sections \mbox{$\sigma(e^+e^-\ra Z^0H^0,
A^0H^0)$} as a function of \sqrts obtained in the  EPA and FDC. 
$\tb=2$, other parameters as given in \protect{Table~\ref{tab:par}}.
\label{fig:esig}
}               
\vskip -7mm
\end{figure}
\begin{figure}[htbp]
\vspace*{8cm}
\vskip -7mm
\begin{tabular}{p{0.48\linewidth}p{0.48\linewidth}}             
\end{tabular}   
\vskip -7mm
\caption{
\small
\baselineskip=12pt
Comparison of the cross sections \mbox{$\sigma(e^+e^- \ra Z^0 H^0, A^0
H^0)$} as a
function of \MH in the EPA and FDC. Parameters as given in
\protect{Table~\ref{tab:par}}, \sqrts = 500 GeV.
\label{fig:tbcr}
}               
\vskip -2mm
\end{figure}
%

\newpage

\setcounter{footnote}{0}
\setcounter{figure}{0}
\setcounter{equation}{0}

\begin{center}

{\large\sc {\bf Radiative corrections to $e^+e^-\to H^+ H^-$}}

\vspace*{0.5cm}

{\sc A. Arhrib$^{1,2}$ and G. Moultaka$^1$} \\
\vspace*{5mm}

{\small $^1$ Physique Math\'ematique et Th\'eorique, E.S.A. du 
CNRS N$^o$ 5032, \\ 
Universit\'e Montpellier II, F-34095 Montpellier France\\
\vspace{3mm}
$^2$ L.P.T.N., Facult\'e des Sciences Semlalia,
B.P. S15, Marrakesh, Morocco} 
\end{center}

\begin{abstract}

We discuss the one-loop electroweak corrections to the pair production of
charged Higgs bosons $e^+ e^- \rightarrow H^+ H^-$ in the Minimal Supersymmetric
Standard Model.  

\end{abstract}


In contrast to hadronic machines, a high energy $e^+ e^-$ collider in the TeV range  will be a rather unique place to discover and study charged higgses in a clean environment. These would be produced either in pairs
\cite{komamiya}, our main concern here, or in associate (rare) production with $W^\pm$. It was first found in \cite{ACM}  that  loop corrections from matter fermions and their susy partners
(mainly the $(t,b), (\tilde{b}, \tilde{t}) $ sector), are likely to change the tree-level result at $\sqrt{s} = 500 $ GeV \cite{komamiya,tree-H}, by asmuch as $10\%$  dip in the cross-section.The effect could even lie between  $-25\%$ and $25\%$ and perhaps grow out of perturbative control, though in a 
reasonable range of the model-parameters. Such a sensitivity to loop effects appears to be related to the fact that at tree-level the $\gamma$ and $Z$ 
mediated process is exclusively controlled by $U_B(1) \times U_{W_3}(1)$ gauge invariance and thus
knows nothing about the non-standard extension whatsoever.

The aim of the present study is to improve on the previous one by including:
{\bf a)} the complete Higgs sector contributions (self-energies, vertices and boxes),
{\bf b)} the  infrared part, including initial and final soft photon radiation as well as $\gamma \gamma$ and $\gamma Z$ boxes,
{\bf c)} The complete set of charginos/neutralinos/$\tilde{e}$/$\tilde{\nu}$  
box diagrams,
and thus to identify the various origins of large effects, whether in the MSSM or in a type II
two-Higgs-doublet model (THDM-II). 

It turns out that besides the sensitivity to the heavy quark-squark sector
there are, on one hand large effects from the soft photon radiation
and on the other, possibly important 
effects in the purely Higgs sector. The latter case
occurs when deviations from
the tree-level supersymmetric $H^+H^- -H^0 ( h^0)$ couplings are allowed bringing in increasingly large effects for increasing values
of $\tan \beta$ at a given $\sqrt{s}$. To quantify such effects we thus 
allow for a general deviation from the supersymmetric relations
among the bare parameters of the Higgs potential as follows :
\begin{eqnarray}
&&\lambda_1= \lambda_2 +\delta_{12} \ \ \ \ \ \ \ \ \ \ ,\ \ \ \ \ \ \ \ \
\lambda_3=  \frac{1}{8}(g^2+g'^2)-\lambda_1 +\delta_{31} \nonumber\\
&& \lambda_4= 2 \lambda_1 -\frac{1}{2}g'^2 + \delta_{41}  \ \ \ \ \ , \
\ \ \ \lambda_5=  -\frac{1}{2}(g^2+g'^2)+ 2 \lambda_1  +
\delta_{51}\\
&&\lambda_6=  -\frac{1}{2}(g^2+g'^2)+ 2 \lambda_1 + \delta_{61}\nonumber
\end{eqnarray}
The $\lambda_i's$ are as defined in \cite{Gunion} and the (softly broken) susy case corresponds to $\delta_i=0$. 
Eq.(1) translates in a definite way into deviations from the MSSM tree-level relations 
among 
 the higgs masses, $\tan 2\alpha$ and $\tan \beta$, as well as the couplings of the higgs
sector, leading to 6 free parameters. 
However, one can find conditions relating the $\delta 's$ in such a way to  {\sl preserve} these relations {\sl even in a non-susy case}.
These conditions which we dub ``quasi-susy'', are certainly not
generic but constitute a good ground
to test minimal deviations from the MSSM in a simple way, since one then has just one extra free parameter (ex. $\lambda_3$) besides $\tan \beta$
and $M_{H^\pm}$ in the higgs sector.
In quasi-susy the only deviations from the MSSM 
tree-level triple Higgs couplings that contribute to one-loop order in our case are in $H^+-H^{-}-(H^0, h^0)$.
For large $\tan \beta$ these couplings behave as:   
\begin{equation}
 (H^+H^-(H^0,h^0))_{susy} - i g M_W(cos(\beta-\alpha) \Delta_{(1,2)} + (1 \mp cos(2 \alpha)) cos(\alpha) {\bf tan(\beta)} \Delta_3
\end{equation}
the $\Delta's$ being functions of the $\delta's$ of eq.(1), and vanish in the
MSSM.
 
Fig.1 illustrates how the Higgs sector contributions can counterbalance those of
the heavy quarks found in \cite{ACM} for large $\tan (\beta)$, but only near threshold.
Far from threshold most of the effects become again negative, except for $WW$ boxes. Furthermore the ``neutral'' model-independent contributions, including soft bremsstrahlung, obtained by adding one
photon (or Z) line to the tree diagrams depend loosely on $M_{H^\pm}$ or $\sqrt{s}$ and contribute at the level of $-17\%$ for $\Delta E_\gamma \sim 0.1 E_{beam}$. 
In Fig.2 we show (excluding those ``neutral'' contributions) the integrated cross-section for two values of $M_{H^\pm}$ and $\tan (\beta)$. In THDM-II the total loop effect increases (negatively) with increasing $\tan (\beta)$, the farther one goes from production threshold.  
In the MSSM ($\lambda_3= {\lambda_3}_{susy}$ ) the leading effects come exclusively from
the heavy quark-squark sector and the conclusions of \cite{ACM} remain unaltered in this case. [For instance the 150 boxes involving
$\chi^\pm/ \chi^0/\tilde{e}$/$\tilde{\nu}$ largely cancel among each other leading at most to $1-3 \%$ effect for a wide range of sparticle masses.]
   

\newpage

\vspace*{7cm}
\noindent {\small Fig.1: Contributions in \% to the integrated cross section 
in quasi-susy, 
$\lambda_3= -0.61, M_{H^\pm}=220$i GeV; a) Higgs sector, $\tan \beta=2$; b) Higgs sector, $\tan \beta=30$; c) virtual Z, $\gamma$ and soft bremsstrahlung; d) virtual W
boxes; e) matter fermion sector, $m_{top}=180 GeV, \tan \beta=30$; f) same as
e) but with $\tan \beta=2$.}

\bigskip

\vspace*{8.1cm}
\noindent {\small Fig.2: a) Tree-level, $M_{H^\pm}=220 GeV$; b) quasi-susy, 
$\lambda_3= -0.61$ 
(MSSM value $-0.71$), $\tan \beta=30$; c) quasi-susy, $\lambda_3= -0.61, \tan \beta=2$;
d) Tree-level, $M_{H^\pm}=430 GeV$; e) quasi-susy, $\lambda_3= -2.6$ (MSSM value 
$-2.84$), $\tan \beta=30$; f) same as e) but with $\tan \beta=2$; $m_{top}=180 GeV$.}

\newpage

\setcounter{footnote}{0}
\setcounter{figure}{0}
\setcounter{equation}{0}

\def\thefootnote{\fnsymbol{footnote}}

\begin{center}

{\large\sc {\bf Multiple Production of \MSSM\ Neutral Higgs Bosons}} 

\vspace*{3mm}

{\large\sc {\bf at High--Energy e$^+$e$^-$ Colliders}}

\vspace{.71cm}

{\sc A.~Djouadi$^{1,2}$\footnote{Supported by Deutsche Forschungsgemeinschaft 
DFG (Bonn).}, H.E.~Haber$^3$, and P.M.~Zerwas$^2$ } 

\vspace{.51cm}

{\small
$^1$ Institute f\"ur Theoretische Physik, Universit\"at Karlsruhe, \\
D--76128 Karlsruhe, FRG. \\
\vspace{0.3cm}

$^2$ Deutsches Elektronen--Synchrotron DESY, D-22603 Hamburg, FRG. \\
\vspace{0.3cm}

$^3$ Santa Cruz Institute for Particle Physics, University of California, \\
Santa Cruz, CA 95064, USA. \\
\vspace{0.3cm}
}
\end{center}

\begin{abstract}
\normalsize
\noindent

\nn The cross sections for the multiple production of the lightest
neutral Higgs boson at high--energy $e^+ e^-$ colliders are presented in
the framework of the Minimal Supersymmetric extension of the Standard
Model (\MSSM). We consider production through Higgs--strahlung,
associated production of the scalar and the pseudoscalar bosons, and the
fusion mechanisms for which we use the effective longitudinal
vector--boson approximation. These cross sections allow one to determine
trilinear Higgs couplings $\lambda_{Hhh}$ and $\lambda_{hhh}$, which are
theoretically determined by the Higgs potential. 

\end{abstract}

\def\thefootnote{\arabic{footnote}}
\setcounter{footnote}{0}

\subsection*{1. Introduction}

The only unknown parameter in the Standard Model (\SM) is the quartic coupling
of the Higgs field in the potential, which determines the value of the Higgs
mass. If the Higgs mass is known, the potential is uniquely fixed. Since the
form of the Higgs potential is crucial for the mechanism of spontaneous
symmetry breaking, {\it i.e.} for the Higgs mechanism {\it per se}, it will be
very important to measure the coefficients in the potential once Higgs
particles have been discovered. \s 

If the mass of the scalar particle is less than about 150 GeV, it very
likely belongs to the quintet of Higgs bosons,
$h,H,A,H^\pm$ predicted in the
two--doublet Higgs sector of supersymmetric theories \cite{R1}
[$h$ and $H$ are the light and heavy CP--even Higgs bosons, $A$ is the CP--odd 
(pseudoscalar) Higgs boson, and $H^\pm$ is the charged Higgs pair]. The
potential of the two doublet Higgs fields, even in the Minimal Supersymmetric
Standard Model (\MSSM), is much more involved than in the Standard Model
\cite{X2}. If CP is conserved by the potential, the most general
two--doublet model contains three mass parameters and seven real
self--couplings. In the \MSSM, the potential automatically conserves CP;
in addition, supersymmetry fixes all the Higgs
self--couplings in terms of gauge couplings.  The remaining three
free mass parameters can be traded in for the two vacuum expectation 
values (VEV's) of the neutral Higgs fields and one of the physical Higgs masses.
The sum of the squares of the VEV's is fixed by the $W$ mass, while
the ratio of VEV's is a free parameter of the model called $\tb$.  It is
theoretically convenient to choose the free parameters of the \MSSM\ Higgs
sector to be $\tb$ and $M_A$, the mass of the CP--odd Higgs boson $A$.
The other Higgs masses and the mixing angle $\alpha$ of the CP--even neutral
sector are then determined. Moreover, since all coefficients in the
Higgs potential are also determined, the trilinear and
quartic self--couplings of the physical Higgs particles can be
predicted theoretically. By measuring these couplings, the Higgs potential can
be reconstructed -- an experimental {\it prima  facie} task to establish the
Higgs mechanism as the basic mechanism for generating the masses of the
fundamental particles. \s 

The endeavor of measuring all Higgs self--couplings in the \MSSM\ is a
daunting task. We will therefore discuss a first step by analyzing
theoretically the production of two light Higgs particles of the \MSSM.
These processes may be studied at the proton collider LHC \cite{X2a} and at a
high--energy $\ee$ linear collider. In this paper we will focus on the
$\ee$ accelerators that are expected to operate in the first phase at an
energy of 500 GeV with a luminosity of about $\int {\cal L} = 20$
fb$^{-1}$, and in a second phase at an energy of about 1.5 TeV with a
luminosity of order $\int {\cal L} =200$ fb$^{-1}$ {\it per annum}
\cite{X3}. They will allow us to eventually study the couplings 
$\lambda_{Hhh}$ and $\lambda_{hhh}$. The measurement of the coupling 
$\lambda_{hAA}$ will be very difficult. \s 

Multiple light Higgs bosons $h$ can [in principle] be generated in the \MSSM~by
four mechanisms\footnote{The production of two light Higgs bosons, $\ee \ra
hh$, through loop diagrams does not involve any trilinear Higgs 
coupling; the production rates are rather small \cite{R3}.}: \s

\noindent (i) \underline{Decay of the heavy CP--even neutral Higgs boson}, 
produced either by $H$--strahlung and associated $AH$ pair production, 
or in the $WW$ fusion mechanisms, Fig.~1a, 
\begin{eqnarray}
\left. 
\begin{array}{l}
\ee \ra ZH, \ AH \\
\ee \ra \nu_e \bar{\nu}_eH 
\end{array}
\right\} \hspace*{.6cm} H \ra hh
\end{eqnarray}
Associated production $\ee \ra hA$ followed by $A \ra hZ$ decays leads to 
$hhZ$ background final states. \s

\nn (ii) \underline{Double Higgs--strahlung in the continuum}, with a final 
state $Z$ boson, Fig.~1b, 
\begin{eqnarray}
\ee \ra Z^* \ra hhZ
\end{eqnarray}
\nn (iii) \underline{Associated production with the pseudoscalar $A$ in the 
continuum}, Fig.~1c, 
\begin{eqnarray}
\ee \ra Z^* \ra hhA
\end{eqnarray}
(iv) \underline{Non--resonant $WW(ZZ)$ fusion in the continuum}, Fig.~1d, 
\begin{eqnarray}
\ee \ra \bar{\nu}_e \nu_e W^* W^* \ra \bar{\nu}_e \nu_e hh 
\end{eqnarray}
The cross sections for $ZZ$ fusion in (1) and (4) are suppressed by an order 
of magnitude.
The largest cross sections can be anticipated for the processes (1), where
heavy on--shell $H$ Higgs bosons decay into pairs of the light Higgs bosons.
[Cross sections of similar size are expected for the backgrounds involving the
pseudoscalar Higgs bosons.] We have derived the cross sections for the four
processes analytically; the fusion process has been treated in the equivalent
particle approximation for longitudinal vector bosons. \s 

We will carry out the analysis in the \MSSM~for the value $\tb=1.5$. [A
summary will be given in the last section for all values of $\tb$]. In
the present exploratory study, squark mixing will be neglected, {\it
i.e.} the supersymmetric Higgs mass parameter $\mu$ and the parameter
$A_t$ in the soft symmetry breaking interaction will be set to zero, and
the radiative corrections will be included in the leading $m_t^4$ one--loop
approximation parameterized by \cite{R4} 
\begin{eqnarray}
\epsilon = \frac{3G_F}{\sqrt{2} \pi^2} \frac{m_t^4}{\sin^2 \beta} 
\log	\left( 1+ \frac{M_S^2}{m_t^2} \right)
\end{eqnarray}
with the common squark  mass fixed to $M_S=1$ TeV. 
In terms of $\tb$ and $M_A$, the trilinear Higgs couplings relevant 
for our analysis are given in this approximation by 
\begin{eqnarray}
\lambda_{hhh} &=& 3 \cos2\alpha \sin (\beta+\alpha) 
+ 3 \frac{\epsilon}{M_Z^2} \frac{\cos^3 \alpha}{\sin\beta}  \\
\lambda_{Hhh} &=& 2\sin2 \alpha \sin (\beta+\alpha) -\cos 2\alpha \cos(\beta
+ \alpha) + 3 \frac{\epsilon}{M_Z^2} \frac{\sin \alpha}{\sin\beta}
\cos^2\alpha \non 
\end{eqnarray}
In addition, the coupling 
\begin{eqnarray}
\lambda_{hAA} &=& \cos 2\beta \sin(\beta + \alpha) + \frac{\epsilon}{M_Z^2} 
\frac{\cos \alpha}{\sin\beta} \cos^2\beta 
\end{eqnarray}
will be needed even though it turned out -- {\it a posteriori} -- that it
cannot be measured using the experimental methods discussed in this 
note\footnote{For small masses the decay $h \ra AA$ could have provided 
an experimental opportunity to measure this coupling. However, for
$\tb >1$, this area of the \MSSM\ parameter space is excluded
by LEP \cite{grivaz}.}. As
usual, these couplings are defined in units of $( 2\sqrt{2} G_F)^{1/2} M_Z^2$;
the $h,H,H^\pm$ masses and the mixing angle $\alpha$ can be expressed in terms
of $M_A$ and $\tb$ [see e.g. Ref.~\cite{R5} for a recent discussion]. \s 

In the decoupling limit \cite{R8} for large $A$, $H$ and $H^\pm$ masses, the
lightest Higgs particle becomes \SM--like and the trilinear $hhh$ coupling
approaches the \SM~value $\lambda_{hhh} \ra M_h^2/M_Z^2$.
In this limit, only the first three diagrams of Fig.~1b and 1d contribute
and the cross-sections for the processes $\ee \ra hhZ$ and $WW \ra hh$
approach the corresponding cross sections of the \SM\ \cite{R6,R7}.

\subsection*{2. H Production and hh Decays}

If kinematically allowed, the most copious source of multiple $h$ 
final states are cascade decays $H \ra hh$, with $H$ produced either
by Higgs--strahlung or associated pair production \cite{R1},
\beq
\sigma (\ee \ra ZH) &=& \frac{G_F^2 M_Z^4}{96 \pi s } (v_e^2+a_e^2) 
\cos^2(\beta-\alpha) \frac{ \lambda^{1/2}_Z [\lambda_Z+12M_Z^2/s]}  
{(1-M_Z^2/s)^2}   \\
\sigma (\ee \ra AH) &=& \frac{G_F^2 M_Z^4}{96 \pi s} (v_e^2+a_e^2)
\sin^2(\beta-\alpha) \frac{ \lambda^{3/2}_A} {(1-M_Z^2/s)^2} 
\eeq
The $Z$ couplings to electrons are given by $a_e=-1, v_e=-1+ 4\sin^2\theta_W$ 
and $\lambda_j$ is the usual two--body phase space function $\lambda_{j}
= (1-M_j^2/s-M_H^2/s)^2-4M_j^2 M_H^2/s^2$. The cross sections (8) and (9) are 
shown in Fig.~2 for the total $\ee$ 
energies $\sqrt{s}=500$ GeV and 1.5 TeV as a 
function of the Higgs mass $M_H$ for a small value of $\tb=1.5$ where 
the $H$ cascade decays are significant over a large mass range. As a 
consequence of the decoupling theorem, associated $AH$ production is dominant
for large Higgs masses. \s

The trilinear $Hhh$ coupling can be measured in the decay process $H\ra hh$
\begin{eqnarray}
\Gamma(H \ra hh ) = \frac{G_F \lambda^2_{Hhh} }{16\sqrt{2} \pi} 
\frac{M_Z^4}{M_H}  \left(1-\frac{4M_h^2}{M_H^2} \right)^{1/2} 
\end{eqnarray}
if the branching ratio is neither too small nor too close to unity. This is 
indeed the case, as shown in Fig.~3a, for $H$ masses between 180 and 350 
GeV
and small to moderate $\tb$ values. The
other important decay modes are $WW^*/ZZ^*$ decays. Since the $H$
couplings to the gauge bosons can be measured through the production
cross sections of the fusion and Higgs--strahlung processes, the
branching ratio BR$(H \ra hh)$ can be exploited to measure the coupling
$\lambda_{Hhh}$. \s 

The $ZH$ final state gives rise to resonant two--Higgs $[hh]$ 
final states.  The $AH$ final state typically yields
three Higgs $h[hh]$ final states 
since the channel $A \ra hZ$ is the dominant decay mode in most of 
the mass range we consider. This is shown in Fig.~3b where the 
branching ratios of the pseudoscalar $A$ are displayed for $\tb=
1.5$. \s

Another type of two--Higgs $hh$ final states is generated in the chain
$\ee \ra Ah \ra [Zh]h$, which does not involve any of the Higgs
self--couplings. However, in this case, the two $h$ bosons do not
resonate while $[Zh]$ does, so that the topology of these background
events is very different from the signal events. The size of the $\ee
\ra hA$ background cross section is shown in Fig.~2 together 
with the signal cross sections; for sufficiently large $M_A$, it becomes
small, in line with the decoupling theorem \cite{R8}. \s 

A second large signal cross section is provided by the $WW$ fusion
mechanism. [Since the NC couplings are smaller compared to the CC
couplings, the cross section for the $ZZ$ fusion processes in (1) and 
(4)  is $\sim
16\cos^4 \theta_W$, {\it i.e.} one order of magnitude smaller than for
$WW$ fusion.] In the effective longitudinal $W$ approximation 
\cite{Wlumi} one obtains 
\beq
\sigma( \ee \ra H \bar{\nu}_e \nu_e ) = \frac{G_F^3 M_W^4}{4 \sqrt{2}\pi} 
\left[ \left(1+\frac{M_H^2}{s} \right) \log \frac{s}{M_H^2} -2 \left(1-
\frac{M_H^2}{s} \right) \right] \cos^2(\beta-\alpha) 
\eeq
The magnitude of the cross section\footnote{In the effective $W$ 
approximation, the cross section may be overestimated by as much as a 
factor of 2 for small masses and/or small c.m. energies.
Therefore we display the exact cross sections \cite{EE500} in Fig.2.}
$\ee \ra H\nu_e \bar{\nu}_e$ is also shown in Fig.~2 for the two energies
$\sqrt{s}=500$ GeV and 1.5 TeV as a function of the Higgs mass $M_H$ and
for $\tb=1.5$. The signals in $\ee \ra [hh]$ + missing energy are very
clear, competing only with $H$--strahlung and subsequent neutrino decays
of the $Z$ boson. Since the lightest Higgs boson will decay mainly into
$b\bar{b}$ pairs, the final states will predominantly include four and
six $b$ quarks. \s 

At $\sqrt{s}=500$ GeV, about 500 signal events are predicted in the
mass range of $M_H \sim 200$ GeV for an integrated luminosity of $\int
{\cal L}=20$ fb$^{-1}$ {\it per annum}; and at $\sqrt{s}=1.5$ TeV, about
8,000 to 1,000 signal events for the prospective integrated luminosity
of $\int {\cal L}=200$ fb$^{-1}$ {\it per annum} in the interesting mass
range between 180 and 350 GeV. Note that for both energies, the $Ah$ 
background cross section is significantly smaller. 

\subsection*{3. Non-Resonant Double hh Production}

The double Higgs--strahlung $\ee \ra Zhh$, the triple Higgs production
process $\ee \ra Ahh$ and the $WW$ fusion mechanism $\ee \ra \nu_e \bar{\nu}_e
hh$ outside the resonant $H \ra hh$ range are disfavored by an
additional power of the electroweak coupling compared to the resonance
processes. Nevertheless, these processes must be analyzed carefully in order to
measure the value of the $hhh$ coupling. 
\subsubsection*{3.1 $e^+ e^- \ra Z h h$}

The double differential cross section of the process $e^+ e^- \ra hh Z$,
Fig.~1b, is given by
\beq 
\frac{d\sigma (\ee \ra hhZ) }{dx_1 dx_2} = \frac{G_F^3 M_Z^6 }{384 
\sqrt{2} \pi^3 s} \, (a_e^2+v_e^2) \, \ \frac{ {\cal A} }{(1-\mu_Z)^2} 
\eeq
The couplings have been defined in the previous section.
$x_{1,2}=2E_{1,2}/ \sqrt{s}$ are the scaled energies of the Higgs
particles, $x_3=2-x_1-x_2$ is the scaled energy of the $Z$ boson;
$y_k=1-x_k$. The scaled masses squared are denoted by $\mu_i=M_i^2/s$.
In terms of these variables, the coefficient ${\cal A}$ in the cross
section may be written as: 
\beq
{\cal A} &=&  \left\{ \frac{a^2}{2} f_0 + 
\frac{ \sin^4(\beta-\alpha)}{4\mu_Z^2(y_1+\mu_h-\mu_Z)} \left[ \frac{f_1}
{y_1+\mu_h -\mu_Z} + \frac{f_2 }{y_2+\mu_h-\mu_Z} \right] + \frac{\cos^4
(\beta-\alpha)}{4\mu_Z^2(y_1+\mu_h-\mu_A)} \right. \non \\
&& \times \left[ \frac{f_3}{y_1+\mu_h-\mu_A}+\frac{f_4}{y_2+\mu_h-\mu_A} 
\right] + \frac{a}{\mu_Z} \left[ \frac{ \sin^2(\beta-\alpha) f_5}{y_1+\mu_h
-\mu_Z}+ \frac{\cos^2(\beta-\alpha) f_6} {y_1+\mu_h-\mu_A} \right]  \non \\
&&\left. + \frac{\sin^2 2(\beta-\alpha)}{8 \mu_Z^2(y_1+\mu_h-\mu_Z)} 
\left[ \frac{f_7}{y_1+\mu_h-\mu_Z} + \frac{f_8}{y_2+\mu_h-\mu_Z} \right] 
\right\} + \left\{ y_1 \leftrightarrow y_2 \right\}
\eeq
with 
\beq 
a = \frac{1}{2} \left[ 
\frac{\sin(\beta-\alpha) \lambda_{hhh} } {y_3+\mu_Z-\mu_h} +
\frac{\cos(\beta-\alpha) \lambda_{Hhh} } {y_3+\mu_Z-\mu_H} \right]
+ \frac{ \sin^2(\beta-\alpha)}{y_1+\mu_h-\mu_Z} + \frac{
\sin^2(\beta-\alpha) }{y_2+\mu_h-\mu_Z} + \frac{1}{2\mu_Z} 
\eeq
[omitting the small decay widths of the Higgs bosons]. Only the coefficient 
$a$ includes the Higgs self--couplings $\lambda_{Hhh}$ and $\lambda_{hhh}$. 
Introducing the notation $y_0=(y_1-y_2)/2$, the coefficients $f_i$ which do 
not involve any Higgs couplings, are defined by
\beq
f_0 & = & (y_1+y_2)^2-4\mu_Z(1-3 \mu_Z)   \\
f_1 & = & \left[ (1+y_1)^2 -4\mu_Z(y_1+\mu_h) \right] \left[
y_1^2+\mu_Z^2-2\mu_Z (y_1+2\mu_h) \right] \non \\
f_2 & = & \left[ 2\mu_Z(\mu_Z-2\mu_h +1)-(1+y_1)(1+y_2) \right] 
\left[ \mu_Z(\mu_Z-y_1-y_2-4\mu_h+2)- y_1 y_2 \right] \non \\
f_3 &=& \left[ y_0^2 + \mu_Z(1-y_1-y_2+\mu_Z-4\mu_h) \right] \left[
1+ y_1+y_2+y_0^2 + \mu_Z (\mu_Z -4\mu_h - 2y_1 ) \right] \non \\
f_4 & =& \left[ y_0^2 + \mu_Z (1-y_1-y_2+\mu_Z-4\mu_h) \right] 
\left[ y_0^2 -1+ \mu_Z(\mu_Z-y_1-y_2-4\mu_h+2) \right]  \non \\
f_5 &=& 2\mu_Z^3-4\mu_Z^2(y_1+2\mu_h)+\mu_Z \left[(1+y_1)(3y_1-y_2) +2 \right] 
-y_1^2(1+y_1+y_2) -y_1y_2 \non \\
f_6 &=& 2\mu_Z^3-\mu_Z^2(y_2+3y_1+8\mu_h-2)+2\mu_Zy_0 \left( 1+y_1+y_0 \right)+ 
2 y_1 y_0-y_0^2(y_1+y_2-2) \non \\
f_7 &=& \left[ \mu_Z(4\mu_h-\mu_Z-1+2y_1-y_0) -y_1y_0 \right] \left[
\mu_Z(4\mu_h-\mu_Z-1+3y_1) -(1+y_0)(1+y_1) \right] \non \\
f_8 &=& \left[ \mu_Z(4\mu_h-\mu_Z-1+2y_1-y_0) -y_1y_0 \right] \left[
\mu_Z(4\mu_h-\mu_Z-2+y_1) +(1-y_0)(1+y_1) \right] \non
\eeq
In the decoupling limit, the cross section is reduced to the $\SM$~cross section
for which 
\beq
{\cal A} = \frac{a^2}{2} f_0 +
\frac{1}{4\mu_Z^2(y_1+ \mu_h-\mu_Z)} \left[ \frac{ f_1}{y_1+\mu_h-\mu_Z}
+ \frac{f_2}{y_2+ \mu_h-\mu_Z} + 4a \mu_Z f_5 \right]  
+ \left\{ y_1 \leftrightarrow y_2 \right\} \non 
\eeq
with the $f_i$'s as given above, and 
\beq 
a = \frac{1}{2} \frac{ \lambda_{hhh} } {y_3+\mu_Z-\mu_h}
+ \frac{ 1} {y_1+\mu_h-\mu_Z} + \frac{1}{y_2+\mu_h-\mu_Z} 
+ \frac{1}{2\mu_Z} \non 
\eeq
The cross section $\sigma(\ee \ra hhZ)$ is shown for $\sqrt{s}=500$ GeV
at $\tb=1.5$ as a function of the Higgs mass $M_h$ in Fig.~4a. For small
masses, the cross section is built up almost exclusively by $H \ra hh$
decays [dashed curve], except close to the point where the
$\lambda_{Hhh}$ coupling accidentally vanishes (cf. Ref.\cite{R5}) and
for masses around $\sim 90$ GeV where additional contributions come from
the decay $A \ra hZ$ [this range of $M_h$ corresponds to $M_A$ values
where BR$(A \ra hZ)$ is large; c.f. Fig.3]. For intermediate masses, the
resonance contribution is reduced and, in particular above 90 GeV where
the decoupling limit is approached, the continuum $hh$ production
becomes dominant, falling finally down to the cross section for double
Higgs production in the Standard Model [dashed line]. After subtracting
the $H \ra hh$ decays [which of course is very difficult], the continuum
cross section is about $0.5$~fb, and is of the same order as the
\SM~cross section at $\sqrt{s}= 500$ GeV. Very high luminosity is
therefore needed to measure the trilinear $hhh$ coupling. At higher
energies, since the cross section for double Higgs--strahlung scales
like $1/s$, the rates are correspondingly smaller, c.f. Fig.4b. \s 

Prospects are similar for large $\tb$ values. The cascade decay $H \ra
hh$ is restricted to a small $M_h$ range of less than 70 GeV, with a 
production cross
section of $\sim 20$ fb at $\sqrt{s}=500$ GeV and $\sim 3$ fb at 1.5 TeV. The
continuum cross sections are of the order of $0.1$ fb at both energies, so that
very high luminosities will be needed to measure the continuum cross sections 
in this case if the background problems can be mastered at all. \s 

We have repeated the analysis for the continuum process $\ee \ra Ahh$
(cf. Fig.1c).
However, it turned out that the cross section is built up almost exclusively
by resonant $AH \ra Ahh$ final states, with a very small continuum
contribution, so that the measurement of the coupling $\lambda_{hAA}$ 
is extremely difficult in this process. 

\subsubsection*{3.2 $W_L W_L \ra hh $}

In the effective longitudinal $W$ approximation\footnote{For qualifying 
comments see footnote 3.}, the total cross 
section for the subprocess $W_L W_L  \ra hh$, Fig.~1d, is given by
\beq
\hat{\sigma}_{LL} &=& \frac{G_F^2 \hat{s}}{64 \pi} \frac{\beta_h}{ \beta_W} 
\left\{ (1+\beta_W^2)^2 
\left[\frac{\mu_Z \sin (\beta-\alpha) }{1-\mu_h} \lambda_{hhh} + 
\frac{\mu_Z \cos (\beta-\alpha) }{1-\mu_H} \lambda_{Hhh} +1 \right]^2 
\right. \\
&+& \frac{\beta_W^2} {\beta_W \beta_h} \left[\frac{\mu_Z \sin 
(\beta-\alpha) }{1-\mu_h} \lambda_{hhh} + \frac{\mu_Z \cos 
(\beta-\alpha) }{1-\mu_H} \lambda_{Hhh} +1 \right] [ 
\sin^2(\beta-\alpha) g_1  \non \\
&+& \cos^2 (\beta -\alpha) g_2]  \left. 
+\frac{1} {\beta_W^2 \beta_h^2} \left[ \sin^4(\beta-\alpha) g_3+
\cos^4(\beta-\alpha) g_4+ \sin^22(\beta-\alpha) g_5 \right] \right\} \non
\eeq
with 
\beq
g_1 &=& 2 [(\beta_W -x_W \beta_h)^2 +1-\beta_W^4] l_W
-4 \beta_h (2\beta_W -x_W \beta_h) \non \\
g_2 &=& 2 (x_C\beta_h -\beta_W)^2 l_C
+ 4 \beta_h (x_C\beta_h -2 \beta_W) \non \\
g_3 &=& \beta_h[\beta_hx_W (3 \beta_h^2 x_W^2+14\beta_W^2+2-2\beta_W^4)-
4\beta_W( 3 \beta_h^2 x_W^2 + \beta_W^2 +1 -\beta_W^4 )][l_W + x_Wy_W] \non \\
&&-  [ \beta_W^4 +(1-\beta_W^4)(1+2 \beta_W^2 -\beta_W^4)] [l_W/x_W-y_W]
- 2 \beta_h^2 y_W (2 \beta_W - \beta_h x_W)^2  \non \\
g_4 &=& \beta_h [ \beta_h x_C (3 \beta_h^2 x_C^2 +14 \beta_W^2) -4 
\beta_W (3 
\beta_h^2 x_C^2 + \beta_W^2) ] [l_C +x_C y_C] \non \\
&& -\beta_W^4 [l_C/x_C -y_C]-2y_C \beta_h^2 (2 \beta_W - \beta_h x_C)^2 \non \\
g_5&=& \frac{ \beta_h \beta_W l_W }{x_W^2-x_C^2} [2 x_W ( 2x_W^2 \beta_h 
\beta_W -x_C x_W^2 \beta_h^2 -x_C \beta_W^2 ) -2x_W^2 ( \beta_h^2 x_W^2 
+\beta_W^2 +1-\beta_W^4) \non \\
&&+ \frac{x_C}{\beta_W \beta_h } ( (\beta_h^2 x_W^2 + 
\beta_W^2)(1-\beta_W^4) 
+( \beta_h^2 x_W^2 +\beta_W^2 )^2)] - 4\beta_h^3 \beta_W (x_W+x_C) \non \\
&&+ \frac{ \beta_h \beta_W l_C }{x_C^2-x_W^2} 
[ 4x_C^3 \beta_h \beta_W -2x_C x_W( \beta_h^2 x_C^2 + \beta_W^2 
+1 -\beta_W^4) -2x_C^2 ( \beta_h^2 x_C^2 +\beta_W^2 ) \non \\
&&+ \frac{x_W}{\beta_W \beta_h } ( (\beta_W^2+\beta_h^2 x_C^2)(1-\beta_W^4) 
+( \beta_h^2 x_C^2 +\beta_W^2 )^2)]+2 \beta_H^2(x_Cx_W\beta_H^2+4\beta_W^2)
\eeq
The scaling variables are defined in the same way as before.  
$\hat{s}^{1/2}$ is the c.m. energy of the subprocess, $\beta_W=(1-4M_W^2/
\hat{s})^{1/2}$ and $\beta_h= (1-4M_h^2/\hat{s})^{1/2}$ are the velocities 
of the $W$ and $h$ bosons, and
\beq
x_W= (1-2 \mu_h)/(\beta_W \beta_h) \ \ ,  \ \ 
x_C= (1-2 \mu_h+2 \mu_{H^\pm} -2\mu_W )/(\beta_W \beta_h) 
\non \\
l_i =\log (x_{i}-1)/(x_i+1) \  \ , \ \ y_i= 2/(x_i^2-1) 
\hspace*{2cm}
\eeq
The value of the charged Higgs boson mass $M_{H^\pm}$ in the $H^\pm$ 
$t$--channel exchange diagram of Fig.1d is given by 
$M_{H^\pm}^2=M_A^2+M_W^2$. \s

In the decoupling limit, the cross section reduces again to the $\SM$ cross
section which in terms of $g_1$ and $g_2$, defined above, is given by: 
\beq
\hat{\sigma}_{LL} = \frac{G_F^2 \hat{s}}{64 \pi} \frac{\beta_h}{ \beta_W} 
\left\{ (1+\beta_W^2)^2 \left[\frac{\mu_Z \lambda_{hhh}}{1-\mu_h} +1\right]^2 
+\frac{1+\beta_W^2} {\beta_W \beta_h} \left[\frac{\mu_Z \lambda_{hhh}}
{1-h_1} +1 \right]g_1 +\frac{g_3} {\beta_W^2 \beta_h^2} \right\}
\eeq

After folding $\hat{\sigma}_{LL}$ with the longitudinal $W_LW_L$ luminosity
\cite{Wlumi}, 
one obtains the total
cross section $\sigma(\ee \ra \nu_e \bar{\nu}_e hh)$ shown in Fig.~4b
as a function of the light Higgs mass $M_h$ for $\tb=1.5$ at $\sqrt{s} =1.5$
TeV. It is significantly larger than for double Higgs--strahlung in the
continuum. Again, for very light Higgs masses, most of the events are $H \ra
hh$ decays [dashed line]. The continuum $hh$ production is of the same size as
pair production of \SM~Higgs bosons [dotted line] which, as anticipated,
is being approached near the upper limit of the $h$ mass in the decoupling
limit. The size of the continuum $hh$ fusion cross section renders this channel
more promising than double Higgs--strahlung for the measurement of the
trilinear $hhh$ coupling. \s 

For large $\tb$ values, strong destructive interference effects reduce the cross
section in the continuum to very small values, of order 10$^{-2}$ fb, before
the \SM~cross section is reached again in the decoupling limit. As before, the
$hh$ final state is almost exclusively built up by the resonance $H\ra hh$
decays. 

\subsection*{4. Summa}

It is convenient to summarize our results by presenting Fig.5, which
displays the areas of the $[M_A, \tb$] plane in which $\lambda_{Hhh}$
[solid lines, 135$^0$ hatching] and $\lambda_{hhh}$ [dashed lines, 
$45^0$ hatching] could eventually be accessible by experiment. 
The size of these areas is based on purely theoretical cuts so that
they are expected to shrink if background processes and detector
effects are taken into account. \s

$(i)$ In the case of $H \ra hh$, we require a lower limit of the cross
section $\sigma(H) \times$BR$(H \ra hh) >0.5$ fb and at the same time
for the decay branching ratio $0.1 < $BR$(H\ra hh)<0.9$, as discussed
earlier. Based on these definitions, $\lambda_{Hhh}$ may become
accessible in two disconnected regions denoted by I and II [$135^0$ 
hatched] in Fig.5. For low
$\tb$, the left boundary of Region I is set by LEP1 data. The gap
between Regions I and II is a result of the nearly vanishing
$\lambda_{Hhh}$ coupling in this strip. The right boundary of Region II
is due to the overwhelming $t\bar{t}$ decay mode for heavy $H$ masses,
as well as due to the small $H$ production cross section. For
moderate values of $\tb$, the left boundary of Region I is defined by
BR$(H\ra hh)>0.9$. In the area between Regions I and II, $H$ cannot
decay into two $h$ bosons, {\it i.e.} $M_H <2M_h$. For large $\tb \gsim
10$, BR$[H\ra hh (AA)]$ is either too large or too small, except in a
very small strip, $M_A \simeq 65$ GeV, towards the top of Region I. [Note that 
$h$ and $A$ are nearly  mass--degenerate in this area.]  \s

$(ii)$ The dashed line in Fig.5 describes the left boundary of the area
[$45^0$ hatched] in which $\lambda_{hhh}$ may become accessible; it is
defined by the requirement that the continuum $W_L W_L \ra hh$ cross
section, $\sigma_{\rm cont}$, is larger than $0.5$ fb. Note that the
resonant $H \ra hh$ events in Region II must be subtracted in order to
extract the $\lambda_{hhh}$ coupling. \s

In conclusion, we have derived the cross sections for the double
production of the lightest neutral Higgs boson in the \MSSM~at $\ee$
colliders: in the Higgs--strahlung process $\ee \ra Zhh$, [in the triple
Higgs production process $\ee \ra Ahh$], and in the $WW$ fusion
mechanism. These cross sections are large for resonant $H \ra hh$ decays
so that the measurement of the triple Higgs coupling $\lambda_{Hhh}$ is
expected to be fairly easy for $H\ra hh$ decays in the $M_H$ mass range
between 150 and 350 GeV for small $\tb$ values. The continuum processes
must be exploited to measure the triple Higgs coupling $\lambda_{hhh}$.
These continuum cross sections, which are of the same size as in the
\SM, are rather small so that high luminosities are needed for the
measurement of the triple Higgs coupling $\lambda_{hhh}$.

\vspace*{0.5cm}

\nn {\bf Acknowledgements:}  \s

\nn Discussions with G. Moultaka and technical
help by T. Plehn are gratefully acknowledged. A.D. thanks the Theory
Group for the warm hospitality  extended to him at DESY, and H.E.H.
acknowledges the partial support of the U.S. Department of Energy.

\newpage


\vspace*{20.cm}
\nn Fig.1: Main mechanisms for the double production of the light MSSM Higgs boson
in $\ee$ collisions: a) $\ee \ra ZH$, $\ee \ra AH$ and $W W \ra H$
followed by $H \ra hh$; (b) $\ee \ra hhZ$, (c) $\ee \ra hhA$ and (d) $W W
\ra hh$.

\newpage

\vspace*{19.2cm}
\nn Fig.~2: Cross sections for the production of the heavy CP--even Higgs boson $H$
in $\ee$ collisions, $\ee \ra ZH/AH$ and $\ee \ra H\nu_e \bar{\nu}_e$, and for
the background process $\ee \ra Ah$ [the dashed curve shows $\frac{1}{2}
\times \sigma(Ah)$ for clarity of the figures]. The c.m.~energies are
chosen $\sqrt{s}=500$ GeV in (a), and 1.5 TeV in (b).

\newpage

\vspace*{19.2cm}
\nn Fig.~3:
The branching ratios of the main decays modes of the heavy CP--even neutral
Higgs boson $H$ in (a), and of the pseudoscalar Higgs boson $A$ in (b).

\newpage

\vspace*{20.2cm}
\nn Fig.~4: The cross sections for $hh$ production in the continuum for $\tb=1.5$:
$\ee \ra hhZ$ at a c.m.~energy of $ \sqrt{s}=500$ GeV (a) and $W_L W_L \ra
hh$ at $\sqrt{s}=1.5$ TeV (b).
\newpage

\vspace*{11.2cm}
\nn Fig.~5: The areas of the $[M_A, \tb$] plane in which the Higgs self--couplings
$\lambda_{Hhh}$  and $\lambda_{hhh}$ could eventually be accessible by
experiment at $\sqrt{s}=1.5$ TeV [see text for further discussions]. 

\newpage

\setcounter{footnote}{0}
\setcounter{figure}{0}
\setcounter{equation}{0}

\def\thefootnote{\fnsymbol{footnote}}

\begin{center}

{\large\sc {\bf  Loop Induced Higgs Boson Pair Production}}

\vspace{.5cm}

{\sc A.~Djouadi, V. Driesen and C. J\"unger}

\vspace{.5mm}

{\small Institut f\"ur Theoretische Physik, Universit\"at Karlsruhe, 
D--76128 Karlsruhe, Germany.} 

\end{center}

\begin{abstract}
\nn We discuss the loop induced production of Higgs boson pairs at
high--energy $\ee$ colliders, both in the Standard Model and in its 
minimal supersymmetric extension. The cross sections are rather small, 
but these processes could be visible with high-enough luminosities 
and if longitudinal polarization is available. 
\end{abstract}

\def\thefootnote{\arabic{footnote}}

\bigskip

\nn {\bf 1. Introduction}

\bigskip

If the genuine supersymmetric particles were too heavy to be
kinematically accessible in collider experiments, the only way to
distinguish between the Standard Model (SM) and the lightest Higgs boson
of its minimal extension (MSSM) in the decoupling limit [where all the
other MSSM Higgs bosons are heavy, and the lightest Higgs boson $h$ has
exactly the same properties [1] as the SM Higgs boson except that its mass
is restricted to be smaller than $M_h \lsim 140$ GeV], is to search for
loop induced contributions of the supersymmetric particles, which could
give rise to sizeable deviations from the predictions of the SM. Well
known examples of this loop induced processes are the $\gamma \gamma$
widths of the Higgs particles [2] or the process $\ee \ra Z$+Higgs
which in the MSSM receive extra contributions from supersymmetric
gaugino and sfermion loops [3]. \s 

Another type of such discriminating processes is the pair production of
Higgs bosons which will be analyzed here. In the SM, where it has been
first discussed in Ref.[4], the process $\ee \ra H^0H^0$ is
mediated only by $W$ and $Z$ boson loops, Fig.1a, while in the Minimal
Supersymmetric extension, additional contributions to the corresponding
process $\ee \ra hh$ will originate from chargino, neutralino, selectron
and sneutrino loops, as well as loops built up by the associated $A$ and
$H^{\pm}$ bosons; Fig.1b. The cross sections for these two processes
[as well as for the production of the heavy MSSM Higgs bosons, 
$\ee \ra HH, AA$ and $hH$] have been derived in [5] and here we 
will summarize the results. 

\bigskip

\nn {\bf 2. SM Higgs Pair Production}

\bigskip

In the SM, non--zero contributions to the process $\ee \ra H^0 
H^0$ can only come from one--loop diagrams, in the limit of vanishing electron 
mass. Among these, the diagrams involving the one--loop $H\ee$ vertex 
[because $m_e \simeq 0$] and those with $\gamma$ and $Z$ boson $s$--channel 
exchanges [because of CP invariance] give zero--contribution; additional 
contributions from vertex diagrams involving the quartic $WWH^0H^0/ZZH^0H^0$ 
couplings are proportional $m_e$ and also negligible. The only contribution 
to Higgs pair production in the SM will therefore come from $W$ and $Z$ box 
diagrams, Fig.1a. The expressions of the cross sections, allowing
for longitudinal polarization of the initial beams are given in [5].\newpage

\begin{figure}[htbp]
\vspace*{5cm} 
\nn {\small Fig.~1:
Feynman diagrams contributing to the Higgs boson pair production process in
$\ee$ collisions in the SM (a) and in the MSSM (b). }
\end{figure}

\bigskip

The cross sections are shown in Fig.2 as a function of the Higgs boson
mass for two center--of--mass energies, $\sqrt{s}=500$ GeV and 1.5 TeV.
Except when approaching the $2M_H$ threshold [and the small dip near the
$WW$ threshold], the cross sections are practically constant for a given
value of the c.m. energy, and amount to $\sigma \sim 0.2$ fb at
$\sqrt{s}=500$ GeV in the unpolarized case. The decrease of the cross
sections with increasing center--of--mass energy is very mild: at $
\sqrt{s}=1.5$ TeV, the cross section is still at the level of $\sigma \sim
0.15$ fb for Higgs boson masses less than $M_H \lsim 350$ GeV. \s

With left--handed polarized electrons, the cross
section $e_L^- e^+ \ra H^0 H^0$ is larger by a factor of two, while for
left--handed electrons and right--handed positrons, the cross section
$e_L^- e_R^+ \ra H^0 H^0$ is larger by a factor of four,  compared to
the unpolarized case. Therefore,
the availability of longitudinal polarization of the initial beams is
very important. 
With integrated luminosities of the order of $\int {\cal L} \sim 100$
fb$^{-1}$ which are expected to be available for future high--energy
linear colliders, one could expect a few hundred events in the course of
a few years, if both initial beams can be longitudinally polarized. \s

For $M_H \lsim 140$ GeV, the signal will mainly
consist of four $b$ quarks in the final state, $\ee \ra H^0 H^0 \ra b
\bar{b} b\bar{b}$, since the dominant decay mode of the Higgs boson
in this mass range is $H^0 \ra b\bar{b}$. This calls for very efficient
$\mu$--vertex detectors to tag the $b$ jets. Since these rare
events will be searched for only after the discovery of the Higgs boson
in the main production processes [5], $M_H$ will be 
precisely known and the two mass constraints $m(b\bar{b})=M_H$, together 
with the large number of final $b$ quarks, give a reasonable
hope to experimentally isolate the signals despite of the low rates. 
For $M_H \gsim 140$ GeV, since $H^0 \ra W^+W^-$ 
and $H^0 \ra ZZ$ will be the dominant decay modes of the Higgs boson, the 
signals will consist of four gauge bosons in the final state, 
$\ee \ra H^0 H^0 \ra VVVV$, leading to eight final fermions. These rather 
spectacular events should also help to experimentally isolate the signal. \\

\begin{figure}[htbp]
\vspace*{7cm} 
\nn {\small Fig.~2: The cross sections for, $\ee \ra H^0 H^0$, as a function of 
$M_H$ for $\sqrt{s}=500$ GeV [dashed lines] and $\sqrt{s}=1.5$ TeV [solid 
lines]. The lower, middle and upper curves correspond to the cross sections 
with unpolarized, $e_L^-$ and $e_R^+ e_L^-$ beams respectively.}
\end{figure}

\smallskip

\nn {\bf 3. MSSM Higgs Pair Production}

\bigskip

For the pair production of the light CP--even Higgs boson of
the MSSM, $\ee \ra hh$, several additional diagrams will
contribute to the process; Fig.1b. Besides the $W$ and $Z$ boson box
diagrams, one has the box diagrams with the exchange of the pseudoscalar
and the charged Higgs bosons, $A$ and $H^\pm$ and the box diagrams built 
up by chargino/sneutrino and neutralino/selectron loops. The analytical 
expressions of the cross sections, allowing for longitudinal polarization 
of the initial beams are also given in [5]. \s

In Fig.3, we show the cross section for the process $\ee {\ra} hh$ as a
function of $M_h$ for two c.m. energies $\sqrt{s}= 500 $ GeV
and 1.5 TeV and two values of $\tb=1.5$ and 50. The solid
lines are for the full cross sections, while the dashed lines are for
the cross sections without the SUSY contributions. To include the latter
we have chosen the parameters $M_2=-\mu=150$ GeV, while the common
slepton and squark masses are taken to be $M_L=300$ GeV and $M_S=500$
GeV; the parameter $A_t$ and $A_b$ are set to zero. Only the unpolarized
cross sections are discussed: as mentioned previously, they are simply
increased by a factor of 2(4) when the initial beam(s) are
longitudinally polarized. \s 

\begin{figure}[htbp]
\vspace*{19cm}
\nn {\small Fig.~3: The cross sections for 
$\ee \ra hh$ in the MSSM, as a function of
$M_h$ for $\sqrt{s}=500$ GeV and
$\sqrt{s}=1.5$ TeV and for $\tb=1.5$ and $50$.
The solid curves correspond to the full cross sections, while the dashed
curves correspond to the cross sections without the SUSY contributions.} 
\end{figure}

Let us first discuss the case where the supersymmetric contributions are 
not included, for small $\tb$ the cross section is of the same order as 
the SM cross section and does not strongly depend on  $M_h$ especially at 
very high--energies. Although the $WWh/Zhh$ couplings
are suppressed by $\sin (\beta-\alpha)$ factors, the suppression is not
very strong and the $W/Z$ box contributions are not much smaller than in
the SM; the diagrams where $A/H^\pm$ are exchanged will give
compensating contributions since the $hAZ/hH^\pm W$ couplings are
proportional to the complementary factor $\cos(\beta- \alpha)$. As
in the SM case, the cross sections slightly decrease with increasing
energy. 
For large $\tb$ values, the factors $\sin /\cos(\beta-\alpha)$ vary
widely when $M_h$ is varied. For small $M_h$, the factor $\sin(\beta-\alpha)
\ra 0$, and the contribution of the diagrams with $A/H^\pm$
exchange dominates. The latter contribution decreases with increasing $M_h$
[i.e. with decreasing $\cos(\beta-\alpha)]$, until the decoupling limit is 
reached for $M_h \simeq 110$ GeV. In this case, the factor $\sin(\beta-\alpha)
\ra 1$ and the $W/Z$ boson loops are not suppressed anymore; one then 
obtains the SM cross section. \s

The contributions of the chargino/selectron and neutralino/sneutrino
loops lead to a destructive interference. At high--energies, the
supersymmetric boxes practically do not contribute; but at low energies,
and especially below the decoupling limit, the SUSY contributions can
be of the order of $\sim 10\%$. We have scanned the SUSY parameter
space, and the maximum contribution of the SUSY loops that we have found was
about $\sim -15\%$. In the decoupling limit, the SUSY contributions are,
at most, of the order of a  few percent. Because of the rather 
low production rates, it will therefore be difficult to experimentally 
see this effect.

\bigskip

\nn {\bf 4. Summary}

\bigskip

We have discussed the one--loop induced production of Higgs boson pairs at
future high--energy $\ee$ colliders  in the SM and the MSSM. 
In the SM, the unpolarized cross
section is rather small, of the order $0.1$--$0.2$ fb. The longitudinal
polarization of both the $e^-$ and $e^+$ beams will increase the cross
section by a factor of 4. With integrated luminosities $\int {\cal L} \gsim 
100$ fb$^{-1}$ as expected to be the case for future high--energy
linear colliders, one could expect a few hundred events in the course of
a few years if longitudinal polarization is available. The final states 
are rather clean, giving a reasonable hope to isolate the signals 
experimentally. 
In the MSSM, additional contributions to the processes $\ee \ra hh$
come from chargino/ neutralino and slepton loops.
For $hh$ production, the contributions of the supersymmetric loops are
in general rather small, being of the order of a few percent; the cross
sections are therefore of the same order as in the SM. 
For the processes
involving heavy Higgs bosons, the cross sections are even smaller than
for $\ee \ra hh$, and the signals will be hard to be detected 
experimentally. 

\bigskip

\nn {\bf References}

\bigskip

\nn [1] For reviews see, J. Gunion, H. Haber, G. Kane and S. Dawson, 
{\it The Higgs Hunter's Guide}, Addison--Wesley, Reading 1990; D. Haidt,
P. Zerwas et al., Proceedings of the workshop {\it $\ee$ collisions at 500 
GeV: The Physics Potential}, Munich--Annecy--Hamburg, DESY 92--123A+C, 
P.M. Zerwas (ed.); A. Djouadi, Int. J. Mod. Phys. A10 (1995)1. 
\s

\nn [2] J. Ellis, M.K. Gaillard and D.V. Nanopoulos, Nucl. Phys. 
B106 (1976) 292; for recent discussions of the Higgs--$\gamma \gamma$ 
coupling in the MSSM see: 
G.L. Kane, G.D. Kribs, S.P. Martin and J.D. Wells, Phys. Rev. D53 (1996) 213; 
B. Kileng, P. Osland and P.N. Pandita, NORDITA-95-48-P (1995). 
\s

\nn [3] For a discussion of the radiative corrections to
the process $\ee \ra hZ$, see V. Driesen, W. Hollik and J. Rosiek, 
Report KA-TP-16-1995, hep-ph/9512441. 
\s

\nn [4] K. Gaemers and F. Hoogeveen, Z. Phys. C26 (1984) 249.
\s

\nn [5] A. Djouadi, V. Driesen and C. J\"unger, hep-ph/9602341,
Phys. Rev. D (in press).

\newpage

\setcounter{footnote}{0}
\setcounter{figure}{0}
\setcounter{equation}{0} 

\renewcommand{\thefootnote}{\fnsymbol{footnote}}
\newcommand{\Str}{\mbox{Str}}
\newcommand{\tr}{\mbox{tr}}
\newcommand{\Log}{\mbox{Log}}
\newcommand{\GeV}{\mbox{GeV}}
\newcommand{\TeV}{\mbox{TeV}}
\newcommand{\simlt}{\stackrel{<}{{}_\sim}}
\newcommand{\simgt}{\stackrel{>}{{}_\sim}}
\newcommand{\ol}{\overline}
\newcommand{\re}{\mbox{Re}}
\newcommand{\im}{\mbox{Im}}
\newcommand{\I}{\mbox{I}}
\renewcommand{\i}{\mbox{\small i}}
\newcommand{\e}{\mbox{e}}
\newcommand{\N}{{\cal N}}
\renewcommand{\H}{{\cal H}}

\begin{center}

{\large\sc {\bf Search for the Higgs Bosons of the NMSSM at Linear Colliders}}

\vspace{.7cm}

{\sc B.~R.~Kim$^1$, G.~Kreyerhoff$^1$ and S.~K.~Oh$^2$} 

\vspace{.5cm}

{\small

$^1$ III. Physikalisches Institut A, RWTH Aachen, D-52056 Aachen, Germany\\
\vspace{0.2cm}

$^2$ Department of Physics, Kon-Kuk University, Seoul, Korea.   }

\end{center}

\begin{abstract}
We show that at least one of the Higgs bosons of the Next to Minimal 
Supersymmetric
Standard Model can be detected at future Linear Colliders of 500, 1000 and 2000 GeV
c.m. energies.
\end{abstract}

\subsection*{1. Introduction}

The Next to Minimal Supersymmetric Standard Model (NMSSM) [1--3]
is a minimal extension to the Minimal Supersymmetric Standard Model \cite{mssm}. The 
NMSSM provides the most economic solution of the so called $\mu$-problem
of the MSSM by introducing an additional Higgs singlet superfield
 $\N=(N,\psi_N,F_N)$ with a Higgs singlet $N$, a higgsino singlet $\psi_N$
and an auxiliary field $F_N$. Together with the two Higgs doublets superfields
 $\H_{1,2} = (H_{1,2},\psi_{1,2},F_{1,2})$ the superpotential of the
NMSSM is given by
\beq
W=\lambda \H^T_1 \epsilon \H_2 \N - \frac{1}{3} k \N^3
\label{superpot}
\eeq
The soft breaking part of the Higgs sector is given by
\beq
V_{soft} = -\lambda A_\lambda H_1^T\epsilon H_2 N - \frac{1}{3}k A_k N^3 + h.c.
\label{softbrk}
\eeq
where $A_\lambda$ and $A_k$ are soft breaking mass parameters.\par
$H_1, H_2$ and $N$ develop vacuum expectation values $v_1, v_2$ and $x$ 
respectively. The NMSSM contains three scalar Higgs bosons $S_1, S_2$ and
 $S_3$ with masses $m_{S_1}\le m_{S_2}\le m_{S_3}$, two pseudoscalar
Higgs bosons $P_1$ and $P_2$ with masses $m_{P_1}\le m_{P_2}$ and a 
charged one $H^+$ with mass $m_C$.\par
The Higgs sector has 6 free parameters $\lambda, k, \tan\beta=v_1/v_2, x, 
A_\lambda$ and $A_k$.

\subsection*{2. Constraints on parameters}

A remarkable result of the MSSM is the tree level bound of 
 $m_{S_1}\le m_Z\cos 2\beta$. This is due to the fact that all quartic terms
have gauge coupling constants. In case of the NMSSM there is a quartic term with
the coupling constant $\lambda$. It turns out that the upper bound of
 $\lambda$ may be relevant for $m_{S_1}$. An effective way of determining this
bound is RG-analysis [1--8].\par
The one loop RG-equation of $\lambda$ is coupled with that of $k$ and $h_t$,
the Yukawa coupling constant of the top quark (neglecting other quarks) and
is given by
\beq
\frac{d\lambda}{dt}&=&\frac{1}{8\pi^2}\left(k^2 + 2\lambda^2 +
\frac{3}{2}h_t^2 + \frac{3}{2}g_2^2 - \frac{1}{2}g_1^2\right)\lambda\nn\\
\frac{dk}{dt} &=& \frac{3}{8\pi^2}\left(k^2+\lambda^2\right)k \\
\frac{dh_t}{dt} &=& \frac{1}{8\pi^2}\left( \frac{1}{2} \lambda^2 +
3h_t^2 - \frac{8}{3}g_3^2 - \frac{3}{2} g_2^2 - \frac{13}{18}g_1^2\right)h_t\nn
\label{rgg}
\eeq
where $t=\ln \mu$ and $\mu$ being the renormalization scale. By demanding no
Landau pole up to the GUT scale one can determine from eq.~(\ref{rgg}) the
upper bound of $\lambda$ and $k$ and the lower bound of $\tan\beta$ at the
electroweak scale. We plot our results in Fig.~1 
(Fig.~2) for $m_t=175\;\GeV\;\;(190\;\GeV)$. They show that
 $\lambda_{max}$ decreases with increasing $k$\footnote{We obtain $\lambda_{max}
= 0.64 - 0.74$ for $m_t = 175 - 190\;\; \GeV$.}. The lower bound of $\tan\beta$
is about $1.24$ for $m_t = 175\GeV$ and $2.6$ for $m_t = 190\GeV$. For 
 $\tan\beta\simgt 3$ $\lambda_{max}$ is almost independent on $\tan\beta$. The
upper bound of $k$ is about 0.7.

\subsection*{3. Mass upper bounds}

The tree level bound of $m_{S_1}$ is given by \cite{nmssm3}
\beq
m^2_{S_1} \le m_Z^2\left(\cos^2 2\beta + \frac{2\lambda^2}{g_1^2+g_2^2}
\sin^2 2\beta\right) = m_{S_1^{max}}^2
\label{tree-upper}
\eeq
The upper bound of $m_{S_2}$ and $m_{S_3}$ can be expressed in terms of
 $m_{S_1^{max}}$ and $m_{S_1}$
\beq
 m_{S_2}^2 &\le& m_{S_2^{max}}^2 = \frac{m_{S_1^{max}}^2 
 - R_1^2 m_{S_1}^2}{1-R_1^2}\nn\\
 m_{S_3}^2 &\le& m_{S_3^{max}}^2 = \frac{m_{S_1^{max}}^2 - (R_1^2 + R_2^2)m_{S_1}^2}
{1-(R_1^2+R_2^2)} 
\eeq
where $R_i = U_{i1} \cos\beta + U_{i2}\sin\beta$ and $U_{ij}$ is the $3\times 3$
orthogonal matrix which diagonalizes the scalar mass matrix.\par
 $R_1$ and $R_2$ satisfy the unitarity condition $0\le R_1^2 + r_2^2 \le 1$. 
The tree level upper bound (\ref{tree-upper}) yields $m_{S_1} \le m_Z$ for
 $\lambda^2\le  (g_1^2+g_2^2)/2 = (0.52)^2$ and $m_{S_1}\le \sqrt{2/(g_1^2+g_2^2)}
\lambda m_Z = 1.92\lambda m_Z$ for $\lambda^2 > (0.52)^2$. Using 
 $\lambda_{max} = 0.64 - 0.74$ from section 2 the tree level relation yields
 $m_{S_1}\le 113 \GeV - 131\GeV$.\par
As in the case of the MSSM the contributions of radiative corrections may change
this result considerably. Several groups calculated higher order contributions
to the mass matrices using the one loop effective potential and determined the
corrected upper bound \cite{nmssm7,rc1,rc2,rc3}.\par
The result in our notation \cite{nmssm7} is given by
\beq
m_{S_1}^2\le m_Z^2\left(\cos^2 2\beta + \frac{2\lambda^2}{g_1^2+g_2^2}
 \sin^2 2\beta\right) + \alpha \cos^2\beta  + \beta \sin 2\beta + 
\gamma \sin^2\beta
\eeq
with ($A_T = -A_t + \lambda x \cot\beta$)
\beq
\alpha &=& -\frac{1}{16\pi^2} {\left(\frac{\lambda x A_T}{v_1}\right)}^2 
 {\left( \frac{m_t}{m_{\tilde{t}}}\right)}^4\nn\\
\beta &=& \frac{3}{8\pi} \lambda x A_T {\left( 
\frac{m_t^2}{m_{\tilde{t}}v_1}\right)}^2
\left( 1+\frac{A_t A_T }{6 m_{\tilde{t}}^2}\right)\nn\\
\gamma &=& \frac{3}{8\pi} {\left(\frac{m_t^2}{v_1^2}\right)}^2 \left[
2\ln\frac{m_{\tilde{t}}^2}{m_t^2} - \frac{2 A_t A_T}{m_{\tilde{t}}^2} -
\frac{A_t^2 A_T^2}{6m_{\tilde{t}}^4}\right]
\eeq
In this result only top and stop contributions were taken into account.
We numerical calculated $m_{S_1^{max}}$ in the region $175\GeV \le m_t
\le 190\GeV $, $250\GeV \le x,A_\lambda,A_t, m_{\tilde{t}}\le 1000 \GeV$ and
 $2\le \tan\beta \le 20$ and obtained \cite{nmssm7}
\beq
120 \le m_{S_1^{max}} \le 156 \GeV
\eeq

\subsection*{4. Production of scalar Higgs bosons at $e^+ e^-$ Colliders}

The upper bound $m_{S_1}\le 120-156\GeV$ suggests that the accessible area
of the parameter space at LEP1 with $\sqrt{s}=m_Z$ might be very small.
Actually we showed that the existing LEP1 data do not exclude the existence
of $S_1$ with $m_{S_1}=0 \GeV$ \cite{kim}.\par
For colliders with $\sqrt{s}=500,1000$ or $2000\GeV$ the situation is different.
In this case the production cross section of one $S_i$ via the Higgsstrahlung
 $e^+e^-\to ZS_i$ with real $Z$ and $S_i$ is always possible as the collider energy
is larger than $E_T= 212-248\GeV$. $E_T$ is a kind of threshold energy and is
an important quantity of a model. \par
In this case it is possible to derive a lower bound for the production cross
section $\sigma_i$ of $S_i$ as a function of the collision energy only. This lower 
bound would give information about how far the model could be tested.\par
In order to derive the lower bound of $\sigma_i$ we consider the production
cross sections of $S_1,S_2,S_3$ via the Higgsstrahlungs process, denoted by
 $\sigma_1, \sigma_2, \sigma_3$, which can be expressed in terms of the 
standard model Higgs production cross section $\sigma_{SM}$ and $R_1$ and $R_2$
defined in section 2:
\beq 
\sigma_1(m_{S_1}) &=& R_1^2\sigma_{SM}(m_{S_1})\nn\\
\sigma_2(m_{S_2}) &=& R_2^2\sigma_{SM}(m_{S_2})\\
\sigma_3(m_{S_3}) &=& (1-R_1^2-R_2^2)\sigma_{SM}(m_{S_3})\nn
\eeq
A useful observation is that $\sigma_i(m_{S_i}^{max})\le \sigma_i(m_{S_i})$
which allows one to derive parameter independent lower limits on $\sigma_i$
as we will see in the following. \par
First we determine at a fixed set of $m_{S_1}, R_1, R_2$ the
cross sections $\sigma_1(R_1,R_2,m_{S_1})$, $\sigma_2(R_1,R_2,m_{S_2^{max}})$
and $\sigma_3(R_1,R_2,m_{S_3^{max}})$. Then we keep $R_1$ and $R_2$ fixed,
but vary $m_{S_1}$ from its minimum to its maximum value and determine the
quantity $\sigma(R_1,R_2)$ defined by
\beq
\sigma(R_1,R_2) = \min_{0\le m_{S_1}\le m_{S_1^{max}}} \left[
  \max(\sigma_1, \sigma_2, \sigma_3)\right]
\eeq
where $\sigma_1=\sigma_1(R_1,R_2,m_{S_1})$, $\sigma_2=\sigma_2(R_1,R_2,
 m_{S_2^{max}})$ and $\sigma_3=\sigma_3(R_1,R_2,m_{S_3^{max}})$. As a last
step we vary $R_1^2$ and $R_2^2$ from 0 to 1 with $R_1^2+R_2^2\le 1$ and
plot $\sigma(R_1,R_2)$ in the $R_1$-$R_2$-plane. It is plausible that
 $\sigma(R_1,R_2)=0$ for $\sqrt{s} < E_T = m_z + m_{S_1^{max}} = 
 212-248 \GeV$ in the entire $R_1$-$R_2$-plane. This is the case for
LEP2 with $\sqrt{s}\le 205\;\GeV$. Therefore this method does not give any results
for LEP2. For $\sqrt{s} > E_T$ $\sigma(R_1,R_2)$ never vanishes and the minimum of
 $\sigma(R_1,R_2)$ is a parameter independent lower limit
of one of the $\sigma_i$. This minimum is thus a characteristic quantity of the
model.\par 
In Fig.~3 we plotted $\sigma(R_1,R_2)$ for $\sqrt{s}=
 500\;\;\GeV$ and $m_{S_1^{max}}=145\;\;\GeV$. The minimum is about 16 fb. 
When the discovery limit is about 30 events, one would need a luminosity of about
 25 fb, which is a realistic one.
Fig.~4 (Fig.~5) shows $\sigma(R_1,R_2)$ for
 $\sqrt{s}=1000 (2000)\;\;\GeV$ with minimum cross section of
 4 fb (1 fb).
Fig.~6 shows the minimum of $\sigma(R_1,R_2)$ as a function of $\sqrt{s}$
and $m_{S_1^{max}}$ as a parameter. We see that the effect of $m_{S_1^{max}}$
on $\sigma_{min}$ is very big around $\sqrt{s}=300\;\;\GeV$, but rather small
for $\sqrt{s}\ge 500\;\;\GeV$. \par
Fig.~7 shows the tree level cross sections $\sigma_1,\sigma_2$ and
 $\sigma_3$ for an exemplary set of parameters with the contributions from 
(i) the Higgsstrahlungsprocess $e^+ e^- \to ZS_i\to b\ol{b}S_i$,
(ii) the process where $S_i$ is radiated off from $b$ or $\ol{b}$ and 
(iii) $e^+ e^- \to Z \to P_j S_i \to b\ol{b}S_i$, where $P_j(j=1,2)$ is
a pseudoscalar Higgs boson.\par
Fig.~8 shows the same as Fig.~7, but with one loop 
contributions via the effective potential. The higher order contribution is rather
important for the energy region around 150 GeV and decreases with $\sqrt{s}$. In 
this parameter region the dominant production is that of $S_2$ at $\sqrt{s}=500\;\; \GeV$
and is about 13 fb.\par
We conclude that the Higgs sector of the NMSSM can most probably be tested conclusively
at the future linear $e^+ e^-$-colliders with 500, 1000 or 2000 GeV c.m. energies.

\newpage
\vspace{.5cm}
\begin{figure}[t]
\vspace{7.5cm}
\caption{Upper bound for $\lambda$ as a funtion of $\tan\beta$ for $m_{top}=175\;\GeV$.}
\label{lambda175}
\end{figure}
\vspace{.5cm}
\begin{figure}[b]
\vspace{7.5cm}
\caption{Upper bound for $\lambda$ as a funtion of $\tan\beta$ for $m_{top}=190\;\GeV$.}
\label{lambda190}
\end{figure}
\clearpage
\begin{figure}[t]
\vspace{.5cm}
\vspace{7.5cm}
\caption{$\sigma(R_1,R_2)$ as definded in the text for $\protect\sqrt{s}=500\;\GeV$.}
\label{r1r2}
\end{figure}
\begin{figure}[b]
\vspace{.5cm}
\vspace{7.5cm}
\caption{$\sigma(R_1,R_2)$ for $\protect\sqrt{s}=1000\;\GeV$.}
\label{r1r2_1000}
\end{figure}
\clearpage
\begin{figure}[t]
\vspace{.5cm}
\vspace{7.5cm}
\caption{$\sigma(R_1,R_2)$ for $\protect\sqrt{s}=2000\;\GeV$.}
\label{r1r2_2000}
\end{figure}
\begin{figure}[b]
\vspace{.5cm}
\vspace{7.5cm}
\caption{Minimal value of $\sigma(R_1,R_2)$ as a function of $\protect\sqrt{s}$ for various values
of $m_{S_1^{max}}$}
\label{smin}
\end{figure}
\clearpage
\begin{figure}[t]
\vspace{.5cm}
\vspace{7.5cm}
\caption{Cross section for $e^+e^-\to Zb\ol{b}$ for $A_\lambda=220\;\GeV, A_k=160\;\GeV,
x=1000\;\GeV, \tan\beta=2, k=0.04, \lambda=0.12$. Masses and mixing angles have calculated 
from the tree level protential.}
\label{sigmatree}
\end{figure}
\begin{figure}[b]
\vspace{7.5cm}
\caption{The same as above, but with masses and mixing angles obtained from the one loop
effective potential. The top mass is 175 GeV, $m_{\tilde{t}_l}=m_{\tilde{t}_R}=1\;\TeV$ and
$A_t=0\;\GeV$.}
\label{sigmarc}
\end{figure}

\end{document}